\documentclass[aps,prd,showpacs, amssymb,superscriptaddress, notitlepage,floats,floatfix,nofootinbib]{revtex4-1}
\usepackage{graphicx}
\usepackage{amssymb}
\usepackage{amsmath}
\usepackage{amsfonts}
\usepackage{bm}
\usepackage{mathrsfs}
\usepackage[colorlinks=true]{hyperref}
\usepackage[all]{hypcap} 
\usepackage[utf8]{inputenc} 
\usepackage{mathptmx}
\usepackage{color}
\usepackage{multirow}
\usepackage{palatino}
\usepackage{float}
\usepackage{journals}
\usepackage{soul}
\usepackage{calrsfs}
\DeclareMathAlphabet{\pazocal}{OMS}{zplm}{m}{n}
\begin{document}
\title{On the Inverse Spectrum Problem of Neutron Stars}
\author{Sebastian H. V\"olkel}
\email{sebastian.voelkel@uni-tuebingen.de}
\affiliation{Theoretical Astrophysics, IAAT, University of T\"ubingen, Germany}
\author{Kostas D. Kokkotas}
\affiliation{Theoretical Astrophysics, IAAT, University of T\"ubingen, Germany}
\date{\today}
\begin{abstract}
In this work we revisit axial perturbations of spherically symmetric and non-rotating neutron stars. Although it has been object of many studies, it still offers new insights that are of potential interest for more realistic scenarios or in the study exotic compact objects, which have drawn much attention recently. By using WKB theory, we first derive a new Bohr-Sommerfeld rule that allows to investigate the quasi-normal mode spectrum and address the inverse spectrum problem. 
The pure analytical treatment of the wave equation is rather involved, because it requires the solution of the TOV equations and the non-trivial tortoise coordinate transformation depending on the underlying space-time. Therefore we provide an easy way to construct potentials that simplifies the analytical treatment, but still captures the relevant physics. The approximated potential can be used for calculations of the axial perturbation spectrum. These results are also useful in the treatment of the inverse problem. We demonstrate this by reconstructing the time-time component of the metric throughout the star and constraining the equation of state in the central region. Our method also provides an analytical explanation of the empirically known asteroseismology relation that connects the fundamental QNM and radius of a neutron star with its compactness.
\end{abstract}
\maketitle
%
\section{Introduction}
%
Neutron stars are among the most versatile and extreme objects being studied in modern physics. These ultra compact stars confront and join different fields of modern physics, ranging from nuclear and particle physics in their center, up to gravitational and mathematical physics to describe their space-time. With the groundbreaking rise of observational gravitational wave physics using sensitive gravitational wave detectors as LIGO and Virgo \cite{LIGO1,PhysRevLett.116.241103,PhysRevLett.118.221101,PhysRevLett.119.141101,PhysRevLett.119.161101}, there are novel ways to extract information about these objects in addition to electromagnetic observations. The characteristic oscillations of neutron stars, which are going to be measured with increasing precision in the future, can be used in asteroseismology to reconstruct their properties. Along with their typical masses and radii, there is even more interestingly the prospect to constrain the still unknown equation of state in their central regions. It is obvious that the full problem is extremely complicated and not traceable with analytical methods only. Rotation and the absence of axial symmetry lead to complicated coupled system of equations and non-linear effects call for state of art numerical relativity techniques to simulate oscillating neutron stars, which can currently only be done on short time scales. The unique reconstruction of the underlying equation of state by matching simulations to current and future observations is one of the biggest open problems in neutron star physics, but in reach with current and future gravitational wave detectors.
\par
In the present work we follow a semi-analytic approach, which does not aim to solve the full problem, but to focus and enhance our understanding of the theoretical minimum. The minimum, but still physically reasonable setup, is to study linear perturbations of spherically symmetric and non-rotating neutron stars within general relativity. In this case one finds two types of perturbations, the so-called axial and polar ones. The axial perturbations decouple from the fluid perturbations, which simplifies the analytical treatment, and are therefore the ideal starting point for such studies.
\par
The first study of non-radial oscillations of neutron stars within general relativity traces back to the seminal work of Thorne and Campolattaro in 1967 \cite{1967ApJ...149..591T}. Since then, extensive efforts have been made to describe and understand these oscillations, among them are pioneering works by Lindblom and Detweiler \cite{1983ApJS...53...73L}, as well as Chandrasekhar and Ferrari \cite{1991RSPSA.432..247C}. Another work demonstrated that the coupled oscillations of fluid perturbations inside a star with space-time perturbations itself, can produce new types of oscillation modes \cite{1986GReGr..18..913K}. Studying such simplified models is rewarding, since they allow for a better understanding of the more complicated problems. Different types of space-time oscillations have been found in various works \cite{1992MNRAS.255..119K,1993PhRvD..48.3467L,1991RSPSA.434..449C,1994MNRAS.268.1015K}. The response of neutron stars to small perturbations introduced by test particles or ingoing radiation has first been studied in \cite{Kokkotas:1995av,PhysRevD.60.024004,2000PhRvD..62j7504F}. The new kind of modes, named w-modes, that appear in ultra compact systems are exposed to renewed interest in the study exotic compact objects, which could be alternatives to black holes and potentially indicate quantum gravitational effects on the horizon scale. We refer to \cite{1999LRR.....2....2K,1999CQGra..16R.159N,2009CQGra..26p3001B} for classical reviews and more recently for exotic compact objects to \cite{Cardoso:2017njb}. Relating axial and polar perturbations to the unknown neutron star equation of state has also first been studied systematically in the 90's \cite{1996PhRvL..77.4134A,Andersson:1997rn,1999MNRAS.310..797B,Kokkotas:1999mn}, which is related to the field of gravitational wave asteroseismology. A lot of work has also been done in a series of papers by Lindblom \cite{1992ApJ...398..569L,2012PhRvD..86h4003L,Lindblom:2013kra,Lindblom:2014sha,Lindblom:2018ntw}, who studied the inverse problem for the stellar structure of neutron stars. His approach is not based on the QNM spectrum itself, but on the assumption that the mass radius curve for many neutron stars will be estimated from future observations.
\par
Axial perturbations of spherically symmetric and non-rotating compact objects, although describing strongly idealized situations, are still among the most commonly studied systems. The reason being that they allow the application of semi-analytic methods that deepen our understanding of the theoretical minimum of the corresponding system. As a result, this can potentially be useful to highlight interesting new physics that are then studied with more refined physics. This is the case for new types of exotic compact objects, such as gravastars, boson stars, and wormholes, or as well if one seeks to study the gravitational wave dynamics in alternative theories of gravity, which impose additional physical and technical difficulties. See \cite{2001gr.qc.....9035M,2004CQGra..21.1135V,PhysRevD.76.024016,2016PhRvD..94h4031C,Price:2017cjr,PhysRevD.96.064033,Barcelo2017,Konoplya:2005et,Bueno:2017hyj,PhysRevD.95.084034,Berthiere:2017tms} for some models that fall into those categories.
\par
This paper is structured as follows.  In Sec. \ref{Bohr-Sommerfeld} we derive a new Bohr-Sommerfeld rule that will fit to our type of potential. Its implications for the axial quasi-normal mode spectrum is then discussed in Sec. \ref{Direct Spectrum Problem}, before we use it to address the inverse spectrum problem in Sec. \ref{Inverse Problem}. The results we obtain from applying the previously derived methods to constant density stars and polytropes are found in Sec. \ref{Results}. These findings are then discussed in Sec. \ref{Discussion}. Our conclusions are presented in Sec. \ref{Conclusions}. The appendix \ref{Appendix} shows the details for some of the calculations carried out in the main paper. Throughout this work we set $G=c=\hbar=2m=1$.
%
\section{The Bohr-Sommerfeld Rule}\label{Bohr-Sommerfeld}
%
The classical Bohr-Sommerfeld rule is a simple but powerful tool for analytic or semi-analytic calculations of the spectrum of bound states $E_n$ in a given potential well $V(x)$, which are described by two classical turning points and appear in the one-dimensional wave equation. Its simple form is given by
\begin{align}\label{classical_BS}
\int_{x_0}^{x_1} \sqrt{E_n - V(x)} \text{d}x = \pi \left(n+\frac{1}{2} \right),
\end{align}
with $n \in \mathbb{N}$ and $(x_0, x_1)$ being the classical turning points defined by $E_n = V(x)$. It can formally be derived by using the WKB method and as well be extended to different types of potentials with more turning points, see \cite{1978amms.book.....B,PhysRevA.38.1747,1991PhLA..157..185P,2013waap.book.....K} for some overview.
As it is well known in the literature \cite{1999LRR.....2....2K,1999CQGra..16R.159N}, the study of axial perturbations of spherically symmetric and non-rotating stars leads to the one-dimensional wave equation
\begin{align}\label{wave equation}
\frac{\text{d}^2}{\text{d}{r^{*}}^2} \Psi({r^{*}}) + \left[\omega_n^2 - V(r) \right]\Psi({r^{*}}) = 0,
\end{align}
where $\omega_n^2 \equiv E_n$ are the corresponding eigenvalues, the so-called quasi-normal modes (QNMs). $V(r)$ is an effective potential being characteristic for the underlying type of object and given by
\begin{align}
V(r) = g_{00}(r) \left(\frac{L}{r^2}  + 4 \pi \left(\rho(r) - P(r)\right)- \frac{6m(r)}{r^3}\right),
\end{align}
with density $\rho(r)$, pressure $P(r)$, integrated mass $m(r)$ and metric function $g_{00}(r) = \exp(2 \nu(r)))$. The factor $L\equiv l(l+1)$ originates from the decomposition of the metric perturbations into tensor spherical harmonics. In the study of gravitational perturbations, the relevant radial coordinate, in which the wave equation takes the simple form shown in eq. \eqref{wave equation}, is called the tortoise coordinate and it is connected with the usual Schwarzschild coordinate $r$ via the relation
\begin{align}\label{tortoisee}
r^{*}(r)  \equiv \int_{}^{r} \sqrt{\frac{g_{11}(r^{\prime})}{g_{00}(r^{\prime})}}\text{d} r^\prime,
\end{align}
where the transformation from $r$ to $r^*$ depends on the metric functions $g_{00}$ and $g_{11}$. We expand on this relation in Sec. \ref{Direct Spectrum Problem}. The typical axial perturbation potential for neutron stars admits only one classical turning point and the spectrum is complex valued. As a consequence, the classical Bohr-Sommerfeld rule in eq. \eqref{classical_BS} is not applicable. The QNM spectrum of black holes, which is described by a two turning point potential barrier in the non-rotating case, has been investigated with a generalized Bohr-Sommerfeld rule involving integration on the complex plane \cite{1991CQGra...8.2217K,0264-9381-10-4-010}. Note that in the case of ultra compact stars and exotic compact objects, three or four turning points potentials are possible. Such potentials admit a new family of semi-trapped modes \cite{1991RSPSA.434..449C,1994MNRAS.268.1015K}, whose real part can be described to good accuracy with the classical Bohr-Sommerfeld rule, while a more complete generalization has been studied \cite{Festuccia:2009:1936-6612:221,2009PhRvD..80l4047P,2014PhRvD..90d4069C,paper1,paper2,paper5}. 
\par
\begin{figure}[H]
\centering
	\includegraphics[width=0.5\linewidth]{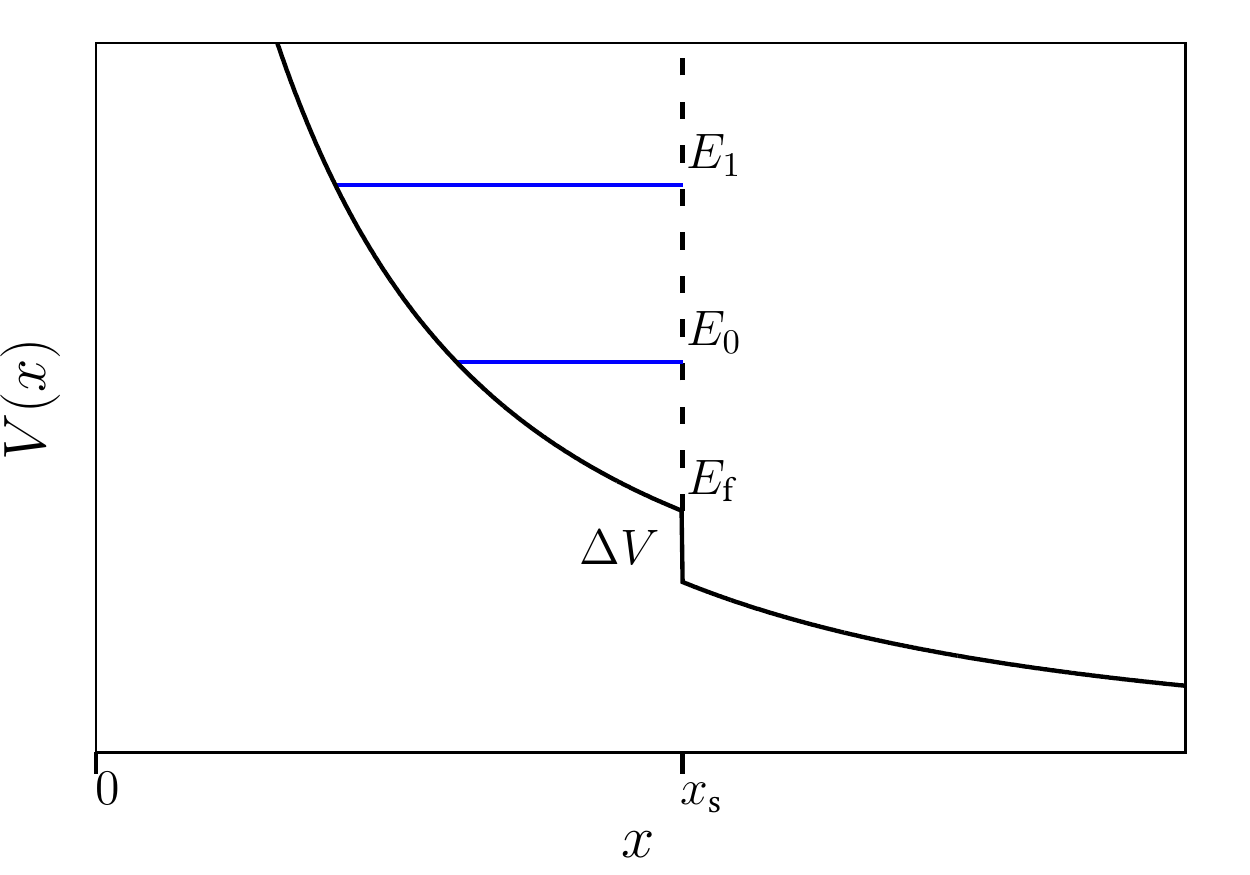}
	\caption{The type of axial perturbation potentials being studied in this work. It appears for simple neutron star models. The potential admits a discontinuity $\Delta V$ at the surface. As we demonstrate in Sec. \ref{Outline of the Derivation}, the real part of $E_n$ can effectively be approximated as bound states inside the star. \label{potential}}
\end{figure}

To determine any type of QNM spectrum, appropriate boundary conditions are required emanating from the specific physics of the problem under study. The common ones used for neutron stars are regular solutions at the center and purely outgoing waves at infinity. The perturbation potential of a neutron star with mass $M$ and radius $R$ with $R/M \geq 3$ can qualitatively be described by a purely repulsive potential that diverges at the center of the star and falls off proportional to $\sim 1/r^2$ far away from the object. We show such a potential in Fig. \ref{potential}, where $x$ acts as tortoise coordinate, for which the divergence is shifted to $x=0$.
For stars with constant density or solid crust, $V(x)$ has a discontinuity at the surface, which plays an important role for the QNM spectrum. The role of discontinuities in the potential has been investigated and pointed out in different works that are related to the QNM spectra of compact stars and black holes \cite{1996PhRvD..53.4397N,1999JMP....40..980N,PhysRevD.83.064012} and has implications for the significance of QNMs as well as the completeness of the obtained solutions to describe arbitrary perturbations.
\par
In the following subsection we outline the derivation for the new Bohr-Sommerfeld rule. The interested reader can find the full derivation in the appendix \ref{AGBS}.
\subsection{Outline of the Derivation}\label{Outline of the Derivation}
The derivation of the new Bohr-Sommerfeld rule can be summarized as follows. The typical potential shown in Fig. \ref{potential} is split up into one internal region $x \leq x_\text{s}$ and one external region $x_\text{s}< x$. We write the standard WKB solution for $\Psi(x)$ separately in the two regions and demand that the Wronskian at $x=x_\text{s}$ vanishes. This condition, together with the QNM condition of purely outgoing waves and regularity at $x=0$ determines discrete eigenvalues, if the potential is discontinuous at $x_\text{s}$. Using this procedure yields the complex valued equation
\begin{align}\label{meq_GBS}
\int_{x_{0}}^{x_\text{s}} \sqrt{E_n-V_1(u)} \text{d}u  
=\pi \left(n+\frac{3}{4}\right) + i \tanh^{-1}\left(\frac{\sqrt{E_n-V_1(x_\text{s})}}{\sqrt{E_n-V_4(x_\text{s})}} \right),
\end{align}
where $V_1(x_\text{s})$ and $V_4(x_\text{s})$ are the values of the potential on each side of the discontinuity and $n = (0,1,2,\dots)$. It is well known for neutron stars that high eigenvalues grow linearly in the real part, but seem to saturate for the imaginary part. Under this condition one can further simplify the above relation. Assuming a much larger real than imaginary part for $E_n$, as well as $\Delta V << E_n$, one can expand eq. \eqref{meq_GBS} into one equation for the real part $E_{0n}$ and a second one for the imaginary part $E_{1n}$
\begin{align}\label{new_BS}
\int_{x_{0}}^{x_\text{s}} \sqrt{E_{0n}-V_1(u)} \text{d}u = \pi \left(n+\frac{3}{4}\right),
\qquad \qquad
E_{1n} = \left(\int_{x_{0, {1}}}^{x_\text{s}}\frac{1}{\sqrt{E_{0n}-V(u)}} \text{d}u \right)^{-1}  \ln\left(\frac{4 (E_{0n}-V_4(x_\text{s}))}{\Delta V} \right).
\end{align}
The relation for the real part looks similar to the classical Bohr-Sommerfeld rule for potential wells, while the second equation predicts a logarithmic scaling for the imaginary part.
In contrast to the classical Bohr-Sommerfeld rule, our type of potential admits only one classical turning point $x_{0}(E_{0n})$ and the upper limit of integration is constant, which followed from the purely outgoing QNM boundary condition. The similarity between both Bohr-Sommerfeld rules is in contrast to the underlying type of potential. This result shows that the real part of the QNMs $E_n = \omega_n^2$ are closely related to the bound states of a potential well contained inside the star, with an infinitely high barrier at the surface.
%
\section{Direct Spectrum Problem}\label{Direct Spectrum Problem}
%
In this section we investigate the new modified Bohr-Sommerfeld rule eq. \eqref{new_BS} and its implications for the axial QNM spectrum of neutron stars.
\subsection{General Implications}\label{General Implications}
The new Bohr-Sommerfeld eq. \eqref{new_BS} rule allows to draw some straight forward conclusions about the QNM spectrum. Since the upper turning point always coincides with the surface of the star, and it is known that the potential diverges in the center, the domain of integration for high eigenvalues is confined on a range given by
\begin{align}
\pazocal{R} \equiv r^{*}(R)- r^{*}(0),
\end{align}
which we call the tortoise radius of the star. For high eigenvalues one can therefore expect a potential independent scaling of the kind \footnote{If one neglects terms of the order $\sim V(x)/E_n$ and higher.}
\begin{align}\label{E_asympt}
E_{0n} \sim \frac{\pi^2 n^2 }{{\pazocal{R}}^2}.
\end{align}
This spectrum looks qualitatively like the one of the infinite box potential with width $\pazocal{R}$. It is also straightforward to see that the spacing of the real part of $\omega_n = \sqrt{E_n}$ should become constant and scale like
\begin{align}\label{Delta_omega}
\Delta \omega \equiv \frac{\pi}{\pazocal{R}}.
\end{align}
Exactly this scaling has been found for the asymptotic high frequency behavior of axial and polar modes in \cite{PhysRevD.83.064012} by using an alternative WKB analysis. In the same work, it was also shown how discontinuities in the derivatives of the potential can affect the QNM spectrum. In contrast to our result, no Bohr-Sommerfeld rule has been reported, but a direct approximation for high eigenvalues. We want to emphasize that the modified Bohr-Sommerfeld rule eq. \eqref{new_BS} can also be applied to lower eigenvalues. In addition, it is very versatile for analytic studies related to the inverse spectrum problem, because it can be inverted to constrain the potential from the QNM spectrum. 
\par
The appearance of $\pazocal{R}$ instead of $R$ calls for two comments. First, for less compact stars, both scale similar and the difference is only logarithmic. Increasing $R$ corresponds to increasing $\pazocal{R}$. This situation has been discussed in the literature \cite{1996ApJ...462..855A} and one might thinks of $\pazocal{R}$ as quantitative correction. Second, the more compact the object is, the stronger is not only their quantitative difference, but also their qualitative behavior. Going to very compact objects, which means decreasing $R$ (fixed $M$), changes the behavior and results in increasing $\pazocal{R}$ at a critical value, which depends on the equation of state (because the tortoise transformation given by eq. \eqref{tortoisee} inside the star depends on it). We discuss this observation in detail in Sec. \ref{maximum_spacing} and appendix \ref{tortoise}, where we also introduce a linear relation for the tortoise transformation $r^{*}(r)$ that turns out to be very useful for normal neutron star compactness.
\subsection{Universal Relation}\label{Universal Relation}
In the literature it is well known that the real part of the fundamental axial QNM follows a simple universal scaling that can be described as
\begin{align}\label{w_empirical}
\text{Re}(\omega_{\text{f}}) R = A - B \frac{M}{R},
\end{align}
where $(A, B)$ are constants whose numerical values do not follow from first principles. Such a linear behavior has been reported for polar and axial QNMs \cite{Andersson:1997rn,1999MNRAS.310..797B}. This has been done for neutron stars using different equations of state, in general relativity, as well as alternative theories of gravity \cite{BlazquezSalcedo:2012pd,Blazquez-Salcedo:2018qyy}. The explanations for the scaling are of qualitative character, e.g. the coupled string toy model presented in \cite{1986GReGr..18..913K} or considering the modes to be trapped in the star, due to the space-time curvature and being partially reflected at the potential discontinuity that appears at the surface \cite{1996ApJ...462..855A}. The new Bohr-Sommerfeld rule can improve the situation a bit and we can provide a more satisfying reason for the scaling. Since the QNM spectrum is obtained from integrating the potential inwards, starting at the surface, one would expect that the real part of the fundamental QNM should be close to the surface value of the potential. Unfortunately our Bohr-Sommerfeld rule is not able to find a precise value for the fundamental QNM. The derivation of the rule only applies for QNMs that are related to the region above the discontinuity, but does not allow to make any conclusion of possible QNMs that exist around this value or below. On the other hand, one can argue that the value of the fundamental mode $\text{Re}(\omega^2_\text{f})$ should also not be too far away from the value of the potential at the surface, because one knows numerically that the eigenvalues of the spectrum follow a systematic pattern. Since one might not trust the WKB method in this delicate situation, we have have treated the internal and external potential regions with two potentials that have the right asymptotic $1/r^2$ behavior for small and very large values of the tortoise coordinate, and might be matched smoothly or discontinuously at the surface. Applying the standard QNM conditions along with the Wronskian at the surface, one finds an involved equation, which however, can be solved numerically once the parameters of the model are provided. With this approach we can locate the fundamental QNM that is found numerically, but is missing within the Bohr-Sommerfeld rule. We can confirm that its real part is close to the surface value of the potential. Making the identification that it should correspond to the value of the Regge-Wheeler potential at the surface, the predicted universal scaling between the fundamental mode and the compactness $M/R$ follows directly and is given by
\begin{align}\label{universal_scaling_eq}
R \sqrt{\text{Re} (\omega_{\text{f}}^2)}  = \sqrt{\left(1- \frac{2 M}{R} \right) \left(L -  \frac{6 M}{R}\right)}
\approx \sqrt{L} - \left( \frac{3+ L}{\sqrt{L}} \right)  \frac{M}{R} + \pazocal{O}\left( \left( \frac{M}{R}\right)^2 \right),
\end{align}
which is functionally closely related to the empirical relation eq. \eqref{w_empirical}. In the linear case and assuming that $\sqrt{E_\text{f}} \approx  \text{Re}(\omega_f)$, the functional form coincides. We show results for this prediction in Sec. \ref{Universal_scaling}.
\subsection{Approximate Potential}\label{approx_potential}
One major difficulty in the pure analytic treatment of the perturbation problem is that the simple form of the wave equation only appears in the tortoise coordinate. Making explicit use of it is almost impossible, since it is related to an integral over the metric functions $g_{00}$  and $g_{11}$, which for most equations of state can only be obtained numerically by solving the TOV equations. In this section we show how the perturbation potential can be simplified significantly, without affecting the spectral properties to a good precision. To do so we make use of three things. First, for the compactness of normal neutron stars, the tortoise coordinate transformation can be approximated very well with a linear function and is explicitly known, once $\pazocal{R}$ has been obtained. Second, only the first few QNMs correspond to the region of the potential far away from the center. Especially the internal classical turning point of the asymptotic part of the spectrum is located very close to the center. Third, in the central region one can expand all functions that appear in the potential and keep only dominant terms. Details for these approximations and calculations are provided in the appendix \ref{tortoise} and \ref{approximate_pot}.
Making use of all three observations allows one to write the wave equation explicitly in a shifted\footnote{The coordinate is shifted such that the potential diverges at $u^{*} = 0$, which simplifies the following calculations.} tortoise coordinate $u^{*}$ in the very simple form
\begin{align}\label{app_pot}
V(u^{*}) =  C_0 + \frac{C_2}{{u^{*}}^2},
\end{align}
with
\begin{align}\label{params}
C_0 = e^{2 \nu_0 } \left[\nu_2 L -4\pi (\rho_0 + P_0 )  \right],\qquad  \text{and} \qquad
C_2 = L e^{2\nu_0}  \left(\frac{\pazocal{R}}{R} \right)^2,
\end{align}
where $\nu_0, \nu_2, \rho_0$ and $P_0 $ come from the expansion of the metric potential $\nu(r)$, defined via $g_{00} = e^{2 \nu(r)}$, the density $\rho$, and the pressure $P$, in the center of the star. We provide details of the expansion in Sec. \ref{approximate_pot}. For central potentials one can use the Langer correction \cite{2013waap.book.....K} to push the precision for small $l$, which means to replace the factor $L = l(l+1)$ with $(l+1/2)^2$. 
Applying the new Bohr-Sommerfeld rule eq. \eqref{new_BS} to the potential eq. \eqref{app_pot} yields
\begin{align}\label{BS_master}
\sqrt{(E_{0n} - C_0) {\pazocal{R}}^2-C_2} + \sqrt{C_2}\left(\arctan\left(\frac{\sqrt{C_2}}{\sqrt{(E_{0n} - C_0) {\pazocal{R}}^2-C_2}} \right) - \frac{\pi}{2} \right) = \pi\left(n+\frac{3}{4} \right).
\end{align}
It can either be solved for $E_{0n}$ numerically for a given value of $n$ or be expanded for high eigenvalues. The latter one predicts a polynomial behavior
\begin{align}\label{E_n_app}
E_{0n} \approx  A n^2 + B n + C,
\end{align}
with 
\begin{align}\label{poly_ABC}
A \equiv \frac{\pi^2}{{\pazocal{R}}^2}, \qquad \qquad
B \equiv A \left(\sqrt{C_2} + \frac{3}{2} \right), \qquad \qquad
C \equiv C_0  -\frac{C_2}{R^2} + A \left(\frac{C_2}{4} + \frac{3 \sqrt{C_2}}{4}+\frac{9}{16} \right).
\end{align}
From this it is straight forward to see that the scaling for very high eigenvalues is to leading order as expected from eq. \eqref{E_asympt}. Especially the relation eq. \eqref{Delta_omega} for the spacing of the real part of the QNMs $\Delta \omega_n = \Delta \sqrt{E_n}$ converges to $\pi/\pazocal{R}$, as found in \cite{PhysRevD.83.064012}. The results of this section applied to polytropes are shown in Sec. \ref{Results}.
%
\section{Inverse Spectrum Problem}\label{Inverse Problem}
%
In this section we discuss the inverse spectrum problem, which aims to reconstruct parameters of the source from the knowledge of the spectrum. It is split up in three parts. In Sec. \ref{Invert_BS} we first discuss the inversion of the new Bohr-Sommerfeld rule to find the underlying perturbation potential. The second part in Sec. \ref{Recovering Fundamental Parameters} uses the results obtained from the approximated potential to reconstruct the fundamental parameters related to the potential. Finally, in \ref{Recovering the Equation of State} we show how the previously obtained results can be used to recover the $g_{00}$ component inside the star, and how the equation of state can be constrained.
\subsection{Inverting the Bohr-Sommerfeld Rule}\label{Invert_BS}
As it is known in the literature \cite{lieb2015studies,MR985100}, the classical Bohr-Sommerfeld rule can be ``inverted'' to reconstruct the potential well from the eigenvalue spectrum $E_n$. For pure potential wells, this two turning point problem can not be solved uniquely, unless one of the two classical turning points is known. Instead, infinitely many so-called WKB spectrum equivalent potentials exist \cite{1992JMP....33.2958B,1991JPhA...24L.795B,paper6}. All of them share the same width, defined as the difference of the classical turning point functions $\pazocal{L}(E) \equiv x_1(E)-x_0(E)$.  Something similar is also true for two turning point potential barriers \cite{1980AmJPh..48..432L,2006AmJPh..74..638G}. In the case of three or four turning point problems, as they appear for ultra compact stars and exotic compact objects, a similar analysis can be found in \cite{paper2,paper4,paper5}.
\par
The inversion of the new Bohr-Sommerfeld rule eq. \eqref{new_BS} is equivalent to its classical pendant shown in \cite{lieb2015studies,MR985100}. The inversion yields the distance $\pazocal{L}(E)$ from the surface of the star $r^{*}_R$, which is the same for all $E$, to the internal turning point $r^{*}(E)$ as
\begin{align}\label{Lwidth}
\pazocal{L}(E) \equiv r_{R}^{*}-r^{*}(E) = 2 \frac{\partial }{\partial E} \int_{E_\text{min}}^{E} \frac{n(E^{\prime})+\frac{3}{4}}{\sqrt{E-E^{\prime}}} \text{d} E^\prime,
\end{align}
where $n(E)$ is the spectrum written as function of $E$. We use the lowest provided eigenvalue for $E_\text{min}$, which corresponds to $n(E_\text{min}) = -3/4$. Once the discrete spectrum is provided, $n(E)$ has to be interpolated. This introduces some non-uniqueness to the problem, but the spectrum is usually well behaved and different interpolations do not cause major differences. For more details we refer to the discussions presented in \cite{paper2,paper5}. This observation is in agreement with the integral character of the Bohr-Sommerfeld rule, which is ``blind'' to small deviations in the potential, if they average out between consecutive $n$.
\par
The associated turning point $r^{*}(R)$ at the surface of the star is uniquely determined from the external tortoise coordinate transformation if $(M,R)$ are known. In this case, the perturbation potential inside the star follows uniquely from inverting eq. \eqref{Lwidth} for $E$, which can also be done numerically. Therefore, within the approximations being done to derive the Bohr-Sommerfeld rule, the perturbation potential is in principle explicitly known, if the QNM spectrum is provided. In Sec. \ref{Results} we show the results for the reconstructed potentials of constant density stars and polytropes.
\subsection{Recovering Fundamental Parameters}\label{Recovering Fundamental Parameters}
The previously presented inversion of the Bohr-Sommerfeld rule transforms consecutive QNMs, starting from the fundamental one, to the potential. Because it was derived on rather general properties of the potential, it requires a precise knowledge of the spectrum, due to the involved interpolation. An alternative approach for applications in neutron stars, where additional properties of the potential can be imposed, is described in the following.
\par
In Sec. \ref{approx_potential} we presented an approximation for the perturbation potential and showed its Bohr-Sommerfeld spectrum. Using this result it is possible to reconstruct the fundamental parameters of the potential ($\pazocal{R}, C_0, C_2$) from the spectrum, if $(M,R)$ are provided. Because the number of parameters is small, it is not necessary to assume that many QNMs are known. Since the Bohr-Sommerfeld spectrum is not exact, one expects that the reconstructed parameters differ slightly. As it is evident from the polynomial expansion of $E_{0n}$ in eq. \eqref{E_n_app}, one would expect that the parameters related to $(A, B, C)$ might not be reconstructed equally well, since the precision of the underlying WKB method is more precise for high eigenvalues, where $A$ dominates, and not very precise for small $n$, where $C$ is important. Using $n(E_{0n})$ to fit a given spectrum, instead of the polynomial expansion, yields better results and is given by
\begin{align}\label{nE_master}
n(E_{0n}) = \frac{1}{\pi} \left(-\frac{3}{4} + \sqrt{(E_{0n}-C_0) {\pazocal{R}}^2-C_2} + \sqrt{C_2}\left(\arctan\left(\frac{\sqrt{C_2}}{\sqrt{(E_{0n}-C_0) {\pazocal{R}}^2-C_2}} \right) - \frac{\pi}{2} \right) \right).
\end{align}
The reconstruction of $(\pazocal{R}, C_0, C_2)$ imposes stringent constraints on the underlying equation of state. It is highly unlikely that for fixed $(M,R)$ two different equation of states, which are intersecting in the mass-radius diagram, will also agree for these reconstructed properties. While $\pazocal{R}$ is an averaged property obtained by integrating $\sqrt{g_{11}/g_{00}}$ throughout the whole star, $C_2$ can directly be used to obtain the $g_{00}$ component of the metric at the center $r=0$. The knowledge of $C_0$ allows in principle to relate the central pressure as function of the central density. We show results for this in in Secs. \ref{Results_reconstruction_param} and \ref{Results_reconstruction_EOS}.
\subsection{Recovering $P(r)$ and $\nu(r)$}\label{Recovering the Equation of State}
In the previous section we have shown how the perturbation potential can approximatively be reconstructed from the QNM spectrum. As it was recently discussed in \cite{Suvorov:2018bvs}, the knowledge of the potential has crucial implications for the stellar structure, which is described by the TOV equations. This system of equations is usually solved by integrating from the center of the star towards the surface, where the metric is matched with the external Schwarzschild solution. This requires the equation of state throughout the star, as well as the central pressure or density. Now, if the perturbation potential is known and used in the rewritten TOV equations, the stellar structure, as well as the equation of state can be determined by integrating the TOV equations inwards (assuming ($M, R$) are also known). Although our method allows to reconstruct the perturbation potential from the spectrum, it might turns out not to be precise enough for such a treatment and then requires a more refined study, which is beyond the scope of this work. Nevertheless, we are able to show that the equation of state in the central region can be constrained by assuming that $(M,R)$ are known and $\nu_0$ has been reconstructed to good precision from the spectrum, as described in the previous section. A close look on $\nu(r)$, obtained from the TOV equations for different neutron stars, suggests that the expansion shown in \cite{1983ApJS...53...73L} could be a good approximation throughout the whole star, as long as there are no local effects. From $(M, R)$ one can directly provide the metric functions $\nu_R \equiv \nu(R)$ and $\nu^{\prime}_R \equiv \nu^{\prime}(R)$ at the surface, by matching it with the external Schwarzschild space-time. Using the reconstructed value of $\nu_0$, one can directly deduce $(\nu_2, \nu_4)$, which imprint information of the equation of state in the central region
\begin{align}
\nu_2 = \frac{4}{R^2}\left(\nu_R-\nu_0 \right) - \frac{\nu^\prime_R}{R}, 
\qquad \qquad
\nu_4 = \frac{2 \nu^\prime_R}{R^3} - \frac{4}{R^4} \left(\nu_R-\nu_0 \right).
\end{align}
Once $\nu_2$ is known, one has an approximation for the central pressure $P_0$ as function of the central density $\rho_0$, as well as the pressure $P(r, P_0)$ in the central region
\begin{align}\label{central_P}
P_0(\rho_0) = \frac{\nu_2}{4 \pi}-\frac{\rho_0}{3},\qquad \qquad
P(r, P_0) = P_0 - r^2 \left(\frac{3 \nu_2^2}{8 \pi} -P_0 \nu_2 \right).
\end{align}
Note that this reconstruction leaves some ambiguity, since it only allows to reconstruct a combination of central pressure and density, not their absolute values. However, it allows to approximate an upper limit for the central pressure, which appears for $\rho_0 = 0$, as well as an upper limit for the central density, which appears for $P_0 = 0$. They are given as
\begin{align}
P_{0, \text{max}} = \frac{\nu_2}{4 \pi}, \qquad \qquad 
\rho_{0, \text{max}} = \frac{3 \nu_2}{4 \pi}.
\end{align}
We show the results for the here discussed approach in Sec. \ref{Results_reconstruction_EOS}.
%
\section{Results}\label{Results}
%
In this section we present the results of the semi-analytic methods that were derived in the previous sections. In most cases we show two different polytropes and provide in some cases constant density stars as toy model for ultra compact stars. The basic parameters of the polytropes can be found in Table \ref{table}. The precise numerical data for the QNM spectra is provided from a code first presented in \cite{1994MNRAS.268.1015K}. The polytropic equation of state is defined as
\begin{align}
P = K \rho^{\gamma}.
\end{align}
\begin{table}[H]
\centering
\caption{Parameters of the studied polytropic neutron star models.\label{table}}
\begin{tabular}{|c||c|c|c|c|c|c|c|}
\hline
Model &		$M$	&	$R$		&	$C$ 	& $K$ & $\gamma$ & $\rho_\text{c}$ \\
\hline
Poly A		&	2.0666\,km	&	14.1466\,km	&	0.1461	& 100 & 2 & $8.925 \times 10^{14}\,\mathrm{gr/cm^3}$\\
Poly B		&	2.4161\,km	&	11.3831\,km	&	0.2123	& 100 & 2 & $2.482 \times 10^{15}\,\mathrm{gr/cm^3}$\\
\hline
\end{tabular}
\end{table}
\subsection{Calculating the Spectrum}\label{Calculating the Spectrum}
\begin{figure}[H]
\centering
	\begin{minipage}{0.45\linewidth}
	\includegraphics[width=1.0\linewidth]{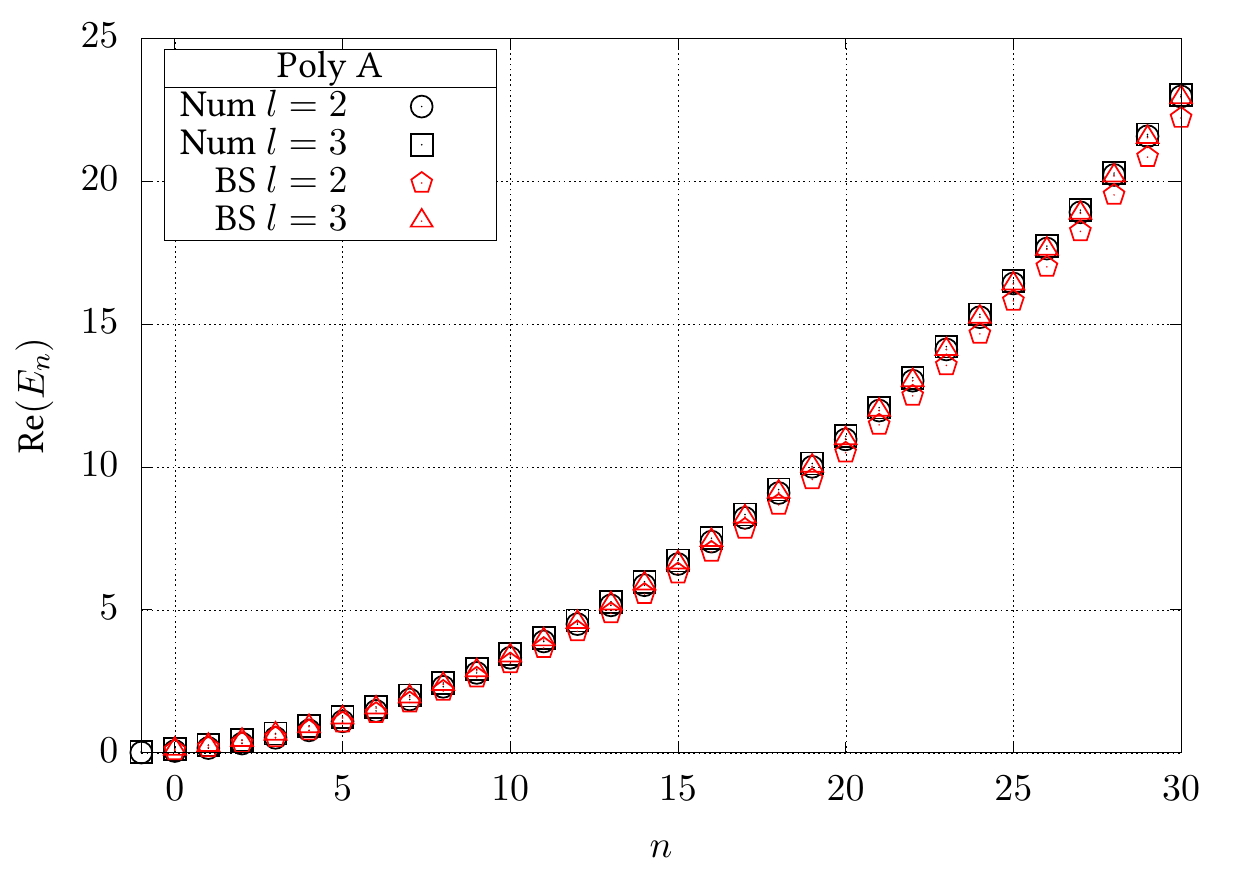}
	\end{minipage}
	\quad
	\begin{minipage}{0.45\linewidth}
	\includegraphics[width=1.0\linewidth]{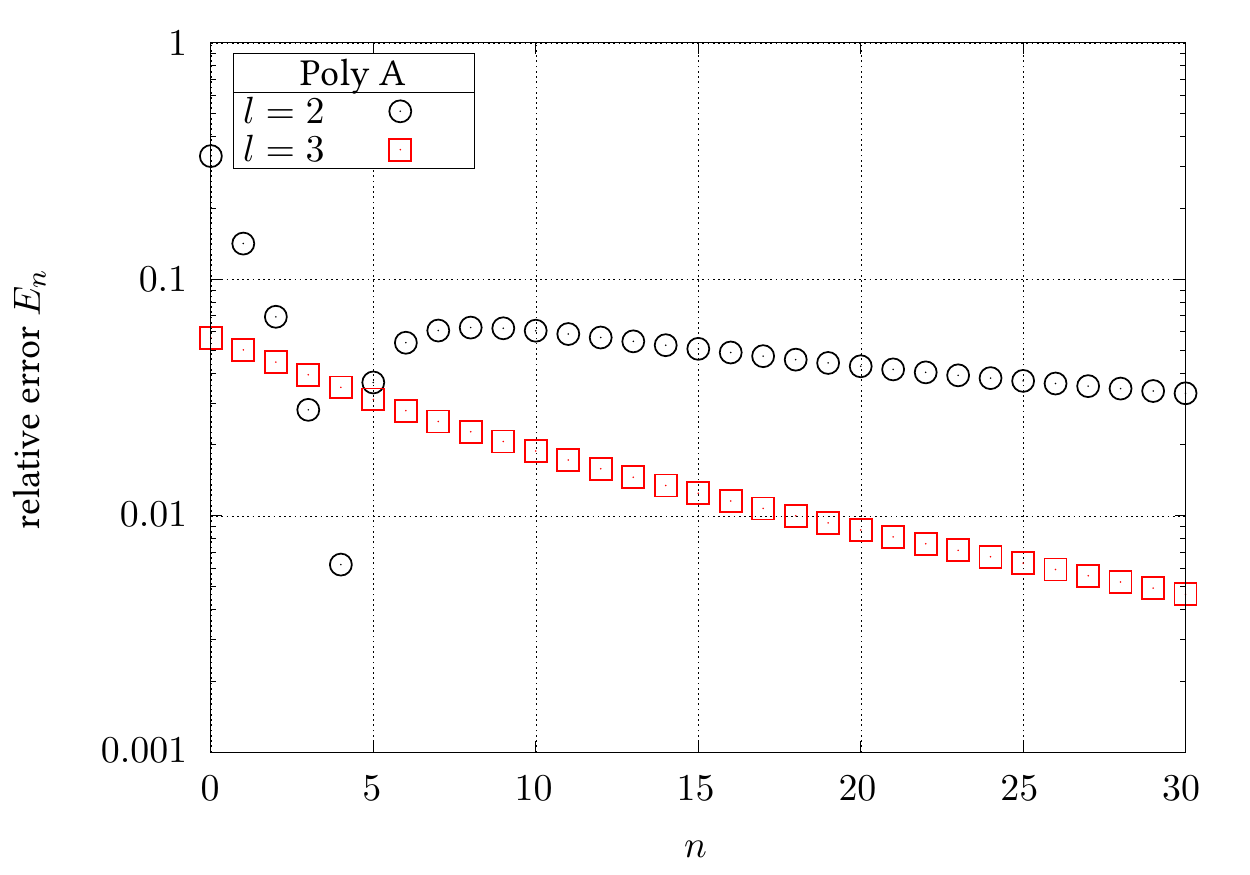}
	\end{minipage}
	\caption{\textbf{Left panel:} Shown is the real part of the spectrum $E_n$ for $l=2$ and $3$ for the polytrope model A, precise numerical values (black symbols) are compared to the Bohr-Sommerfeld prediction (red symbols). \textbf{Right panel:} The relative error of the spectrum is shown for $l=2$ (black circles) and $l=3$ (red boxes).\label{En587}}
\end{figure}

\begin{figure}[H]
\centering
	\begin{minipage}{0.45\linewidth}
	\includegraphics[width=1.0\linewidth]{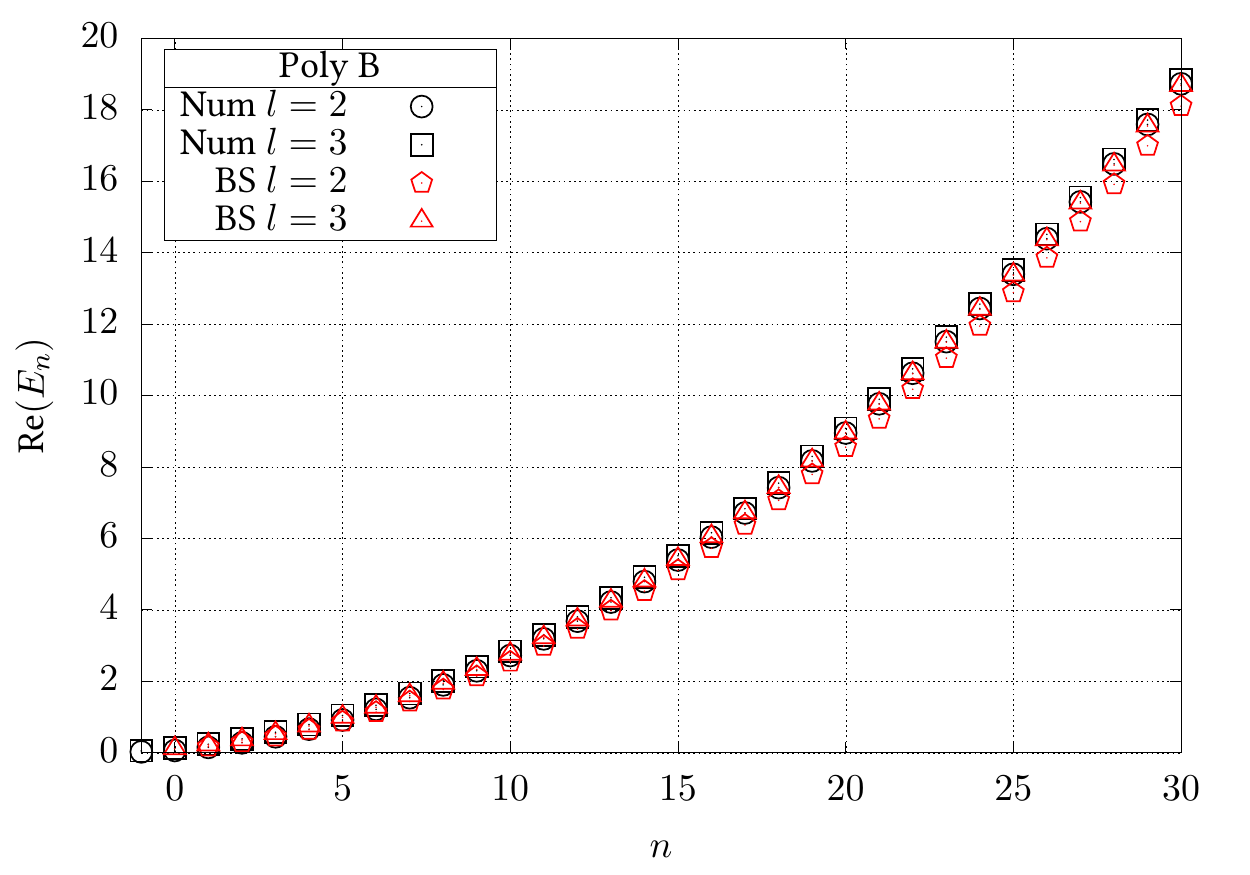}
	\end{minipage}
	\quad
	\begin{minipage}{0.45\linewidth}
	\includegraphics[width=1.0\linewidth]{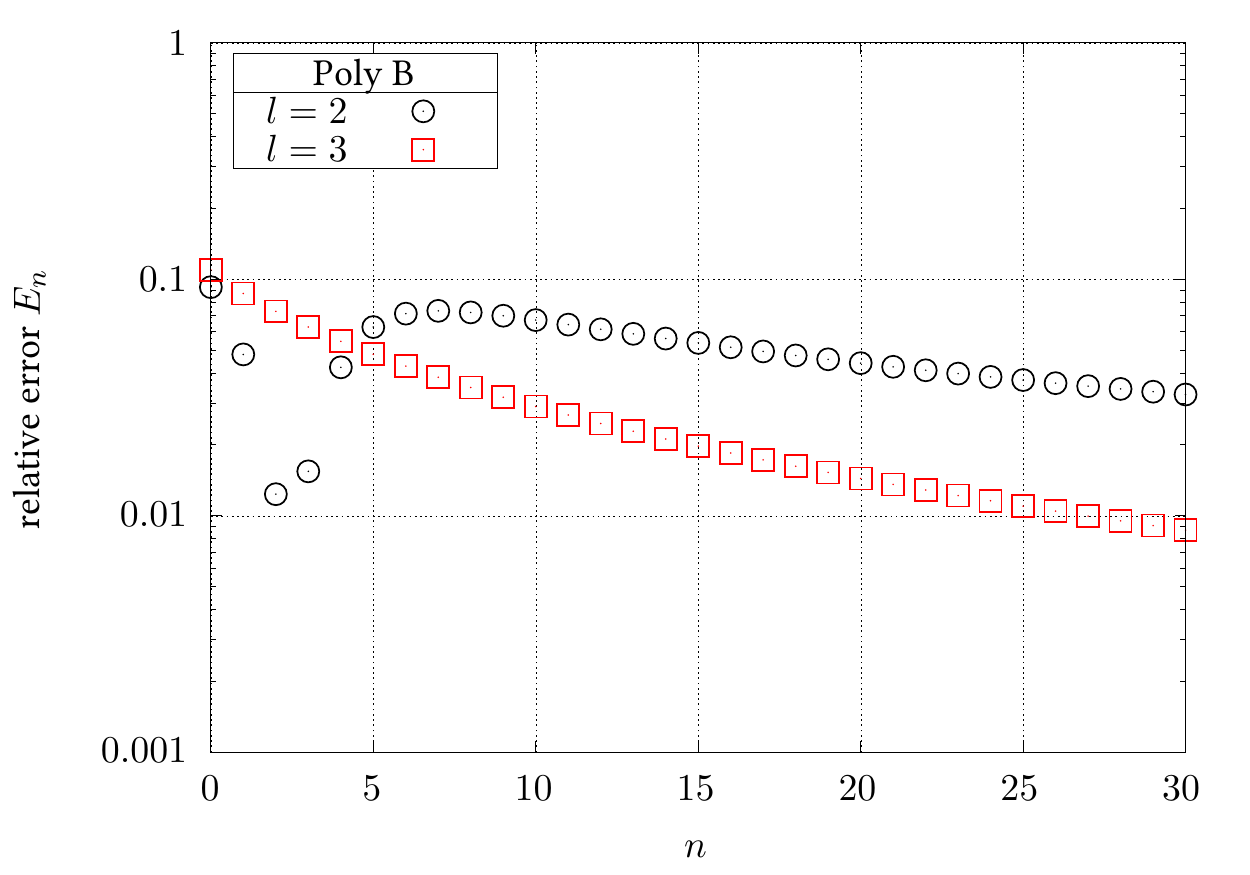}
	\end{minipage}
	\caption{\textbf{Left panel:} Shown is the real part of the spectrum $E_n$ for $l=$ 2 and 3 for the polytrope model B, precise numerical values (black symbols) are compared to the Bohr-Sommerfeld prediction (red symbols). \textbf{Right panel:} The relative error of the spectrum is shown for $l=2$ (black circles) and $l=3$ (red boxes).\label{En311}}
\end{figure}

In the following we compare the predicted spectrum of the new Bohr-Sommerfeld rule eq. \eqref{new_BS} by using the approximated potential eq. \eqref{BS_master} with precise numerical data. The left panels of the subsequent Figs. \ref{En587}, \ref{En311} show the spectra of two different polytropes, each for $l=2$ and $l=3$. The relative errors are defined as usually and shown in the right panel of the same figures. As it is expected from WKB theory, the overall accuracy improves for increasing $n$. The approximated potential resembles the true potential more and more for large $l$, thus the precision also increases for larger $l$. It is surprising that the method provides quite accurate predictions even for small $n$ and $l$ where the performance of the method is worst.
\subsection{Universal Relation}\label{Universal_scaling}
Here we compare the proposed universal relation for the fundamental axial QNM eq. \eqref{universal_scaling_eq} with exact values for different compact stars. We recall that the distribution of numerical values around the empirically known fitting formula shows that the relation depends slightly on the equation of state and is not exact. For $l=2$ we compare the empirical fit \cite{1999MNRAS.310..797B} with some examples for constant density stars, polytropes and the realistic equation of states used in the same work. Considering that the exact values are not exactly described by the empirical linear fit, our analytic estimate yields a good prediction and also takes into account the non-linear slope for large compactness, which is shown for constant density stars. This non-linear scaling with respect to the compactness, described in eq. \eqref{universal_scaling_eq}, becomes more significant for higher $l$, both in our analytic estimate and for constant density stars. Overall, our relation slightly underestimates the true values. However, the estimate goes beyond the linear approximation, which was found only empirically in the literature \cite{Andersson:1997rn,1999MNRAS.310..797B,BlazquezSalcedo:2012pd,Blazquez-Salcedo:2018qyy}, and explains that the fundamental axial QNM can be approximatively ``identified'' with the value of the perturbation potential at the stellar surface.
\begin{figure}[H]
\centering
\includegraphics[width=0.45\linewidth]{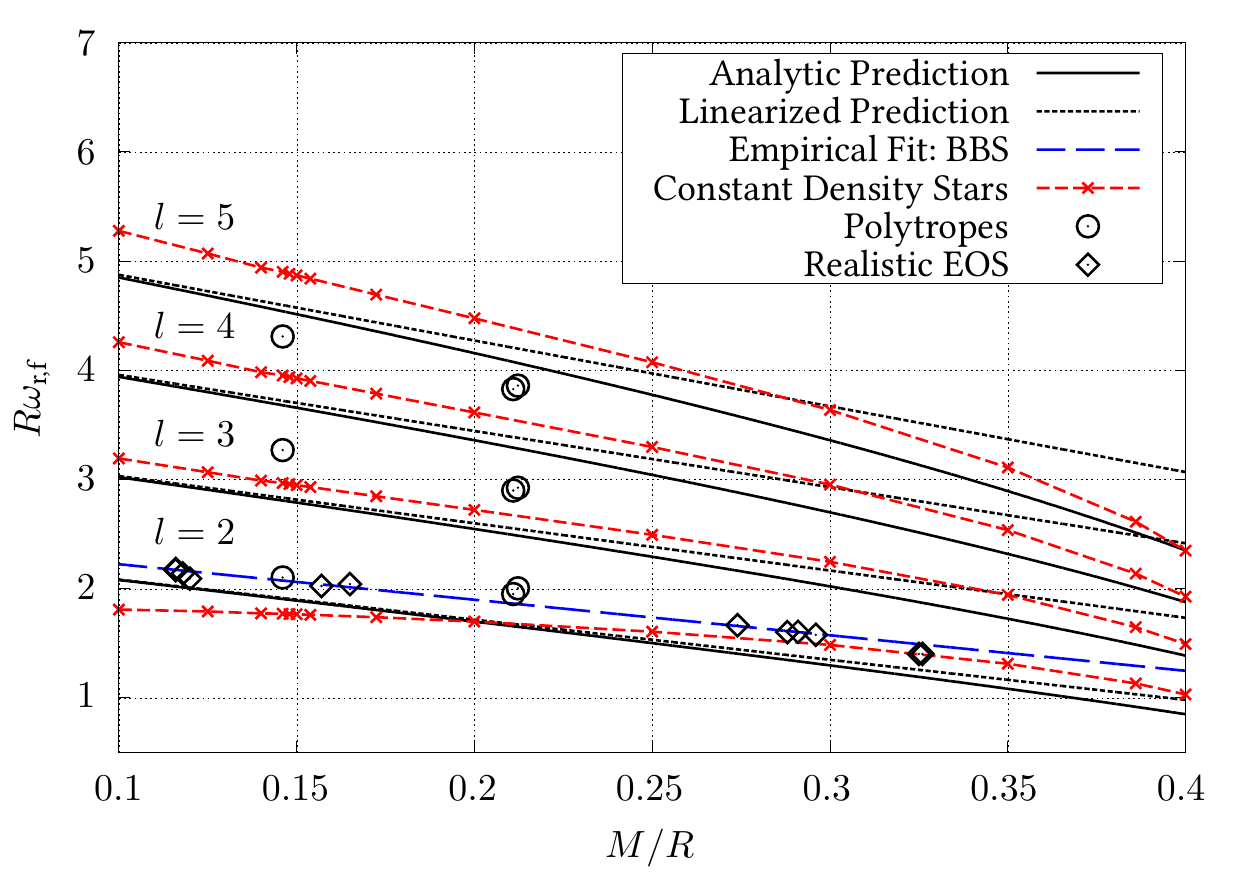}
\caption{In this figure we show the universal relation (black solid) described in eq. \eqref{universal_scaling_eq} as well as its linearized form (black dashed) for $l=$ 2, 3, 4 and 5. The data points correspond to precise numerical values for different constant density stars (red dashed), polytropes (black circles), and different realistic equation of states (black diamonds) taken from \cite{1999MNRAS.310..797B}. The empirical fit for $l=2$ (blue dashed), known from \cite{1999MNRAS.310..797B}, is shown with the corresponding realistic equation of state results (black diamonds) used in the same work.\label{rec_RM}}
\end{figure}
\subsection{Spacing of Modes}\label{Spacing}
The $\pi/\pazocal{R}$ spacing of modes is an interesting property and was first found in \cite{PhysRevD.83.064012} using another WKB approach. After we verify the asymptotic spacing of the polytrope QNM spectra by using numerical data, we show that the spacing of modes should first increase with increasing compactness, as it is known in the literature, but then shrinks again for ultra compact stars.
\subsubsection{Test of the Constant Spacing}
\begin{figure}[H]
\centering
	\begin{minipage}{0.45\linewidth}
	\includegraphics[width=1.0\linewidth]{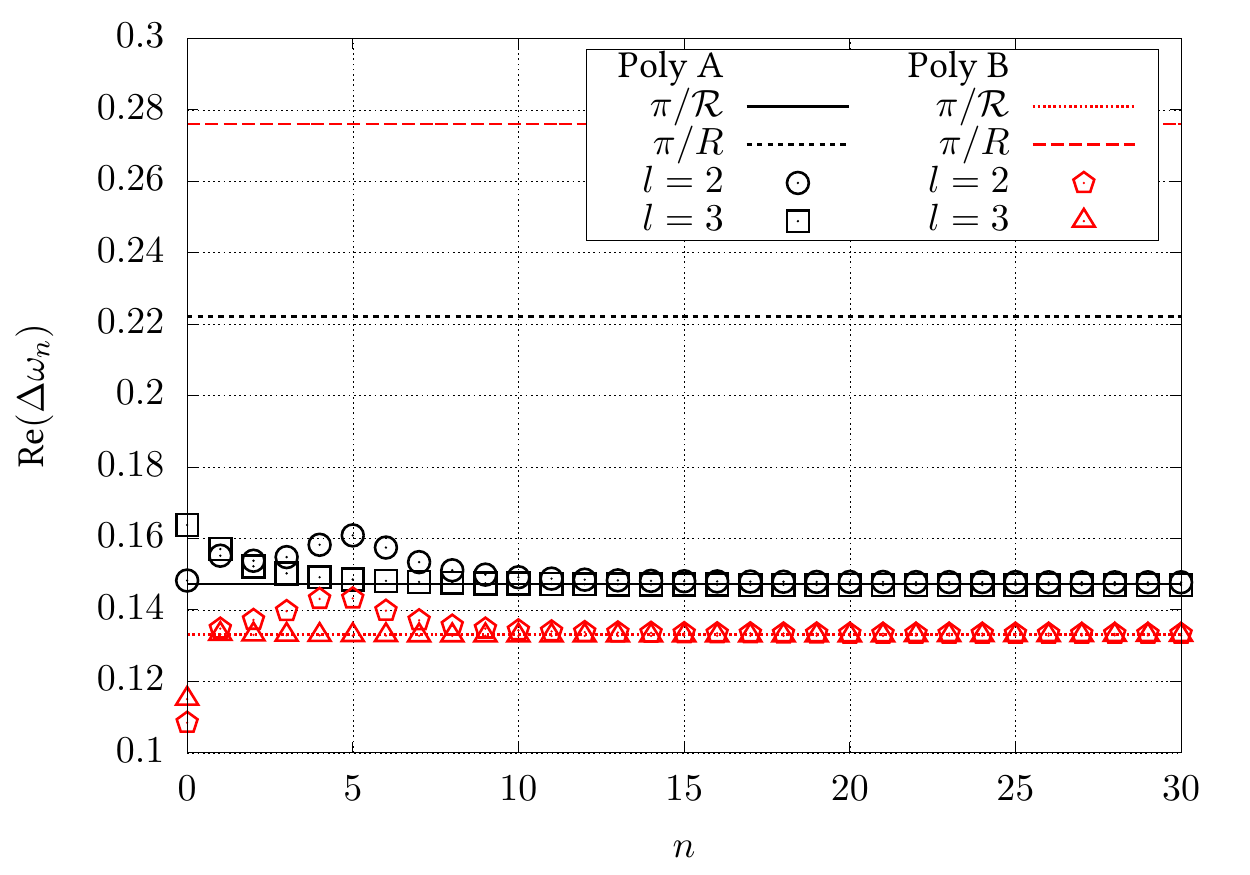}
	\end{minipage}
	\quad
	\begin{minipage}{0.45\linewidth}
	\includegraphics[width=1.0\linewidth]{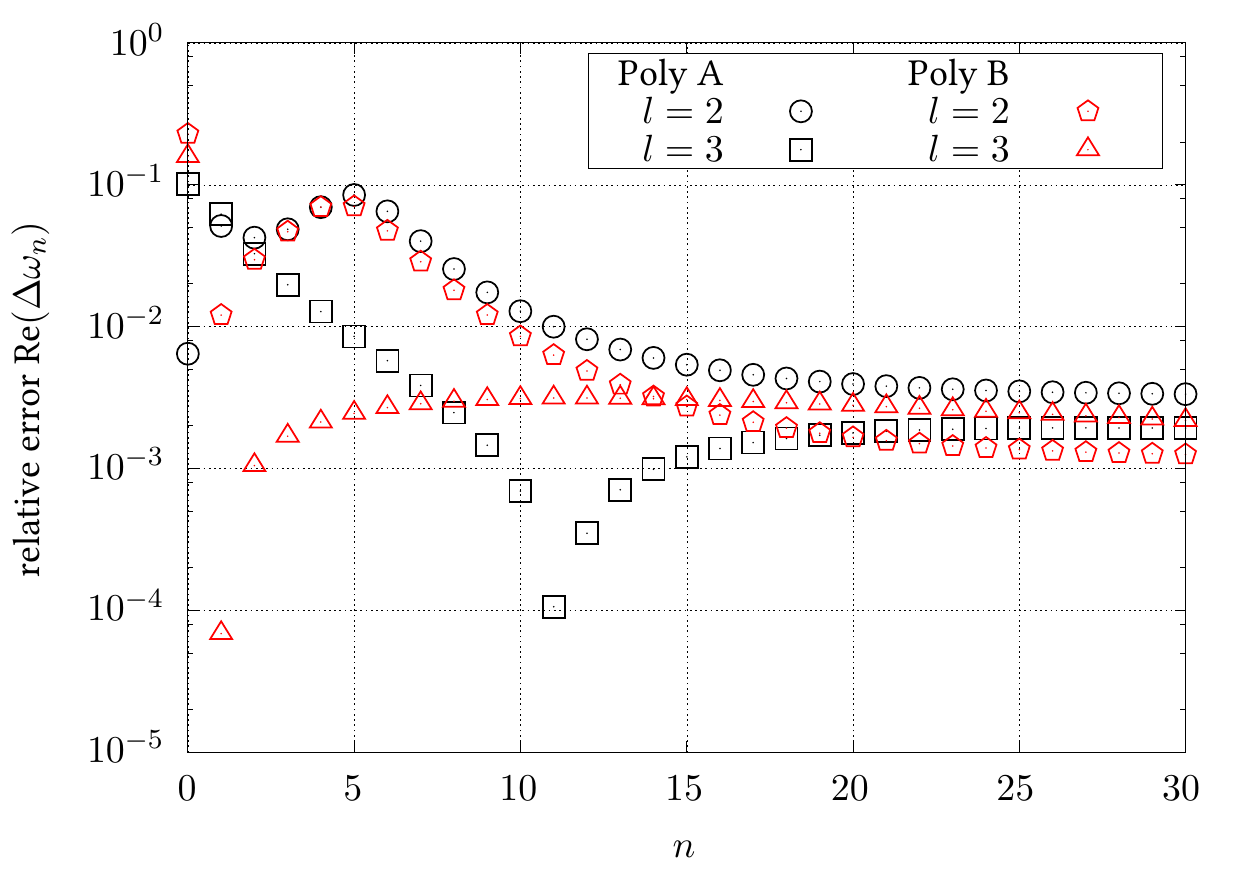}
	\end{minipage}
	\caption{Here we compare the predicted spacing of neighboring modes for two polytrope models A and B for $l=$ 2 and 3 modes. \textbf{Left panel:} The Bohr-Sommerfeld prediction (solid lines) is compared with the rough estimate from the literature (dashed lines) and numerical data (symbols).  \textbf{Right panel:} The relative error between the Bohr-Sommerfeld prediction and the numerical data} \label{mode_spacing}
\end{figure}

In Fig. \ref{mode_spacing} we verify the constant spacing relation $\Delta \omega = \pi/\pazocal{R}$ for high eigenvalues. The left panel shows the absolute numbers for two different polytropes, while the right panel contains the relative errors. For comparison we include the rough estimate $\pi/R$, which is known in the literature \cite{1996ApJ...462..855A} for less compact systems. As it is evident, the $\pi/R$ spacing of modes is by far only a rough estimate, the deviations compared to $\pazocal{R}$ are significant. Note that it predicts a larger spacing for Poly B compared to Poly A, while the opposite is true. Since the $\pi/R$ estimate is valid for less compact objects, it is not surprising that it fails completely for compact objects. However, the $\pi/\pazocal{R}$ prediction is extremely precise fulfilled. The saturation of the relative errors is around $0.1\,\%$ for $n=30$ and might indicate the precision for the high eigenvalues that were obtained numerically. Note that the spacing for the $l=3$ modes is already around $0.1-1\,\%$ for $n=5$.
\subsubsection{Predicted Maximum Spacing of Modes}\label{maximum_spacing}
\begin{figure}[H]
\centering
	\begin{minipage}{0.45\linewidth}
	\includegraphics[width=1.0\linewidth]{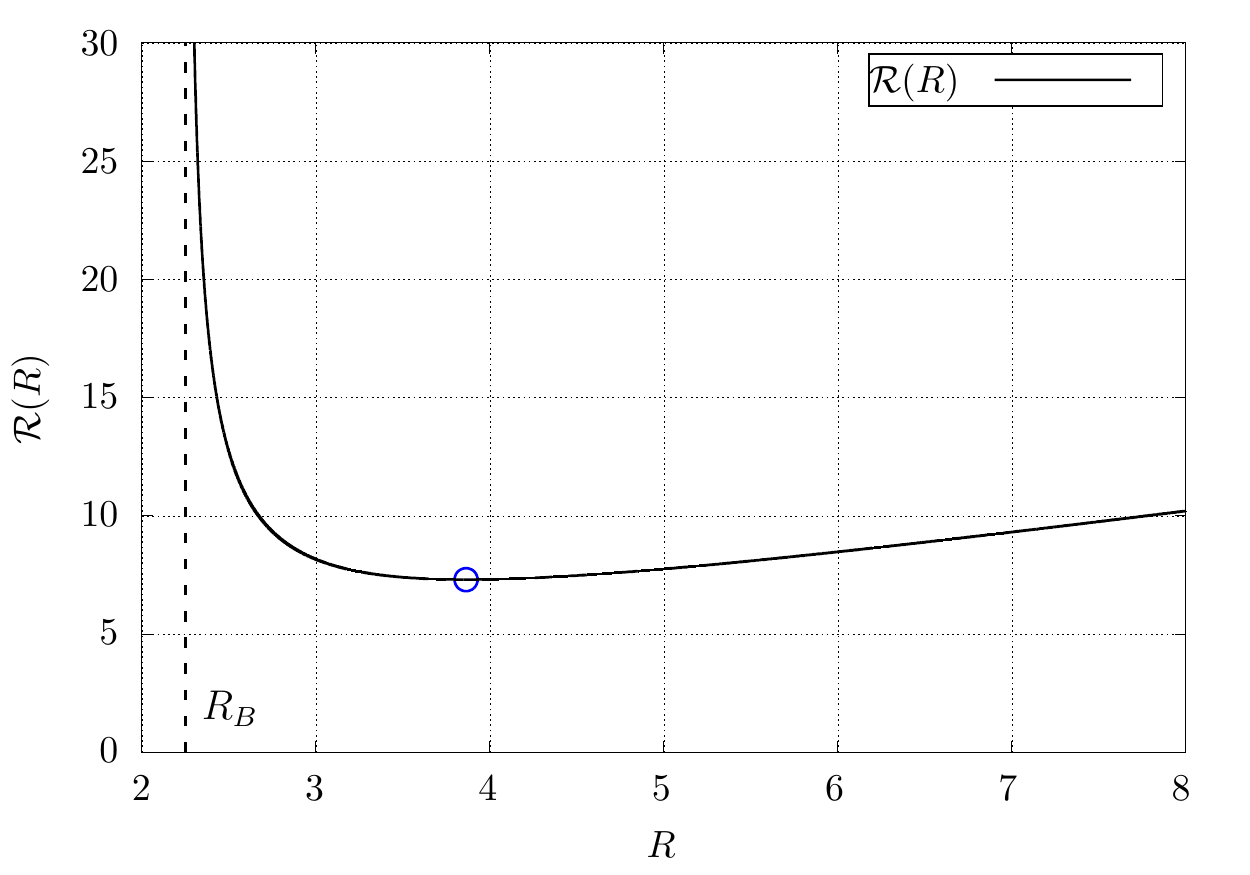}
	\end{minipage}
	\quad
	\begin{minipage}{0.45\linewidth}
	\includegraphics[width=1.0\linewidth]{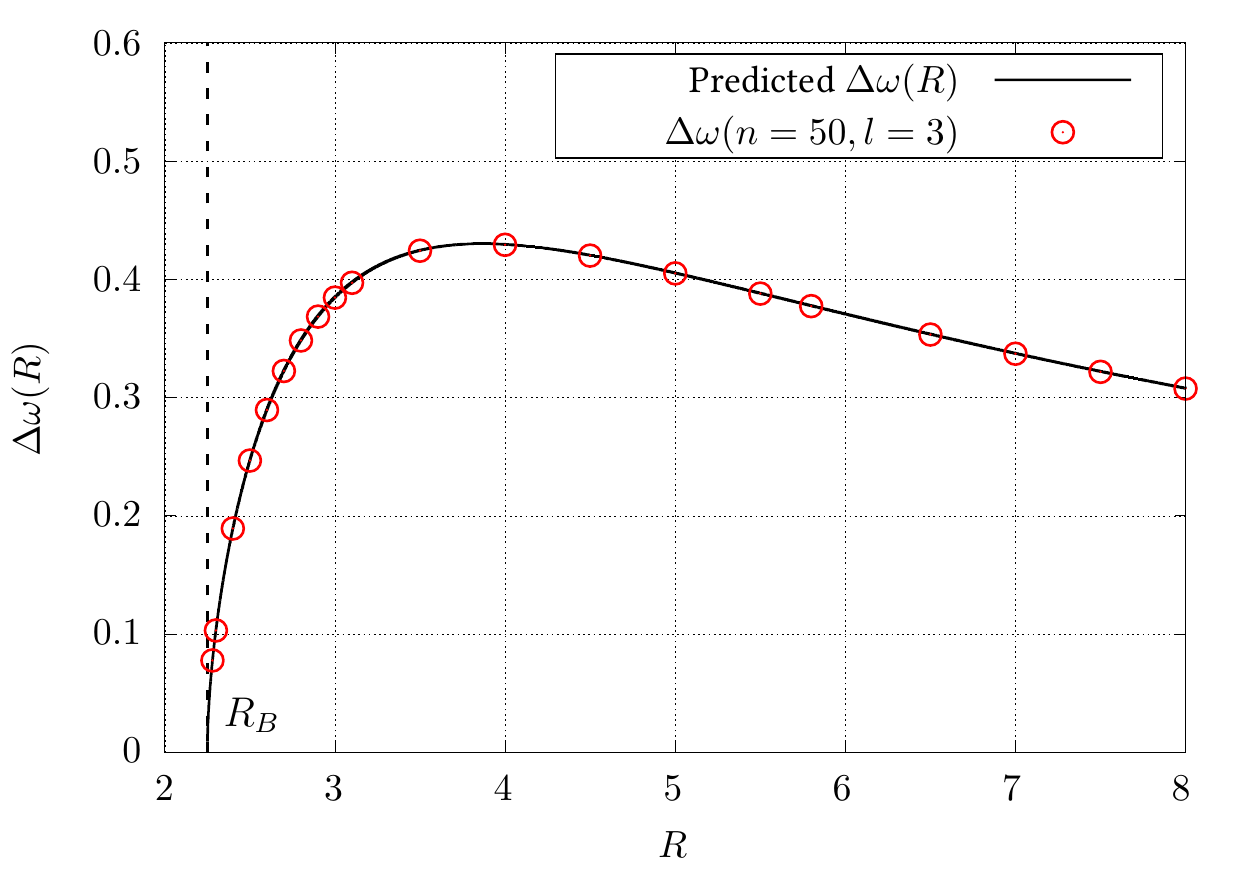}
	\end{minipage}
	\caption{In this figure we show a sequence of constant density stars with $M=1$. The Buchdahl limit $R_\text{B}$ is indicated in both panels as dashed line. \textbf{Left panel:} The exact relation for $\pazocal{R}(R)$ is provided along with the minimum value of $\pazocal{R}$ as blue dot at $R\approx 3.86$.  \textbf{Right panel:} Here we compare the predicted spacing $\Delta \omega = \pi/\pazocal{R}$ with numerical values for different $R$ and $l=3$, evaluated at $n=50$. \label{Dw}}
\end{figure}

As discussed in Sec. \ref{General Implications}, we expect that the asymptotic spacing as function of $\pazocal{R}$ should have a maximum, because for fixed mass, $\pazocal{R}(R)$ is likely to have a minimum for very compact configurations. Verifying this expectation in a general case is not easy, since $\pazocal{R}$ depends on the solutions of the TOV equations, which have to be obtained numerically in most cases. However, for constant density stars, which can exist for any radius larger than the Buchdahl limit of $R/M = 9/4 $, it can easily be confirmed. We show the explicit result for constant density stars in Fig. \ref{Dw}. In this case one can analytically find the relation for $\pazocal{R}(R)$, which is shown in the left panel and follows from $r^{*}(r)$ derived in \cite{2000PhRvD..62h4020P}. The right panel shows the predicted spacing for large modes, along with numerical data for different compactnesses. The asymptotic behavior for non-compact stars is already known in the literature as $\Delta \omega \approx \pi/R$ and was reported in \cite{1996ApJ...462..855A}. The agreement in the ultra compact case is excellent and beyond the approximative nature that might be expected. The maximum value appears for $R\approx 3.86$, which seems not to be related to any special physical property of the star or the potential. Trapped modes, for which the potential admits a quasi-bound region, start to appear for more compact stars. Note that the spacing alone can not be used to uniquely determine the compactness of a star, since there are two possibilities, but their fundamental modes are significantly different.
\subsection{Reconstruction of the Potential}
\begin{figure}[H]
\centering
	\begin{minipage}{0.45\linewidth}
	\includegraphics[width=1.0\linewidth]{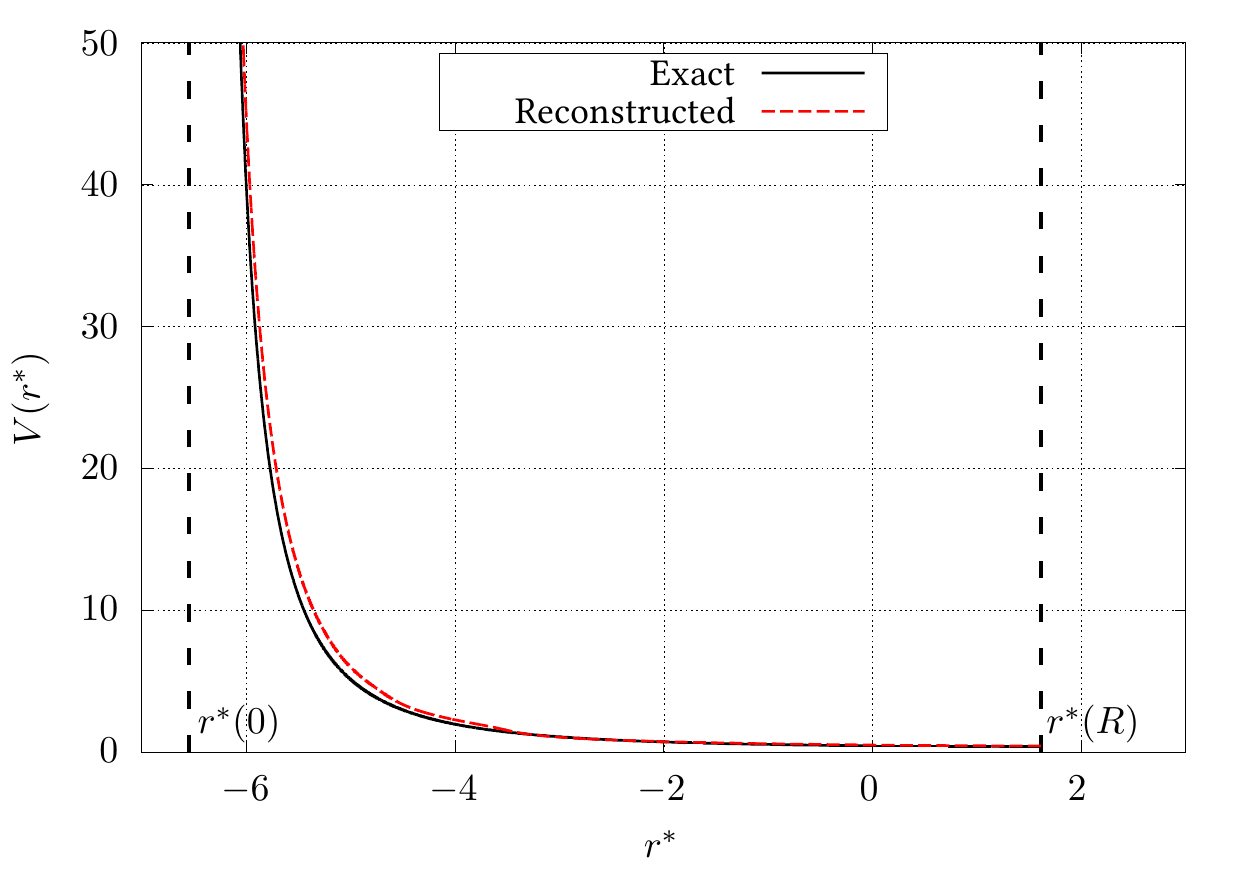}
	\end{minipage}
	\quad
	\begin{minipage}{0.45\linewidth}
	\includegraphics[width=1.0\linewidth]{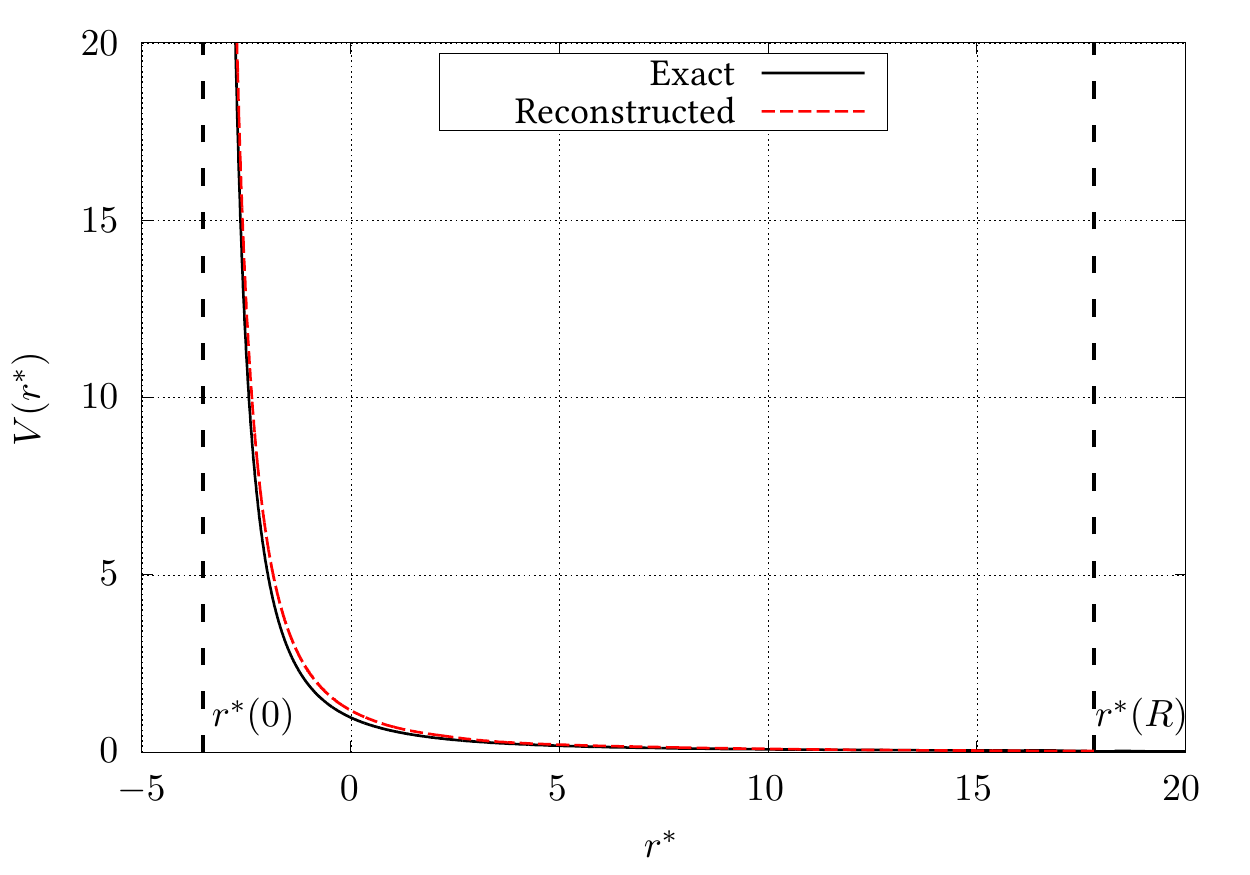}
	\end{minipage}
	\caption{Reconstructed axial perturbation potentials (red dashed) are compared to the exact ones (black solid) for $l=3$. The center and surface value of $r^*$ is provided as well (black dashed). \textbf{Left panel:} The case of a constant density star with $R=3M$. \textbf{Right panel:}   The polytrope model A. \label{CS_rec}}
\end{figure}

By inverting the Bohr-Sommerfeld rule it is possible to reconstruct the internal perturbation potential once the spectrum is known. Here we show the results for an ultra compact constant density star and one of the polytropes. In the left panel of Fig. \ref{CS_rec} we show the reconstructed potential for the constant density star with $R=3\,M$. The right panel of the same figure contains the reconstructed potential of Poly A. Deviations to the true potential are visible, but small. The steep rise at $r^{*}(0)$ is precisely captured.
\subsection{Reconstruction of $\pazocal{R}$ and $\nu_0$ from the Spectrum}\label{Results_reconstruction_param}
\begin{figure}[H]
\centering
	\begin{minipage}{0.45\linewidth}
	\includegraphics[width=1.0\linewidth]{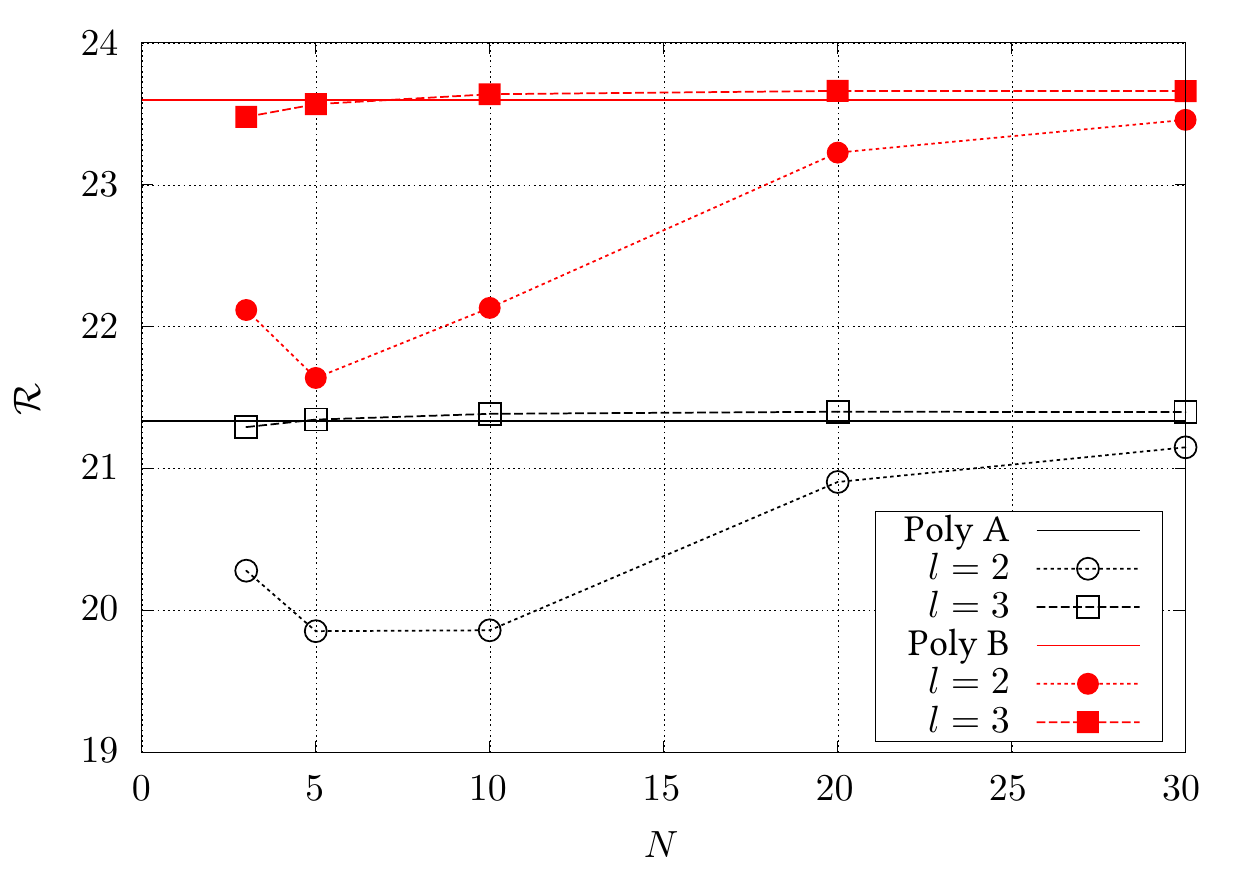}
	\end{minipage}
	\quad
	\begin{minipage}{0.45\linewidth}
	\includegraphics[width=1.0\linewidth]{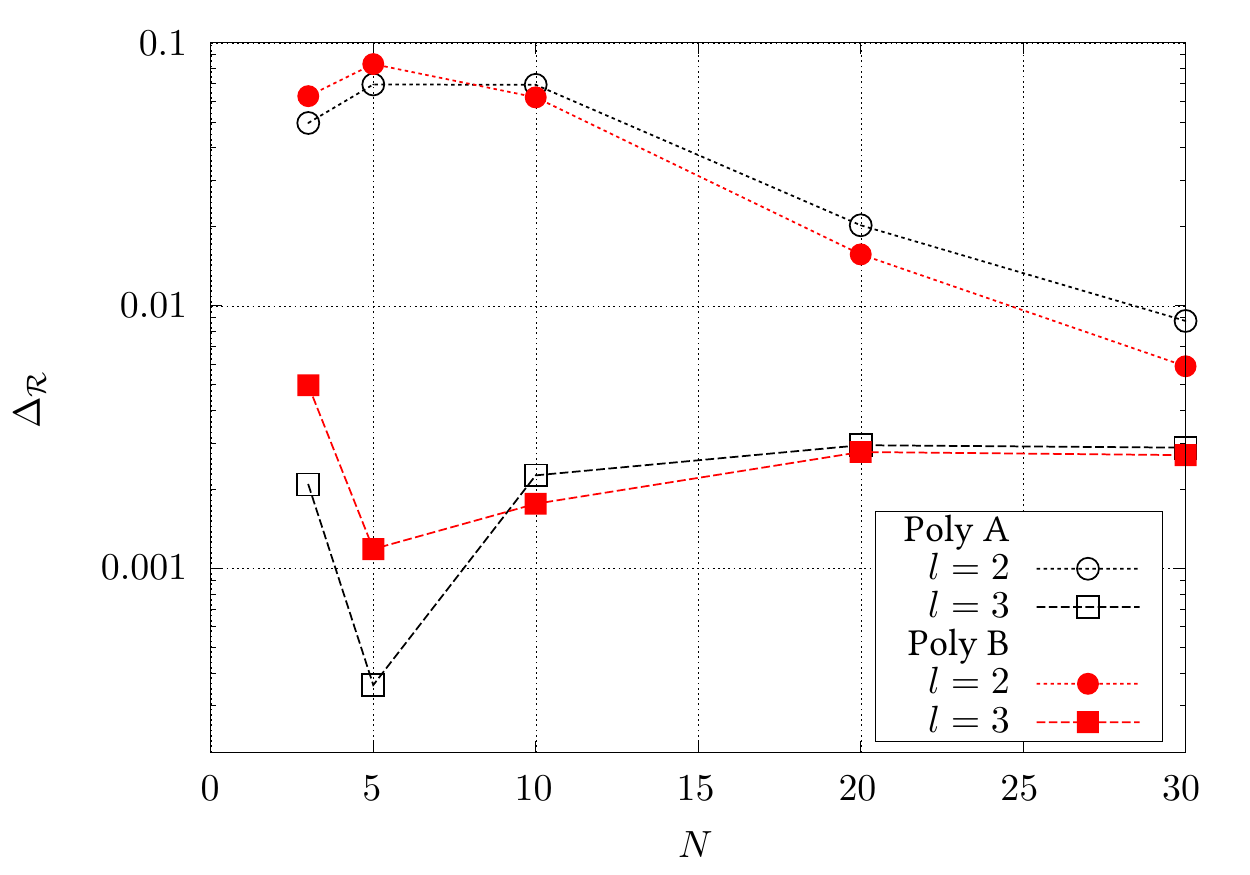}
	\end{minipage}
	\caption{Here we present the reconstruction of $\pazocal{R}$ as function of the number of included eigenvalues $N$ for the spectrum fitting of the two polytrope models A and B for $l=2$ and $3$. \textbf{Left panel:}  The reconstruction of $\pazocal{R}$ (black and red symbols) compared to the true values (black and red solid line). \textbf{Right panel:} The relative errors for the reconstruction of $\pazocal{R}$. \label{Rt_rec}}
\end{figure}

\begin{figure}[H]
\centering
	\begin{minipage}{0.45\linewidth}
	\includegraphics[width=1.0\linewidth]{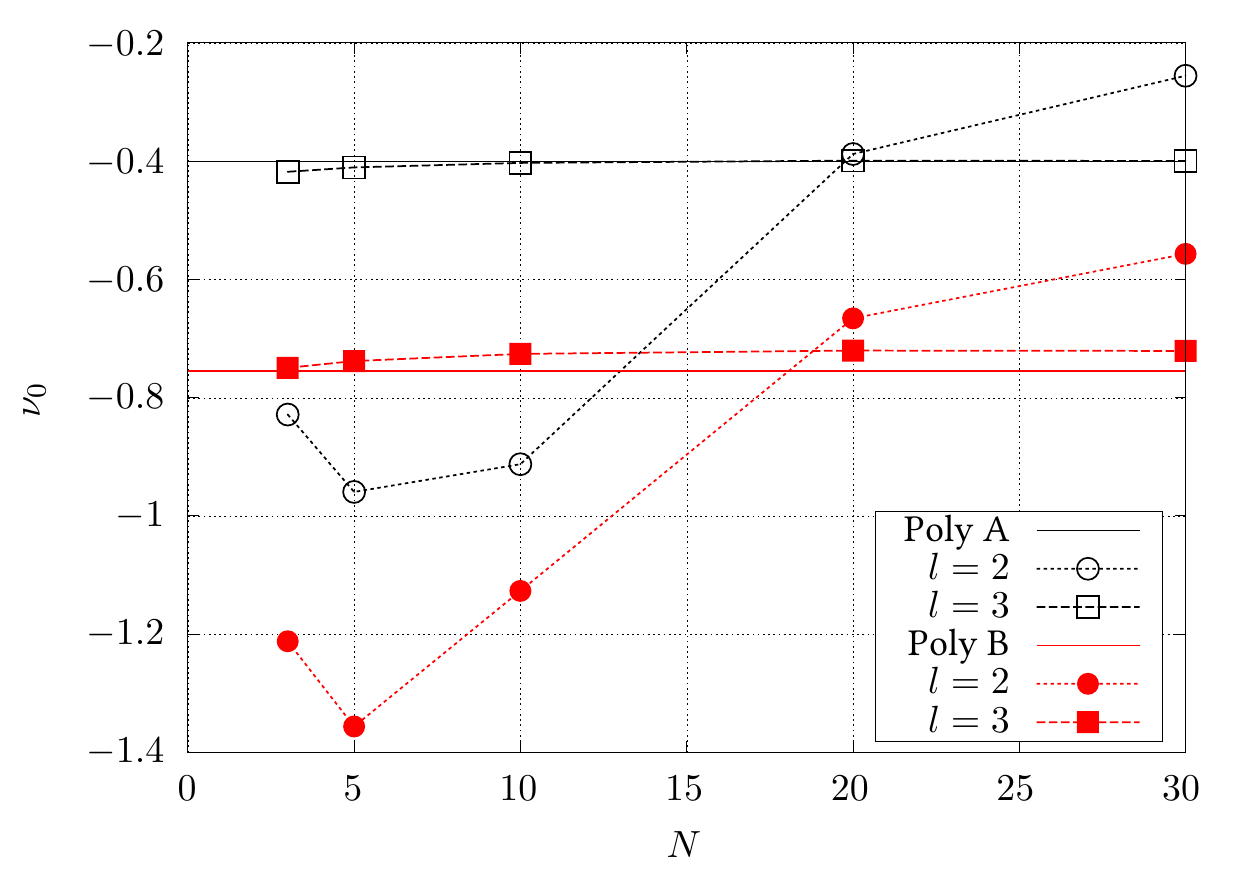}
	\end{minipage}
	\quad
	\begin{minipage}{0.45\linewidth}
	\includegraphics[width=1.0\linewidth]{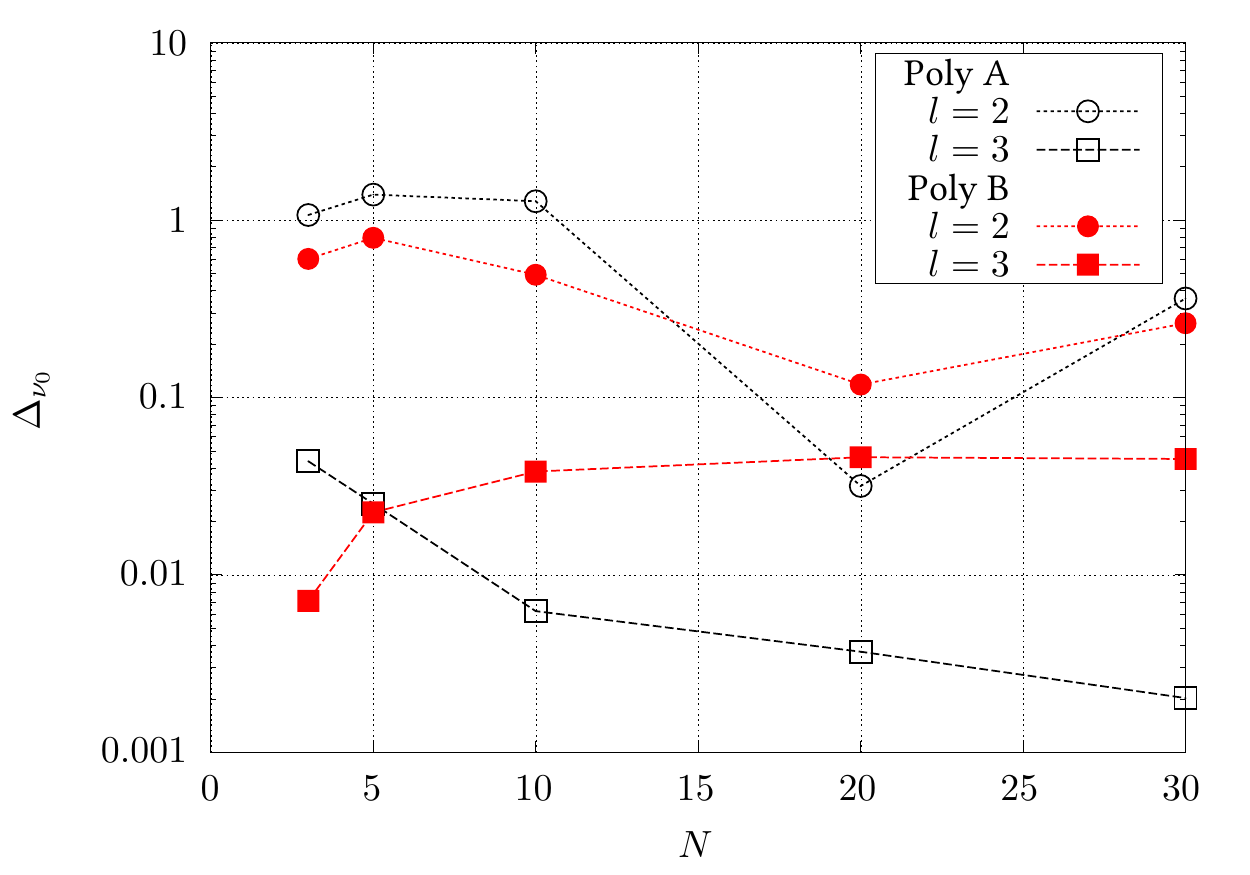}
	\end{minipage}
	\caption{Here we present the reconstruction of $\nu_0$ as function of the number of used eigenvalues $N$, for fitting the spectra of the two polytrope models A and B for $l=$ 2 and 3. \textbf{Left panel:}  The reconstruction of $\nu_0$ (black and red symbols) compared to the true values (black and red solid line). \textbf{Right panel:} The relative errors for the reconstruction of $\nu_0$.  \label{nu0_rec}}
\end{figure}

As introduced in Sec. \ref{Recovering Fundamental Parameters}, one can use the analytic result for $n(E)$ in eq. \eqref{nE_master} to find the parameters $(\pazocal{R}, \nu_0)$, as well as potentially $P_0 = P (\rho_0)$, from spectrum fitting.  Since the analytic Bohr-Sommerfeld result for $n(E_{0n})$ does not include the fundamental mode, we start from the $n=1$ mode, both for $l=2$ and $l=3$. Fig. \ref{Rt_rec} shows the reconstruction of $\pazocal{R}$ in the left panel, while the relative errors are provided in the right panel. In Fig. \ref{nu0_rec} we present the corresponding results for $\nu_0$. On the x-axis we report the number of used modes for the fitting of the spectrum. The relative errors for $\pazocal{R}$ decrease when more modes are included. This is expected since it appears as only parameter that determines the $\sim n^2$ scaling for high eigenvalues. Something similar is initially found for $\nu_0$, however the precision for $l=2$ drops again. Note that the overall accuracy, both for $\pazocal{R}$ and $\nu_0$, is crucially better for $l=3$ modes compared to the ones for $l=2$. This is in agreement with the previous results for the direct spectrum prediction in Sec. \ref{Calculating the Spectrum}.
\par
Although it is possible to fit the parameter $C_0$ within a few percent error for $l=3$, it is not close to the true value, calculated from eq. \eqref{params} and the method thus fails to reconstruct a useful relation for $P(\rho_0)$. Looking at the numerical values of $(\pazocal{R}, C_2, C_0)$, it turns out that $C_0$ is typically much smaller and contributes only to low eigenvalues, where the present method is least precise. However, note that $(\pazocal{R}, \nu_0)$ are implicitly related to the equation of state through the TOV equations and our results thus still impose strong constraints on it.
\subsection{Constraining the Equation of State in the Center}\label{Results_reconstruction_EOS}
\begin{figure}[H]
\centering
	\begin{minipage}{0.45\linewidth}
	\includegraphics[width=1.0\linewidth]{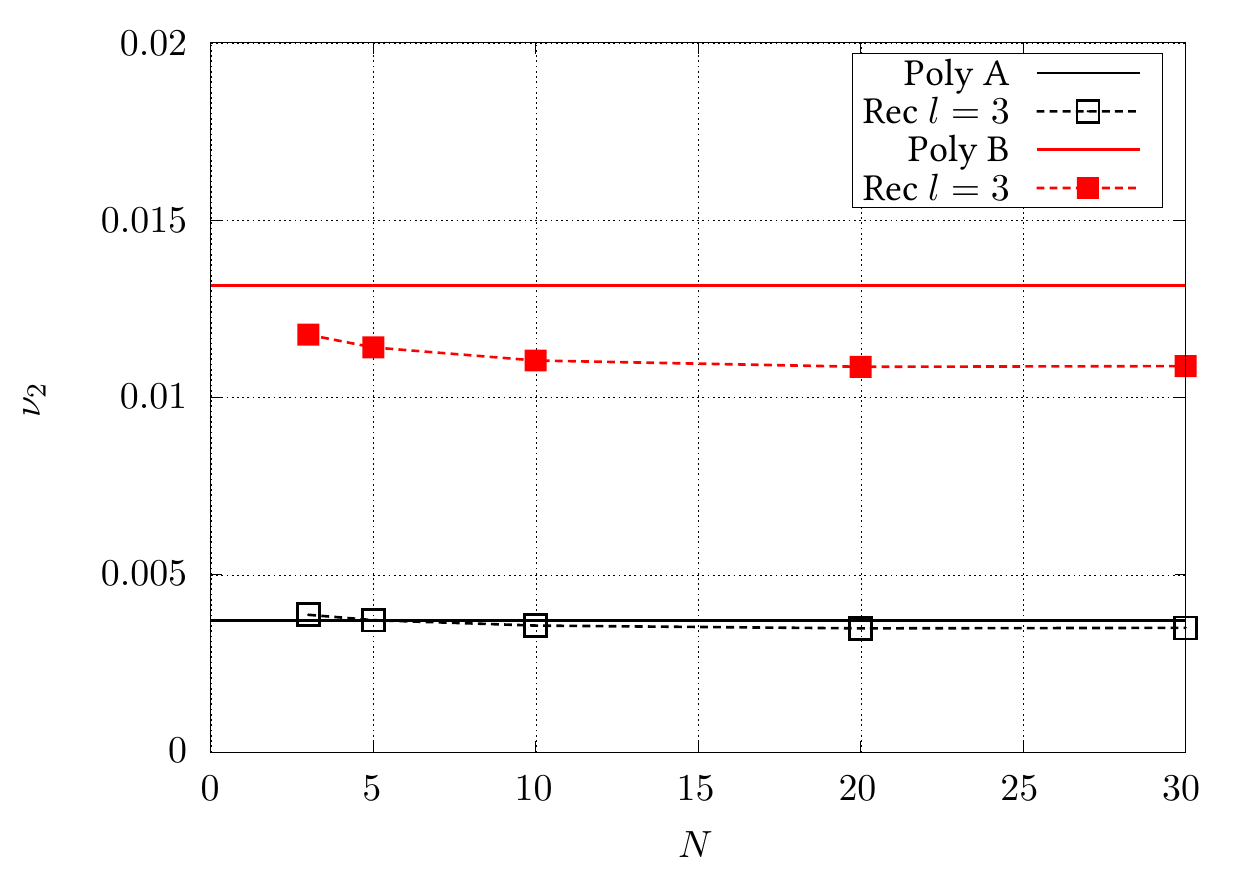}
	\end{minipage}
	\quad
	\begin{minipage}{0.45\linewidth}
	\includegraphics[width=1.0\linewidth]{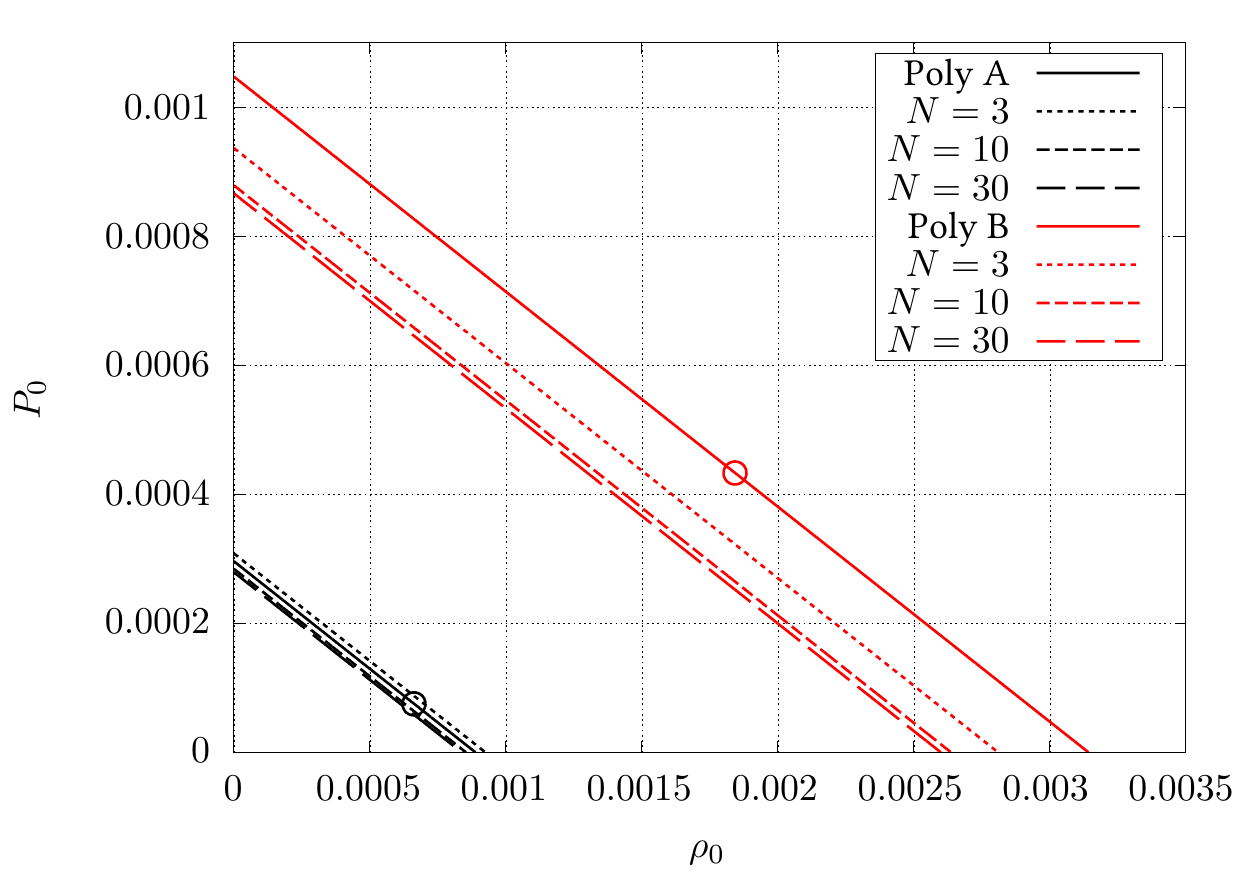}
	\end{minipage}
	\caption{Here we show the results for the reconstruction of $\nu_2$ and $P_0(\rho_0)$ for the two polytrope models A and B. \textbf{Left panel:} The reconstructed values for $\nu_2$, by using different numbers of modes (boxes), are compared to the exact value (solid lines). \textbf{Right panel:}  The relation for $P(\rho_0)$ obtained from the true relation (solid lines) along with the reconstructed ones using different number of modes (dashed lines). The exact values $(P_0,\rho_0)$ are shown as circles. \label{nu2_eos_rec}}
\end{figure}

\begin{figure}[H]
\centering
	\begin{minipage}{0.45\linewidth}
	\includegraphics[width=1.0\linewidth]{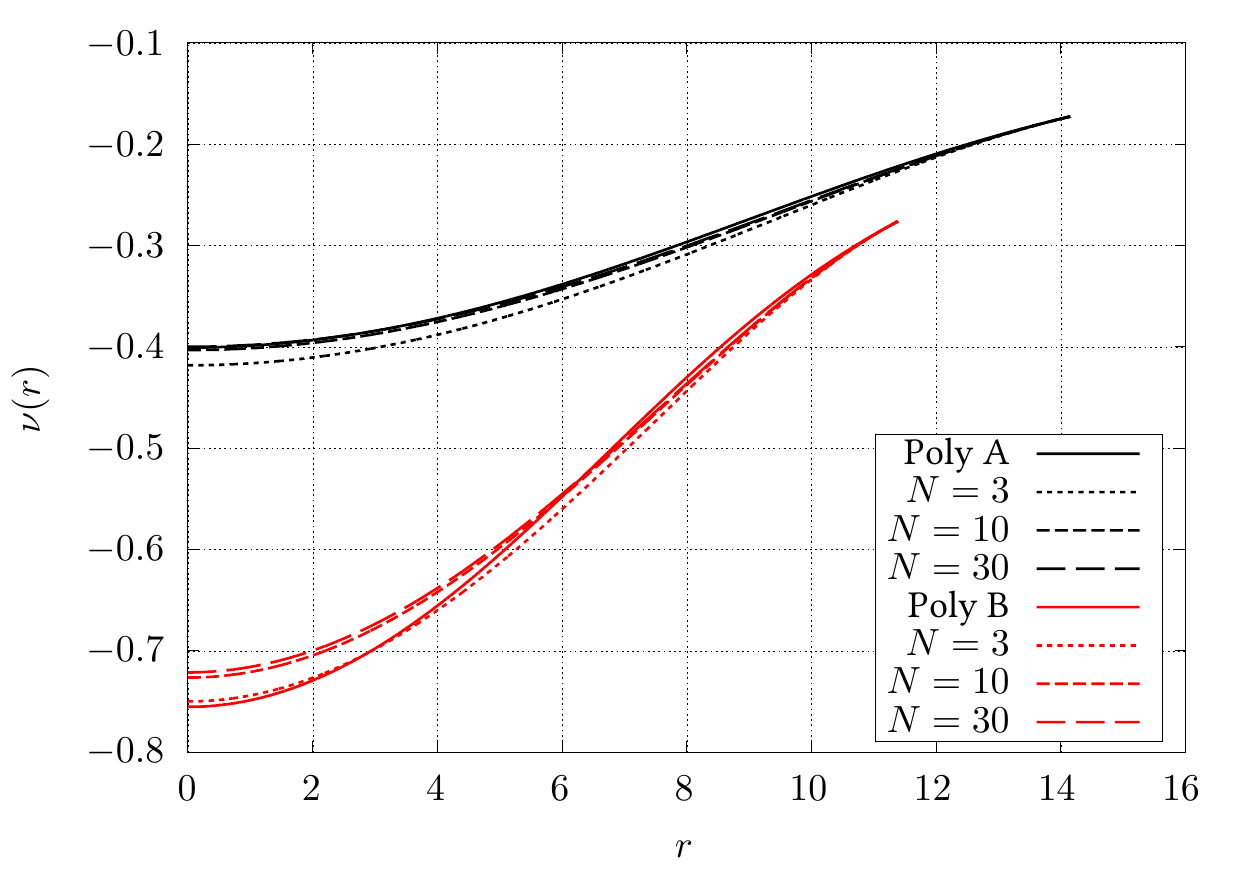}
	\end{minipage}
	\quad
	\begin{minipage}{0.45\linewidth}
	\includegraphics[width=1.0\linewidth]{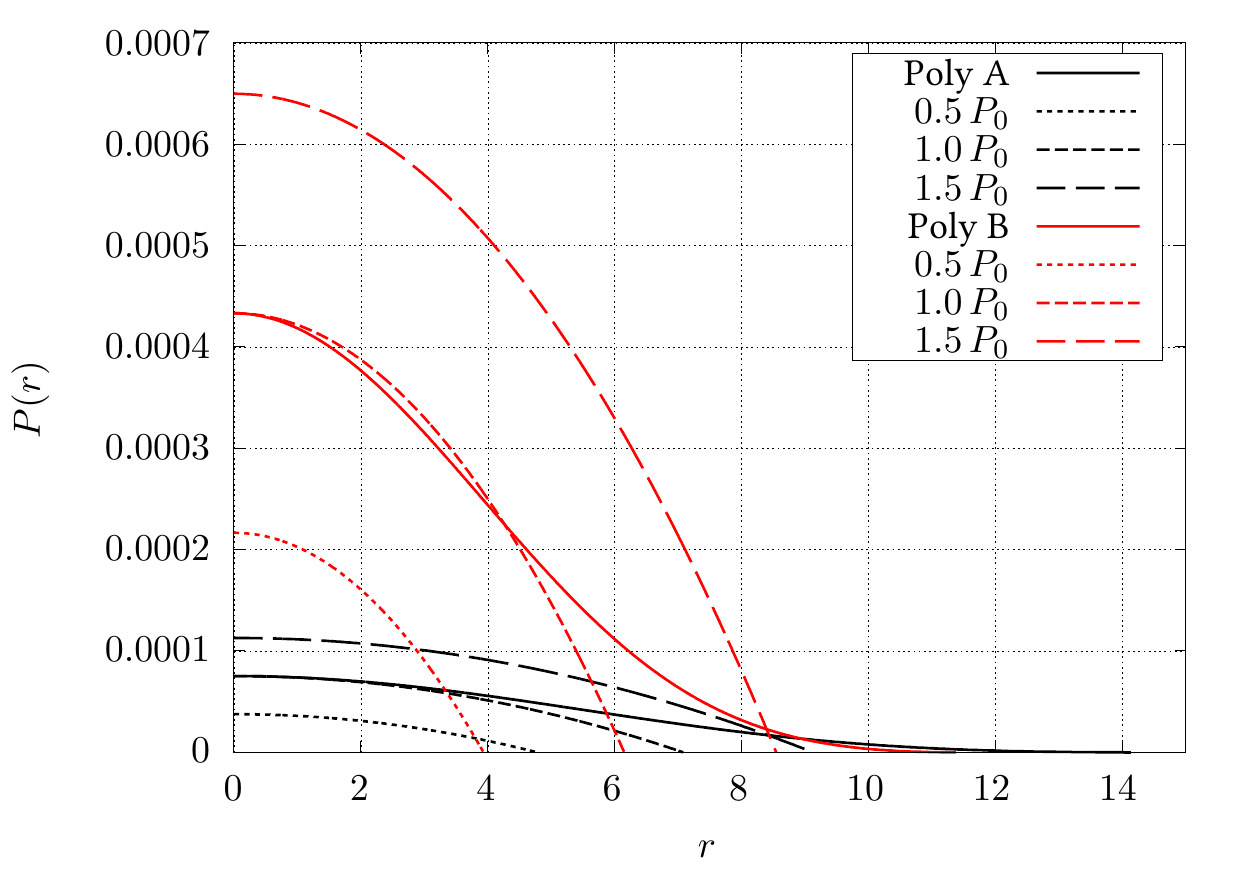}
	\end{minipage}
	\caption{Here we show the reconstructed functions for $\nu(r)$ and $P(r, P_0)$ for the two polytrope models A and B. \textbf{Left panel:} The reconstructed metric functions $\nu(r)$ using a different number of modes (dashed lines) are compared to the exact functions (solid lines). \textbf{Right panel:}{ The reconstructed pressures parameterized by $P_0$ (dashed lines) are compared to the exact ones (solid lines).}\label{nur_rec}}
\end{figure}

Using the spectral fitting result for $\nu_0$ of the previous Sec. \ref{Results_reconstruction_param} allows one to approximate the equation of state at the center, as described in Sec. \ref{Recovering the Equation of State}. Although the approach is approximate, the actual reconstruction yields useful results, as we demonstrate in Fig. \ref{nu2_eos_rec}. In the left panel we show the reconstructed value of $\nu_2$, while in the right panel the relation for the central pressure $P_0$ as function of central density $\rho_0$ is presented. For both polytropes we have used the $l=3$ modes, since $\nu_0$ obtained from the $l=2$ modes has too large errors, as long as not many modes are included. We show the number of used modes in the reconstruction of $\nu_0$ by using different dashed lines. Surprisingly, we find a quite good approximation if a small sample of modes is used, for both polytropes, while the asymptotic value is better for the less compact Poly A model. It does not converge to the right value for the Poly B model, but is still clearly distinguishable from the other model.
From the reconstruction of $\nu_0$ and $\nu_2$ it is possible to approximate $\nu(r)$ throughout the whole star and $P(r, P_0)$ in the central region. Our results for this are shown in the left and right panel of Fig. \ref{nur_rec} for both polytropes. While the reconstruction of $\nu(r)$ is quite robust, considering the involved approximations, $P(r,P_0)$ is only valid in the central region. The parameterization with respect to $P_0$ has negligible influence on the slope, but clearly shifts the absolute values. The relation for the expanded density $\rho(r)$, as it was derived in \cite{1983ApJS...53...73L}, also includes the adiabatic index, which introduces another unknown parameter to the problem, thus we do not consider it here.
%

%
\section{Discussion}\label{Discussion}
%
In this section we discuss our previously shown results and their implications. Sec. \ref{disc_direct} covers the direct spectrum problem, while Sec. \ref{disc_inverse} addresses the inverse problem.
\subsection{Direct Problem}\label{disc_direct}
We start our discussion with the findings that are directly related to the precision of the Bohr-Sommerfeld description of the spectrum, before we continue with the universal relation of the fundamental axial QNM mode. The maximum value for the spacing of neighboring QNMs and the role of the discontinuity follow afterwards.
\subsubsection{Accuracy of the Method}\label{Accuracy of the Method}
We have calculated the real part of $\omega_n^2$ for the QNM spectrum of different polytropes and constant density stars with the new Bohr-Sommerfeld rule derived in Sec. \ref{Bohr-Sommerfeld}. Using a linearized tortoise coordinate transformation and an approximation of the axial perturbation potential, we were able to derive an analytic formula for the spectrum eq. \eqref{nE_master} and found a polynomial expansion for moderate and high eigenvalues eq. \eqref{E_n_app}. Keeping in mind the various approximations being made to derive this result and the approximations for the potential that was used in Sec. \ref{Direct Spectrum Problem}, the method turns out to be a good approximation for the real part of the spectrum. Even low eigenvalues, for which the underlying WKB method and the approximated potential can not be expected to work reliable, are approximatively captured. The accuracy improves significantly for increasing $n$ and $l$, which is expected from WKB theory and shows that our semi-analytic approach could be a powerful tool to determine  very high eigenvalues, where numerical codes might face problems. For moderate $n$ we find relative errors around $\sim 1 - 10\,\%$, while this improves below $1\, \%$ for $l=3$ and high eigenvalues. Since the method was derived for QNMs that are ``located'' above the discontinuity, it missed the fundamental mode, but works for all other QNMs.
\subsubsection{Universal Relation}
The universal asteroseismology relation for the fundamental axial and polar QNM, known empirically from \cite{1996ApJ...462..855A,Andersson:1997rn,1999MNRAS.310..797B}, has been motivated analytically for the axial case in Sec. \ref{Universal Relation}. The proposed relation was compared with the actual QNMs of polytropes, constant density stars and gravastars in  Sec. \ref{Universal_scaling}. Although the universal relation is only approximate, the analytical form provides a satisfying explanation of the scaling, since it relates the fundamental mode with the value of the Regge-Wheeler potential at the surface of the star, which is universal for all objects. The relation also includes ultra compact objects, which are effected by higher order corrections in $M/R$ that also follow from our result.
Since the relation does only depend on the star's mass and radius, it is blind to the details of the underlying equation of state, as long as they produce a star with similar mass and radius. As a consequence, only higher modes should embed details of the equation of state, since they correspond to potential regions inside the star that are sensitive to it. Following the integral character of the Bohr-Sommerfeld rule, it is evident that only higher modes, which are explicitly related to the potential inside the star, can in principle contribute significantly to the reconstruction of the detailed equation of state.
\subsubsection{Maximum Spacing of Neighboring QNMs}
The predicted spacing of moderate and large neighboring QNMs as $\Delta \omega_n \approx  \sqrt{\Delta E_n} = \pi/\pazocal{R}$, has first been found in \cite{PhysRevD.83.064012} using another WKB approach to approximate the high frequency regime of axial and polar QNMs. The spacing of modes is not only extremely simple, but also very precise throughout different polytropes and constant density stars. We have verified this result for our neutron star models by comparing numerically obtained QNM spectra with the predicted spacing. Afterwards, in Sec. \ref{maximum_spacing} we have shown that the scaling also holds for ultra compact constant density stars being close to the Buchdahl limit. This has to be surprising since these potentials are qualitatively different. We recovered the known result that the spacing first grows (for fixed mass), if one considers increasingly compact stars, but then has a maximum at $R \approx 3.86$, before it rapidly decreases again. We do not find any special properties of the star or the potential at this value. Trapped modes appear for more compact configurations. For general neutron stars, the exact position depends on the equation of state. This is because $\pazocal{R}$ depends on the integrated metric functions $g_{00}$ and $g_{11}$, which have to be determined from the TOV equations. For the internal tortoise coordinate $r^{*}(r=0)$ to take large negative numbers, configurations close or above realistic neutron star compactnesses are needed.
\subsubsection{The Role of the Surface Discontinuity}
The discontinuity of the perturbation potential at the stellar surface is necessary to derive discrete eigenvalues in our approach. Without the discontinuity, the surface of the star plays no special role and demanding outgoing QNM at this matching point is arbitrary. Although the potential of constant density stars and the ones of neutron stars with a crust are not continuous, the ones for polytropes can be. From this point of view one would not expect the Bohr-Sommerfeld rule to be valid, but the predicted QNMs are found with similar precision compared to constant density stars. Note that the expanded Bohr-Sommerfeld rule for the real part of $E_n$ does not explicitly depend on the discontinuity value $\Delta V$, but the imaginary part of $E_{n}$ does. It depends explicitly on $\Delta V$ and diverges in the limit of a continuous potential. A detailed discussion on the importance of discontinuities in higher derivatives of the potential can be found in \cite{PhysRevD.83.064012}, where also rapid changes of the equation of state close to the surface can be found and potentially be applied for further studies.
\par
Here we also want to recall related results for the gravitational perturbations of black holes, where introducing small discontinuities in the potential changed the overall QNM spectrum significantly. In \cite{1996PhRvD..53.4397N} the Regge-Wheeler barrier has been replaced by fitted step potentials with different numbers of steps, which extended earlier work \cite{Chandrasekhar441}. It was found that the related QNM spectrum did not approach the exact one, when the number of potential steps being used was increased. The qualitative behavior of the obtained QNM spectrum is always much closer to the ones for neutron stars. Increasing the number of steps increased the density of QNMs, but not significantly their alignment on the complex plane. Somehow counterintuitive or at least remarkable, it was found that the time-evolution of scattered waves at the step potentials becomes close to the ones scattered at the exact potential. The actual QNMs only contributed weakly to the scattering. A more technical discussion on the significance of QNMs has been presented in \cite{1999JMP....40..980N}. The role of environmental effects around black holes lead in a related toy model to similar effects \cite{Barausse:2014tra,2018arXiv181001295K}.
\subsection{Inverse Problem}\label{disc_inverse}
Here we discuss our results related to the inverse problem. We start with the reconstruction of the perturbation potential from the inversion of the Bohr-Sommerfeld rule, before we continue with the reconstruction of the fundamental properties by fitting our analytic result to the spectrum.
\subsubsection{Reconstruction of the Potential}
By inverting the new Bohr-Sommerfeld rule, we have shown that it is possible to reconstruct the internal perturbation potential from the knowledge of the QNM spectrum. The reconstruction is approximate and not exact, because of the underlying WKB method and approximations being involved. In contrast to the non-uniqueness inherited in the inversion of the classical Bohr-Sommerfeld rule for pure potential wells, the reconstruction of the internal potential for typical neutron stars turns out to be unique. This is because the upper limit of integration corresponds to the surface of the star and the problem reduces to the one turning point problem. The method recovers the internal potential by using consecutive modes and approaches the central region for high eigenvalues. In practice one will be limited by a finite number of modes, which means that the potential can only be known from the surface inwards close to the center. Fortunately it turns out that one approaches the central region quite fast, while large modes are only interesting for the region extremely close to the center. However, in the reconstruction procedure it was assumed that the QNM spectrum is known with pristine accuracy and other effects are neglected.
\par
As it was pointed out recently in \cite{Suvorov:2018bvs}, knowing the perturbation potential can be used to determine the equation of state and solve the inverse problem for the stellar structure. The here presented methods allows a simple, but approximate reconstruction of the potential. Whether its accuracy is sufficient for the suggested approach could be investigated in future work. 
\subsubsection{Reconstruction of Crucial Parameters and Metric Function}
With the analytic result for $n(E)$ eq. \eqref{nE_master}, which we obtained by applying the new Bohr-Sommerfeld rule to an approximated version of the perturbation potential, it was possible to reconstruct with very good precision two of the three parameters that describe the potential by fitting the exact QNM spectrum. Compared to the inversion of the Bohr-Sommerfeld rule, only a minimum number of three modes is required. The recovered parameters translate to the tortoise radius of the star $\pazocal{R}$, as well as $\nu_0$, which is the central value of the metric function $\nu(r)$. By using the expanded version of $\nu(r)$, which was derived in \cite{1983ApJS...53...73L}, we were able to approximate it to good accuracy throughout the whole star. Also we were able to approximate $P(r, P_0)$ in the central region, which is not unique, but parameterized by the central pressure $P_0$. However, the slope of $P(r)$ is only slightly influenced by the choice of $P_0$ and can be clearly distinguished for both polytropes.
%
\section{Conclusions}\label{Conclusions}
%
In this work we have studied several aspects of the axial QNM spectrum of spherically symmetric and non-rotating neutron stars and used it to constrain their space-times and equations of state. We also motivated an analytic approximation being closely related to the empirically known universal scaling relation for the fundamental axial QNM. Some of our results even extend to ultra compact stars, which can be seen as a toy model for exotic compact objects.
\par
By using standard WKB theory, we first derived a new Bohr-Sommerfeld rule which looks similar to the classical one for potential wells, but is modified by an additional imaginary part. In contrast to potential wells, the relevant potential for neutron stars admits only one classical turning point, which changes the problem significantly. We demonstrated how the new Bohr-Sommerfeld rule can be used to approximate the real part of the axial QNM spectrum of different polytropes with good precision. Especially for large $n$ and $l$, where numerical codes might face problems, it could be of general interest.
\par
Making use of several well justified approximations, we were able to show that the spacing of moderate and large neighboring QNMs has to be a constant, which our method predicts to be $\pi/\pazocal{R}$, where $\pazocal{R}=r^{*}(R)-r^{*}(0)$ is the ``tortoise radius'' of the star, and coincides with the result reported in \cite{PhysRevD.83.064012}. We have also concluded that this scaling predicts a maximum possible spacing, which depends on the radius (for fixed mass). For constant density stars this was shown explicitly. Furthermore, as a useful approximation in the study of the gravitational perturbations of normal neutron stars, we have demonstrated that the tortoise coordinate transformation can be approximated very well with a linear function. It only depends on the mass, the radius and the tortoise radius of the star and simplifies the perturbation equations significantly.
\par
Even more interesting are the implications of our results for the inverse spectrum problem. The new Bohr-Sommerfeld rule can be inverted to reconstruct approximatively the perturbation potential from a known spectrum and potentially be used to determine the equation of state throughout the star \cite{Suvorov:2018bvs}. This approach requires the knowledge of many modes. As an alternative and more robust solution being tailored to the perturbation potential of neutron stars, we have also shown how a minimum number of 3 modes, for the same $l$, are enough to reconstruct fundamental parameters of the star. These are the tortoise radius and the value of the time-time component of the metric in the center. Both parameters can be obtained with good precision and play the dominant role in characterizing the spectrum. We were also able to approximate the time-time component of the metric throughout the whole star and provide a linear relation between central pressure and central density, which constrains any equation of state significantly. Future work in this direction will include a direct reconstruction of the underlying equation of state in these systems.
\acknowledgements
SV wants to thank Andreas Boden, Christian Kr\"uger and Kostas Glampedakis for useful discussions. SV receives the PhD scholarship Landesgraduiertenf\"orderung. The authors acknowledge support from the COST Action GWverse CA16104. 

\bibliography{literatur1}

\begin{thebibliography}{71}%
\makeatletter
\providecommand \@ifxundefined [1]{%
 \@ifx{#1\undefined}
}%
\providecommand \@ifnum [1]{%
 \ifnum #1\expandafter \@firstoftwo
 \else \expandafter \@secondoftwo
 \fi
}%
\providecommand \@ifx [1]{%
 \ifx #1\expandafter \@firstoftwo
 \else \expandafter \@secondoftwo
 \fi
}%
\providecommand \natexlab [1]{#1}%
\providecommand \enquote  [1]{``#1''}%
\providecommand \bibnamefont  [1]{#1}%
\providecommand \bibfnamefont [1]{#1}%
\providecommand \citenamefont [1]{#1}%
\providecommand \href@noop [0]{\@secondoftwo}%
\providecommand \href [0]{\begingroup \@sanitize@url \@href}%
\providecommand \@href[1]{\@@startlink{#1}\@@href}%
\providecommand \@@href[1]{\endgroup#1\@@endlink}%
\providecommand \@sanitize@url [0]{\catcode `\\12\catcode `\$12\catcode
  `\&12\catcode `\#12\catcode `\^12\catcode `\_12\catcode `\%12\relax}%
\providecommand \@@startlink[1]{}%
\providecommand \@@endlink[0]{}%
\providecommand \url  [0]{\begingroup\@sanitize@url \@url }%
\providecommand \@url [1]{\endgroup\@href {#1}{\urlprefix }}%
\providecommand \urlprefix  [0]{URL }%
\providecommand \Eprint [0]{\href }%
\providecommand \doibase [0]{http://dx.doi.org/}%
\providecommand \selectlanguage [0]{\@gobble}%
\providecommand \bibinfo  [0]{\@secondoftwo}%
\providecommand \bibfield  [0]{\@secondoftwo}%
\providecommand \translation [1]{[#1]}%
\providecommand \BibitemOpen [0]{}%
\providecommand \bibitemStop [0]{}%
\providecommand \bibitemNoStop [0]{.\EOS\space}%
\providecommand \EOS [0]{\spacefactor3000\relax}%
\providecommand \BibitemShut  [1]{\csname bibitem#1\endcsname}%
\let\auto@bib@innerbib\@empty
\bibitem [{\citenamefont {{Abbott}}\ \emph {et~al.}(2016)\citenamefont
  {{Abbott}} \emph {et~al.}}]{LIGO1}%
  \BibitemOpen
  \bibfield  {author} {\bibinfo {author} {\bibfnamefont {B.~P.}\ \bibnamefont
  {{Abbott}}} \emph {et~al.} (\bibinfo {collaboration} {LIGO Scientific
  Collaboration and Virgo Collaboration}),\ }\href {\doibase
  10.1103/PhysRevLett.116.061102} {\bibfield  {journal} {\bibinfo  {journal}
  {Phys. Rev. Lett.}\ }\textbf {\bibinfo {volume} {116}},\ \bibinfo {pages}
  {061102} (\bibinfo {year} {2016})}\BibitemShut {NoStop}%
\bibitem [{\citenamefont {Abbott}\ \emph {et~al.}(2016)\citenamefont {Abbott}
  \emph {et~al.}}]{PhysRevLett.116.241103}%
  \BibitemOpen
  \bibfield  {author} {\bibinfo {author} {\bibfnamefont {B.~P.}\ \bibnamefont
  {Abbott}} \emph {et~al.} (\bibinfo {collaboration} {LIGO Scientific
  Collaboration and Virgo Collaboration}),\ }\href {\doibase
  10.1103/PhysRevLett.116.241103} {\bibfield  {journal} {\bibinfo  {journal}
  {Phys. Rev. Lett.}\ }\textbf {\bibinfo {volume} {116}},\ \bibinfo {pages}
  {241103} (\bibinfo {year} {2016})}\BibitemShut {NoStop}%
\bibitem [{\citenamefont {Abbott}\ \emph
  {et~al.}(2017{\natexlab{a}})\citenamefont {Abbott} \emph
  {et~al.}}]{PhysRevLett.118.221101}%
  \BibitemOpen
  \bibfield  {author} {\bibinfo {author} {\bibfnamefont {B.~P.}\ \bibnamefont
  {Abbott}} \emph {et~al.} (\bibinfo {collaboration} {LIGO Scientific and Virgo
  Collaboration}),\ }\href {\doibase 10.1103/PhysRevLett.118.221101} {\bibfield
   {journal} {\bibinfo  {journal} {Phys. Rev. Lett.}\ }\textbf {\bibinfo
  {volume} {118}},\ \bibinfo {pages} {221101} (\bibinfo {year}
  {2017}{\natexlab{a}})}\BibitemShut {NoStop}%
\bibitem [{\citenamefont {Abbott}\ \emph
  {et~al.}(2017{\natexlab{b}})\citenamefont {Abbott} \emph
  {et~al.}}]{PhysRevLett.119.141101}%
  \BibitemOpen
  \bibfield  {author} {\bibinfo {author} {\bibfnamefont {B.~P.}\ \bibnamefont
  {Abbott}} \emph {et~al.} (\bibinfo {collaboration} {LIGO Scientific
  Collaboration and Virgo Collaboration}),\ }\href {\doibase
  10.1103/PhysRevLett.119.141101} {\bibfield  {journal} {\bibinfo  {journal}
  {Phys. Rev. Lett.}\ }\textbf {\bibinfo {volume} {119}},\ \bibinfo {pages}
  {141101} (\bibinfo {year} {2017}{\natexlab{b}})}\BibitemShut {NoStop}%
\bibitem [{\citenamefont {Abbott}\ \emph
  {et~al.}(2017{\natexlab{c}})\citenamefont {Abbott} \emph
  {et~al.}}]{PhysRevLett.119.161101}%
  \BibitemOpen
  \bibfield  {author} {\bibinfo {author} {\bibfnamefont {B.~P.}\ \bibnamefont
  {Abbott}} \emph {et~al.} (\bibinfo {collaboration} {LIGO Scientific
  Collaboration and Virgo Collaboration}),\ }\href {\doibase
  10.1103/PhysRevLett.119.161101} {\bibfield  {journal} {\bibinfo  {journal}
  {Phys. Rev. Lett.}\ }\textbf {\bibinfo {volume} {119}},\ \bibinfo {pages}
  {161101} (\bibinfo {year} {2017}{\natexlab{c}})}\BibitemShut {NoStop}%
\bibitem [{\citenamefont {{Thorne}}\ and\ \citenamefont
  {{Campolattaro}}(1967)}]{1967ApJ...149..591T}%
  \BibitemOpen
  \bibfield  {author} {\bibinfo {author} {\bibfnamefont {K.~S.}\ \bibnamefont
  {{Thorne}}}\ and\ \bibinfo {author} {\bibfnamefont {A.}~\bibnamefont
  {{Campolattaro}}},\ }\href {\doibase 10.1086/149288} {\bibfield  {journal}
  {\bibinfo  {journal} {\apj}\ }\textbf {\bibinfo {volume} {149}},\ \bibinfo
  {pages} {591} (\bibinfo {year} {1967})}\BibitemShut {NoStop}%
\bibitem [{\citenamefont {{Lindblom}}\ and\ \citenamefont
  {{Detweiler}}(1983)}]{1983ApJS...53...73L}%
  \BibitemOpen
  \bibfield  {author} {\bibinfo {author} {\bibfnamefont {L.}~\bibnamefont
  {{Lindblom}}}\ and\ \bibinfo {author} {\bibfnamefont {S.~L.}\ \bibnamefont
  {{Detweiler}}},\ }\href {\doibase 10.1086/190884} {\bibfield  {journal}
  {\bibinfo  {journal} {\apjs}\ }\textbf {\bibinfo {volume} {53}},\ \bibinfo
  {pages} {73} (\bibinfo {year} {1983})}\BibitemShut {NoStop}%
\bibitem [{\citenamefont {{Chandrasekhar}}\ and\ \citenamefont
  {{Ferrari}}(1991{\natexlab{a}})}]{1991RSPSA.432..247C}%
  \BibitemOpen
  \bibfield  {author} {\bibinfo {author} {\bibfnamefont {S.}~\bibnamefont
  {{Chandrasekhar}}}\ and\ \bibinfo {author} {\bibfnamefont {V.}~\bibnamefont
  {{Ferrari}}},\ }\href {\doibase 10.1098/rspa.1991.0016} {\bibfield  {journal}
  {\bibinfo  {journal} {Proceedings of the Royal Society of London Series A}\
  }\textbf {\bibinfo {volume} {432}},\ \bibinfo {pages} {247} (\bibinfo {year}
  {1991}{\natexlab{a}})}\BibitemShut {NoStop}%
\bibitem [{\citenamefont {{Kokkotas}}\ and\ \citenamefont
  {{Schutz}}(1986)}]{1986GReGr..18..913K}%
  \BibitemOpen
  \bibfield  {author} {\bibinfo {author} {\bibfnamefont {K.~D.}\ \bibnamefont
  {{Kokkotas}}}\ and\ \bibinfo {author} {\bibfnamefont {B.~F.}\ \bibnamefont
  {{Schutz}}},\ }\href {\doibase 10.1007/BF00773556} {\bibfield  {journal}
  {\bibinfo  {journal} {General Relativity and Gravitation}\ }\textbf {\bibinfo
  {volume} {18}},\ \bibinfo {pages} {913} (\bibinfo {year} {1986})}\BibitemShut
  {NoStop}%
\bibitem [{\citenamefont {{Kokkotas}}\ and\ \citenamefont
  {{Schutz}}(1992)}]{1992MNRAS.255..119K}%
  \BibitemOpen
  \bibfield  {author} {\bibinfo {author} {\bibfnamefont {K.~D.}\ \bibnamefont
  {{Kokkotas}}}\ and\ \bibinfo {author} {\bibfnamefont {B.~F.}\ \bibnamefont
  {{Schutz}}},\ }\href {\doibase 10.1093/mnras/255.1.119} {\bibfield  {journal}
  {\bibinfo  {journal} {{Mon. Not. R. Astron. Soc.}}\ }\textbf {\bibinfo
  {volume} {255}},\ \bibinfo {pages} {119} (\bibinfo {year}
  {1992})}\BibitemShut {NoStop}%
\bibitem [{\citenamefont {{Leins}}\ \emph {et~al.}(1993)\citenamefont
  {{Leins}}, \citenamefont {{Nollert}},\ and\ \citenamefont
  {{Soffel}}}]{1993PhRvD..48.3467L}%
  \BibitemOpen
  \bibfield  {author} {\bibinfo {author} {\bibfnamefont {M.}~\bibnamefont
  {{Leins}}}, \bibinfo {author} {\bibfnamefont {H.-P.}\ \bibnamefont
  {{Nollert}}}, \ and\ \bibinfo {author} {\bibfnamefont {M.~H.}\ \bibnamefont
  {{Soffel}}},\ }\href {\doibase 10.1103/PhysRevD.48.3467} {\bibfield
  {journal} {\bibinfo  {journal} {\prd}\ }\textbf {\bibinfo {volume} {48}},\
  \bibinfo {pages} {3467} (\bibinfo {year} {1993})}\BibitemShut {NoStop}%
\bibitem [{\citenamefont {{Chandrasekhar}}\ and\ \citenamefont
  {{Ferrari}}(1991{\natexlab{b}})}]{1991RSPSA.434..449C}%
  \BibitemOpen
  \bibfield  {author} {\bibinfo {author} {\bibfnamefont {S.}~\bibnamefont
  {{Chandrasekhar}}}\ and\ \bibinfo {author} {\bibfnamefont {V.}~\bibnamefont
  {{Ferrari}}},\ }\href {\doibase 10.1098/rspa.1991.0104} {\bibfield  {journal}
  {\bibinfo  {journal} {Proceedings of the Royal Society of London Series A}\
  }\textbf {\bibinfo {volume} {434}},\ \bibinfo {pages} {449} (\bibinfo {year}
  {1991}{\natexlab{b}})}\BibitemShut {NoStop}%
\bibitem [{\citenamefont {{Kokkotas}}(1994)}]{1994MNRAS.268.1015K}%
  \BibitemOpen
  \bibfield  {author} {\bibinfo {author} {\bibfnamefont {K.~D.}\ \bibnamefont
  {{Kokkotas}}},\ }\href {\doibase 10.1093/mnras/268.4.1015} {\bibfield
  {journal} {\bibinfo  {journal} {{Mon. Not. R. Astron. Soc.}}\ }\textbf
  {\bibinfo {volume} {268}},\ \bibinfo {pages} {1015} (\bibinfo {year}
  {1994})}\BibitemShut {NoStop}%
\bibitem [{\citenamefont {Kokkotas}(1995)}]{Kokkotas:1995av}%
  \BibitemOpen
  \bibfield  {author} {\bibinfo {author} {\bibfnamefont {K.~D.}\ \bibnamefont
  {Kokkotas}},\ }in\ \href@noop {} {\emph {\bibinfo {booktitle} {{Relativistic
  gravitation and gravitational radiation. Proceedings, School of Physics, Les
  Houches, France, September 26-October 6, 1995}}}}\ (\bibinfo {year} {1995})\
  pp.\ \bibinfo {pages} {89--102},\ \Eprint
  {http://arxiv.org/abs/gr-qc/9603024} {arXiv:gr-qc/9603024 [gr-qc]}
  \BibitemShut {NoStop}%
\bibitem [{\citenamefont {Tominaga}\ \emph {et~al.}(1999)\citenamefont
  {Tominaga}, \citenamefont {Saijo},\ and\ \citenamefont
  {Maeda}}]{PhysRevD.60.024004}%
  \BibitemOpen
  \bibfield  {author} {\bibinfo {author} {\bibfnamefont {K.}~\bibnamefont
  {Tominaga}}, \bibinfo {author} {\bibfnamefont {M.}~\bibnamefont {Saijo}}, \
  and\ \bibinfo {author} {\bibfnamefont {K.-i.}\ \bibnamefont {Maeda}},\ }\href
  {\doibase 10.1103/PhysRevD.60.024004} {\bibfield  {journal} {\bibinfo
  {journal} {Phys. Rev. D}\ }\textbf {\bibinfo {volume} {60}},\ \bibinfo
  {pages} {024004} (\bibinfo {year} {1999})}\BibitemShut {NoStop}%
\bibitem [{\citenamefont {{Ferrari}}\ and\ \citenamefont
  {{Kokkotas}}(2000)}]{2000PhRvD..62j7504F}%
  \BibitemOpen
  \bibfield  {author} {\bibinfo {author} {\bibfnamefont {V.}~\bibnamefont
  {{Ferrari}}}\ and\ \bibinfo {author} {\bibfnamefont {K.~D.}\ \bibnamefont
  {{Kokkotas}}},\ }\href {\doibase 10.1103/PhysRevD.62.107504} {\bibfield
  {journal} {\bibinfo  {journal} {\prd}\ }\textbf {\bibinfo {volume} {62}},\
  \bibinfo {eid} {107504} (\bibinfo {year} {2000})},\ \Eprint
  {http://arxiv.org/abs/gr-qc/0008057} {gr-qc/0008057} \BibitemShut {NoStop}%
\bibitem [{\citenamefont {{Kokkotas}}\ and\ \citenamefont
  {{Schmidt}}(1999)}]{1999LRR.....2....2K}%
  \BibitemOpen
  \bibfield  {author} {\bibinfo {author} {\bibfnamefont {K.~D.}\ \bibnamefont
  {{Kokkotas}}}\ and\ \bibinfo {author} {\bibfnamefont {B.~G.}\ \bibnamefont
  {{Schmidt}}},\ }\href {\doibase 10.12942/lrr-1999-2} {\bibfield  {journal}
  {\bibinfo  {journal} {Living Reviews in Relativity}\ }\textbf {\bibinfo
  {volume} {2}},\ \bibinfo {eid} {2} (\bibinfo {year} {1999})},\ \Eprint
  {http://arxiv.org/abs/gr-qc/9909058} {gr-qc/9909058} \BibitemShut {NoStop}%
\bibitem [{\citenamefont {{Nollert}}(1999)}]{1999CQGra..16R.159N}%
  \BibitemOpen
  \bibfield  {author} {\bibinfo {author} {\bibfnamefont {H.-P.}\ \bibnamefont
  {{Nollert}}},\ }\href {\doibase 10.1088/0264-9381/16/12/201} {\bibfield
  {journal} {\bibinfo  {journal} {Classical and Quantum Gravity}\ }\textbf
  {\bibinfo {volume} {16}},\ \bibinfo {pages} {R159} (\bibinfo {year}
  {1999})}\BibitemShut {NoStop}%
\bibitem [{\citenamefont {{Berti}}\ \emph {et~al.}(2009)\citenamefont
  {{Berti}}, \citenamefont {{Cardoso}},\ and\ \citenamefont
  {{Starinets}}}]{2009CQGra..26p3001B}%
  \BibitemOpen
  \bibfield  {author} {\bibinfo {author} {\bibfnamefont {E.}~\bibnamefont
  {{Berti}}}, \bibinfo {author} {\bibfnamefont {V.}~\bibnamefont {{Cardoso}}},
  \ and\ \bibinfo {author} {\bibfnamefont {A.~O.}\ \bibnamefont
  {{Starinets}}},\ }\href {\doibase 10.1088/0264-9381/26/16/163001} {\bibfield
  {journal} {\bibinfo  {journal} {Classical and Quantum Gravity}\ }\textbf
  {\bibinfo {volume} {26}},\ \bibinfo {eid} {163001} (\bibinfo {year}
  {2009})},\ \Eprint {http://arxiv.org/abs/0905.2975} {arXiv:0905.2975 [gr-qc]}
  \BibitemShut {NoStop}%
\bibitem [{\citenamefont {Cardoso}\ and\ \citenamefont
  {Pani}(2017)}]{Cardoso:2017njb}%
  \BibitemOpen
  \bibfield  {author} {\bibinfo {author} {\bibfnamefont {V.}~\bibnamefont
  {Cardoso}}\ and\ \bibinfo {author} {\bibfnamefont {P.}~\bibnamefont {Pani}},\
  }\href@noop {} {\  (\bibinfo {year} {2017})},\ \Eprint
  {http://arxiv.org/abs/1707.03021} {arXiv:1707.03021 [gr-qc]} \BibitemShut
  {NoStop}%
\bibitem [{\citenamefont {{Andersson}}\ and\ \citenamefont
  {{Kokkotas}}(1996)}]{1996PhRvL..77.4134A}%
  \BibitemOpen
  \bibfield  {author} {\bibinfo {author} {\bibfnamefont {N.}~\bibnamefont
  {{Andersson}}}\ and\ \bibinfo {author} {\bibfnamefont {K.~D.}\ \bibnamefont
  {{Kokkotas}}},\ }\href {\doibase 10.1103/PhysRevLett.77.4134} {\bibfield
  {journal} {\bibinfo  {journal} {Physical Review Letters}\ }\textbf {\bibinfo
  {volume} {77}},\ \bibinfo {pages} {4134} (\bibinfo {year} {1996})},\ \Eprint
  {http://arxiv.org/abs/gr-qc/9610035} {gr-qc/9610035} \BibitemShut {NoStop}%
\bibitem [{\citenamefont {Andersson}\ and\ \citenamefont
  {Kokkotas}(1998)}]{Andersson:1997rn}%
  \BibitemOpen
  \bibfield  {author} {\bibinfo {author} {\bibfnamefont {N.}~\bibnamefont
  {Andersson}}\ and\ \bibinfo {author} {\bibfnamefont {K.~D.}\ \bibnamefont
  {Kokkotas}},\ }\href {\doibase 10.1046/j.1365-8711.1998.01840.x} {\bibfield
  {journal} {\bibinfo  {journal} {Mon. Not. Roy. Astron. Soc.}\ }\textbf
  {\bibinfo {volume} {299}},\ \bibinfo {pages} {1059} (\bibinfo {year}
  {1998})},\ \Eprint {http://arxiv.org/abs/gr-qc/9711088} {arXiv:gr-qc/9711088
  [gr-qc]} \BibitemShut {NoStop}%
\bibitem [{\citenamefont {{Benhar}}\ \emph {et~al.}(1999)\citenamefont
  {{Benhar}}, \citenamefont {{Berti}},\ and\ \citenamefont
  {{Ferrari}}}]{1999MNRAS.310..797B}%
  \BibitemOpen
  \bibfield  {author} {\bibinfo {author} {\bibfnamefont {O.}~\bibnamefont
  {{Benhar}}}, \bibinfo {author} {\bibfnamefont {E.}~\bibnamefont {{Berti}}}, \
  and\ \bibinfo {author} {\bibfnamefont {V.}~\bibnamefont {{Ferrari}}},\ }\href
  {\doibase 10.1046/j.1365-8711.1999.02983.x} {\bibfield  {journal} {\bibinfo
  {journal} {\mnras}\ }\textbf {\bibinfo {volume} {310}},\ \bibinfo {pages}
  {797} (\bibinfo {year} {1999})},\ \Eprint
  {http://arxiv.org/abs/gr-qc/9901037} {gr-qc/9901037} \BibitemShut {NoStop}%
\bibitem [{\citenamefont {Kokkotas}\ \emph {et~al.}(2001)\citenamefont
  {Kokkotas}, \citenamefont {Apostolatos},\ and\ \citenamefont
  {Andersson}}]{Kokkotas:1999mn}%
  \BibitemOpen
  \bibfield  {author} {\bibinfo {author} {\bibfnamefont {K.~D.}\ \bibnamefont
  {Kokkotas}}, \bibinfo {author} {\bibfnamefont {T.~A.}\ \bibnamefont
  {Apostolatos}}, \ and\ \bibinfo {author} {\bibfnamefont {N.}~\bibnamefont
  {Andersson}},\ }\href {\doibase 10.1046/j.1365-8711.2001.03945.x} {\bibfield
  {journal} {\bibinfo  {journal} {Mon. Not. Roy. Astron. Soc.}\ }\textbf
  {\bibinfo {volume} {320}},\ \bibinfo {pages} {307} (\bibinfo {year}
  {2001})},\ \Eprint {http://arxiv.org/abs/gr-qc/9901072} {arXiv:gr-qc/9901072
  [gr-qc]} \BibitemShut {NoStop}%
\bibitem [{\citenamefont {{Lindblom}}(1992)}]{1992ApJ...398..569L}%
  \BibitemOpen
  \bibfield  {author} {\bibinfo {author} {\bibfnamefont {L.}~\bibnamefont
  {{Lindblom}}},\ }\href {\doibase 10.1086/171882} {\bibfield  {journal}
  {\bibinfo  {journal} {\apj}\ }\textbf {\bibinfo {volume} {398}},\ \bibinfo
  {pages} {569} (\bibinfo {year} {1992})}\BibitemShut {NoStop}%
\bibitem [{\citenamefont {{Lindblom}}\ and\ \citenamefont
  {{Indik}}(2012)}]{2012PhRvD..86h4003L}%
  \BibitemOpen
  \bibfield  {author} {\bibinfo {author} {\bibfnamefont {L.}~\bibnamefont
  {{Lindblom}}}\ and\ \bibinfo {author} {\bibfnamefont {N.~M.}\ \bibnamefont
  {{Indik}}},\ }\href {\doibase 10.1103/PhysRevD.86.084003} {\bibfield
  {journal} {\bibinfo  {journal} {\prd}\ }\textbf {\bibinfo {volume} {86}},\
  \bibinfo {eid} {084003} (\bibinfo {year} {2012})},\ \Eprint
  {http://arxiv.org/abs/1207.3744} {arXiv:1207.3744 [astro-ph.HE]} \BibitemShut
  {NoStop}%
\bibitem [{\citenamefont {Lindblom}\ and\ \citenamefont
  {Indik}(2014)}]{Lindblom:2013kra}%
  \BibitemOpen
  \bibfield  {author} {\bibinfo {author} {\bibfnamefont {L.}~\bibnamefont
  {Lindblom}}\ and\ \bibinfo {author} {\bibfnamefont {N.~M.}\ \bibnamefont
  {Indik}},\ }\href {\doibase 10.1103/PhysRevD.89.064003,
  10.1103/PhysRevD.93.129903} {\bibfield  {journal} {\bibinfo  {journal} {Phys.
  Rev.}\ }\textbf {\bibinfo {volume} {D89}},\ \bibinfo {pages} {064003}
  (\bibinfo {year} {2014})},\ \bibinfo {note} {[Erratum: Phys.
  Rev.D93,no.12,129903(2016)]},\ \Eprint {http://arxiv.org/abs/1310.0803}
  {arXiv:1310.0803 [astro-ph.HE]} \BibitemShut {NoStop}%
\bibitem [{\citenamefont {Lindblom}(2014)}]{Lindblom:2014sha}%
  \BibitemOpen
  \bibfield  {author} {\bibinfo {author} {\bibfnamefont {L.}~\bibnamefont
  {Lindblom}},\ }\bibfield  {booktitle} {\emph {\bibinfo {booktitle}
  {{Proceedings, 5th Mexican Meeting on Mathematical and Experimental Physics:
  Mexico City, Mexico, September 9-13, 2013}}},\ }\href {\doibase
  10.1063/1.4861951} {\bibfield  {journal} {\bibinfo  {journal} {AIP Conf.
  Proc.}\ }\textbf {\bibinfo {volume} {1577}},\ \bibinfo {pages} {153}
  (\bibinfo {year} {2014})},\ \Eprint {http://arxiv.org/abs/1402.0035}
  {arXiv:1402.0035 [astro-ph.HE]} \BibitemShut {NoStop}%
\bibitem [{\citenamefont {Lindblom}(2018)}]{Lindblom:2018ntw}%
  \BibitemOpen
  \bibfield  {author} {\bibinfo {author} {\bibfnamefont {L.}~\bibnamefont
  {Lindblom}},\ }\href@noop {} {\  (\bibinfo {year} {2018})},\ \Eprint
  {http://arxiv.org/abs/1807.02538} {arXiv:1807.02538 [astro-ph.HE]}
  \BibitemShut {NoStop}%
\bibitem [{\citenamefont {{Mazur}}\ and\ \citenamefont
  {{Mottola}}(2001)}]{2001gr.qc.....9035M}%
  \BibitemOpen
  \bibfield  {author} {\bibinfo {author} {\bibfnamefont {P.~O.}\ \bibnamefont
  {{Mazur}}}\ and\ \bibinfo {author} {\bibfnamefont {E.}~\bibnamefont
  {{Mottola}}},\ }\href@noop {} {\bibfield  {journal} {\bibinfo  {journal}
  {ArXiv General Relativity and Quantum Cosmology e-prints}\ } (\bibinfo {year}
  {2001})},\ \Eprint {http://arxiv.org/abs/gr-qc/0109035} {gr-qc/0109035}
  \BibitemShut {NoStop}%
\bibitem [{\citenamefont {{Visser}}\ and\ \citenamefont
  {{Wiltshire}}(2004)}]{2004CQGra..21.1135V}%
  \BibitemOpen
  \bibfield  {author} {\bibinfo {author} {\bibfnamefont {M.}~\bibnamefont
  {{Visser}}}\ and\ \bibinfo {author} {\bibfnamefont {D.~L.}\ \bibnamefont
  {{Wiltshire}}},\ }\href {\doibase 10.1088/0264-9381/21/4/027} {\bibfield
  {journal} {\bibinfo  {journal} {Classical and Quantum Gravity}\ }\textbf
  {\bibinfo {volume} {21}},\ \bibinfo {pages} {1135} (\bibinfo {year}
  {2004})},\ \Eprint {http://arxiv.org/abs/gr-qc/0310107} {gr-qc/0310107}
  \BibitemShut {NoStop}%
\bibitem [{\citenamefont {Damour}\ and\ \citenamefont
  {Solodukhin}(2007)}]{PhysRevD.76.024016}%
  \BibitemOpen
  \bibfield  {author} {\bibinfo {author} {\bibfnamefont {T.}~\bibnamefont
  {Damour}}\ and\ \bibinfo {author} {\bibfnamefont {S.~N.}\ \bibnamefont
  {Solodukhin}},\ }\href {\doibase 10.1103/PhysRevD.76.024016} {\bibfield
  {journal} {\bibinfo  {journal} {Phys. Rev. D}\ }\textbf {\bibinfo {volume}
  {76}},\ \bibinfo {pages} {024016} (\bibinfo {year} {2007})},\ \Eprint
  {http://arxiv.org/abs/0704.2667} {arXiv:0704.2667} \BibitemShut {NoStop}%
\bibitem [{\citenamefont {{Cardoso}}\ \emph {et~al.}(2016)\citenamefont
  {{Cardoso}}, \citenamefont {{Hopper}}, \citenamefont {{Macedo}},
  \citenamefont {{Palenzuela}},\ and\ \citenamefont
  {{Pani}}}]{2016PhRvD..94h4031C}%
  \BibitemOpen
  \bibfield  {author} {\bibinfo {author} {\bibfnamefont {V.}~\bibnamefont
  {{Cardoso}}}, \bibinfo {author} {\bibfnamefont {S.}~\bibnamefont {{Hopper}}},
  \bibinfo {author} {\bibfnamefont {C.~F.~B.}\ \bibnamefont {{Macedo}}},
  \bibinfo {author} {\bibfnamefont {C.}~\bibnamefont {{Palenzuela}}}, \ and\
  \bibinfo {author} {\bibfnamefont {P.}~\bibnamefont {{Pani}}},\ }\href
  {\doibase 10.1103/PhysRevD.94.084031} {\bibfield  {journal} {\bibinfo
  {journal} {\prd}\ }\textbf {\bibinfo {volume} {94}},\ \bibinfo {eid} {084031}
  (\bibinfo {year} {2016})},\ \Eprint {http://arxiv.org/abs/1608.08637}
  {arXiv:1608.08637 [gr-qc]} \BibitemShut {NoStop}%
\bibitem [{\citenamefont {Price}\ and\ \citenamefont
  {Khanna}(2017)}]{Price:2017cjr}%
  \BibitemOpen
  \bibfield  {author} {\bibinfo {author} {\bibfnamefont {R.~H.}\ \bibnamefont
  {Price}}\ and\ \bibinfo {author} {\bibfnamefont {G.}~\bibnamefont {Khanna}},\
  }\href {\doibase 10.1088/1361-6382/aa8f29} {\bibfield  {journal} {\bibinfo
  {journal} {Class. Quant. Grav.}\ }\textbf {\bibinfo {volume} {34}},\ \bibinfo
  {pages} {225005} (\bibinfo {year} {2017})},\ \Eprint
  {http://arxiv.org/abs/1702.04833} {arXiv:1702.04833 [gr-qc]} \BibitemShut
  {NoStop}%
\bibitem [{\citenamefont {Brustein}\ \emph {et~al.}(2017)\citenamefont
  {Brustein}, \citenamefont {Medved},\ and\ \citenamefont
  {Yagi}}]{PhysRevD.96.064033}%
  \BibitemOpen
  \bibfield  {author} {\bibinfo {author} {\bibfnamefont {R.}~\bibnamefont
  {Brustein}}, \bibinfo {author} {\bibfnamefont {A.~J.~M.}\ \bibnamefont
  {Medved}}, \ and\ \bibinfo {author} {\bibfnamefont {K.}~\bibnamefont
  {Yagi}},\ }\href {\doibase 10.1103/PhysRevD.96.064033} {\bibfield  {journal}
  {\bibinfo  {journal} {Phys. Rev. D}\ }\textbf {\bibinfo {volume} {96}},\
  \bibinfo {pages} {064033} (\bibinfo {year} {2017})}\BibitemShut {NoStop}%
\bibitem [{\citenamefont {Barcel{\'o}}\ \emph {et~al.}(2017)\citenamefont
  {Barcel{\'o}}, \citenamefont {Carballo-Rubio},\ and\ \citenamefont
  {Garay}}]{Barcelo2017}%
  \BibitemOpen
  \bibfield  {author} {\bibinfo {author} {\bibfnamefont {C.}~\bibnamefont
  {Barcel{\'o}}}, \bibinfo {author} {\bibfnamefont {R.}~\bibnamefont
  {Carballo-Rubio}}, \ and\ \bibinfo {author} {\bibfnamefont {L.~J.}\
  \bibnamefont {Garay}},\ }\href {\doibase 10.1007/JHEP05(2017)054} {\bibfield
  {journal} {\bibinfo  {journal} {Journal of High Energy Physics}\ }\textbf
  {\bibinfo {volume} {2017}},\ \bibinfo {pages} {54} (\bibinfo {year}
  {2017})}\BibitemShut {NoStop}%
\bibitem [{\citenamefont {Konoplya}\ and\ \citenamefont
  {Molina}(2005)}]{Konoplya:2005et}%
  \BibitemOpen
  \bibfield  {author} {\bibinfo {author} {\bibfnamefont {R.~A.}\ \bibnamefont
  {Konoplya}}\ and\ \bibinfo {author} {\bibfnamefont {C.}~\bibnamefont
  {Molina}},\ }\href {\doibase 10.1103/PhysRevD.71.124009} {\bibfield
  {journal} {\bibinfo  {journal} {Phys. Rev.}\ }\textbf {\bibinfo {volume}
  {D71}},\ \bibinfo {pages} {124009} (\bibinfo {year} {2005})},\ \Eprint
  {http://arxiv.org/abs/gr-qc/0504139} {arXiv:gr-qc/0504139 [gr-qc]}
  \BibitemShut {NoStop}%
\bibitem [{\citenamefont {Bueno}\ \emph {et~al.}(2018)\citenamefont {Bueno},
  \citenamefont {Cano}, \citenamefont {Goelen}, \citenamefont {Hertog},\ and\
  \citenamefont {Vercnocke}}]{Bueno:2017hyj}%
  \BibitemOpen
  \bibfield  {author} {\bibinfo {author} {\bibfnamefont {P.}~\bibnamefont
  {Bueno}}, \bibinfo {author} {\bibfnamefont {P.~A.}\ \bibnamefont {Cano}},
  \bibinfo {author} {\bibfnamefont {F.}~\bibnamefont {Goelen}}, \bibinfo
  {author} {\bibfnamefont {T.}~\bibnamefont {Hertog}}, \ and\ \bibinfo {author}
  {\bibfnamefont {B.}~\bibnamefont {Vercnocke}},\ }\href {\doibase
  10.1103/PhysRevD.97.024040} {\bibfield  {journal} {\bibinfo  {journal} {Phys.
  Rev.}\ }\textbf {\bibinfo {volume} {D97}},\ \bibinfo {pages} {024040}
  (\bibinfo {year} {2018})},\ \Eprint {http://arxiv.org/abs/1711.00391}
  {arXiv:1711.00391 [gr-qc]} \BibitemShut {NoStop}%
\bibitem [{\citenamefont {Holdom}\ and\ \citenamefont
  {Ren}(2017)}]{PhysRevD.95.084034}%
  \BibitemOpen
  \bibfield  {author} {\bibinfo {author} {\bibfnamefont {B.}~\bibnamefont
  {Holdom}}\ and\ \bibinfo {author} {\bibfnamefont {J.}~\bibnamefont {Ren}},\
  }\href {\doibase 10.1103/PhysRevD.95.084034} {\bibfield  {journal} {\bibinfo
  {journal} {{Phys. Rev. D}}\ }\textbf {\bibinfo {volume} {95}},\ \bibinfo
  {pages} {084034} (\bibinfo {year} {2017})}\BibitemShut {NoStop}%
\bibitem [{\citenamefont {Berthiere}\ \emph {et~al.}(2018)\citenamefont
  {Berthiere}, \citenamefont {Sarkar},\ and\ \citenamefont
  {Solodukhin}}]{Berthiere:2017tms}%
  \BibitemOpen
  \bibfield  {author} {\bibinfo {author} {\bibfnamefont {C.}~\bibnamefont
  {Berthiere}}, \bibinfo {author} {\bibfnamefont {D.}~\bibnamefont {Sarkar}}, \
  and\ \bibinfo {author} {\bibfnamefont {S.~N.}\ \bibnamefont {Solodukhin}},\
  }\href {\doibase 10.1016/j.physletb.2018.09.027} {\bibfield  {journal}
  {\bibinfo  {journal} {Phys. Lett.}\ }\textbf {\bibinfo {volume} {B786}},\
  \bibinfo {pages} {21} (\bibinfo {year} {2018})},\ \Eprint
  {http://arxiv.org/abs/1712.09914} {arXiv:1712.09914 [hep-th]} \BibitemShut
  {NoStop}%
\bibitem [{\citenamefont {{Bender}}\ and\ \citenamefont
  {{Orszag}}(1978)}]{1978amms.book.....B}%
  \BibitemOpen
  \bibfield  {author} {\bibinfo {author} {\bibfnamefont {C.~M.}\ \bibnamefont
  {{Bender}}}\ and\ \bibinfo {author} {\bibfnamefont {S.~A.}\ \bibnamefont
  {{Orszag}}},\ }\href@noop {} {\emph {\bibinfo {title} {Advanced Mathematical
  Methods for Scientists and Engineers, New York: McGraw-Hill, 1978}}}\
  (\bibinfo  {publisher} {New York: McGraw-Hill},\ \bibinfo {year}
  {1978})\BibitemShut {NoStop}%
\bibitem [{\citenamefont {Gurvitz}(1988)}]{PhysRevA.38.1747}%
  \BibitemOpen
  \bibfield  {author} {\bibinfo {author} {\bibfnamefont {S.~A.}\ \bibnamefont
  {Gurvitz}},\ }\href {\doibase 10.1103/PhysRevA.38.1747} {\bibfield  {journal}
  {\bibinfo  {journal} {Phys. Rev. A}\ }\textbf {\bibinfo {volume} {38}},\
  \bibinfo {pages} {1747} (\bibinfo {year} {1988})}\BibitemShut {NoStop}%
\bibitem [{\citenamefont {{Popov}}\ \emph {et~al.}(1991)\citenamefont
  {{Popov}}, \citenamefont {{Mur}},\ and\ \citenamefont
  {{Sergeev}}}]{1991PhLA..157..185P}%
  \BibitemOpen
  \bibfield  {author} {\bibinfo {author} {\bibfnamefont {V.~S.}\ \bibnamefont
  {{Popov}}}, \bibinfo {author} {\bibfnamefont {V.~D.}\ \bibnamefont {{Mur}}},
  \ and\ \bibinfo {author} {\bibfnamefont {A.~V.}\ \bibnamefont {{Sergeev}}},\
  }\href {\doibase 10.1016/0375-9601(91)90048-D} {\bibfield  {journal}
  {\bibinfo  {journal} {Physics Letters A}\ }\textbf {\bibinfo {volume}
  {157}},\ \bibinfo {pages} {185} (\bibinfo {year} {1991})}\BibitemShut
  {NoStop}%
\bibitem [{\citenamefont {{Karnakov}}\ and\ \citenamefont
  {{Krainov}}(2013)}]{2013waap.book.....K}%
  \BibitemOpen
  \bibfield  {author} {\bibinfo {author} {\bibfnamefont {B.~M.}\ \bibnamefont
  {{Karnakov}}}\ and\ \bibinfo {author} {\bibfnamefont {V.~P.}\ \bibnamefont
  {{Krainov}}},\ }\href {\doibase 10.1007/978-3-642-31558-9} {\emph {\bibinfo
  {title} {WKB Approximation in Atomic Physics: , ISBN
  978-3-642-31557-2.~Springer-Verlag Berlin Heidelberg, 2013}}}\ (\bibinfo
  {publisher} {Springer-Verlag Berlin Heidelberg},\ \bibinfo {year}
  {2013})\BibitemShut {NoStop}%
\bibitem [{\citenamefont {{Kokkotas}}(1991)}]{1991CQGra...8.2217K}%
  \BibitemOpen
  \bibfield  {author} {\bibinfo {author} {\bibfnamefont {K.~D.}\ \bibnamefont
  {{Kokkotas}}},\ }\href {\doibase 10.1088/0264-9381/8/12/006} {\bibfield
  {journal} {\bibinfo  {journal} {Classical and Quantum Gravity}\ }\textbf
  {\bibinfo {volume} {8}},\ \bibinfo {pages} {2217} (\bibinfo {year}
  {1991})}\BibitemShut {NoStop}%
\bibitem [{\citenamefont {Andersson}\ \emph {et~al.}(1993)\citenamefont
  {Andersson}, \citenamefont {Araujo},\ and\ \citenamefont
  {Schutz}}]{0264-9381-10-4-010}%
  \BibitemOpen
  \bibfield  {author} {\bibinfo {author} {\bibfnamefont {N.}~\bibnamefont
  {Andersson}}, \bibinfo {author} {\bibfnamefont {M.~E.}\ \bibnamefont
  {Araujo}}, \ and\ \bibinfo {author} {\bibfnamefont {B.~F.}\ \bibnamefont
  {Schutz}},\ }\href {http://stacks.iop.org/0264-9381/10/i=4/a=010} {\bibfield
  {journal} {\bibinfo  {journal} {Classical and Quantum Gravity}\ }\textbf
  {\bibinfo {volume} {10}},\ \bibinfo {pages} {757} (\bibinfo {year}
  {1993})}\BibitemShut {NoStop}%
\bibitem [{\citenamefont {{Festuccia}}\ and\ \citenamefont
  {{Liu}}(2009)}]{Festuccia:2009:1936-6612:221}%
  \BibitemOpen
  \bibfield  {author} {\bibinfo {author} {\bibfnamefont {G.}~\bibnamefont
  {{Festuccia}}}\ and\ \bibinfo {author} {\bibfnamefont {H.}~\bibnamefont
  {{Liu}}},\ }\href {\doibase {doi:10.1166/asl.2009.1029}} {\bibfield
  {journal} {\bibinfo  {journal} {{Advanced Science Letters}}\ }\textbf
  {\bibinfo {volume} {{2}}} (\bibinfo {year} {{2009}}),\
  {doi:10.1166/asl.2009.1029}}\BibitemShut {NoStop}%
\bibitem [{\citenamefont {{Pani}}\ \emph {et~al.}(2009)\citenamefont {{Pani}},
  \citenamefont {{Berti}}, \citenamefont {{Cardoso}}, \citenamefont {{Chen}},\
  and\ \citenamefont {{Norte}}}]{2009PhRvD..80l4047P}%
  \BibitemOpen
  \bibfield  {author} {\bibinfo {author} {\bibfnamefont {P.}~\bibnamefont
  {{Pani}}}, \bibinfo {author} {\bibfnamefont {E.}~\bibnamefont {{Berti}}},
  \bibinfo {author} {\bibfnamefont {V.}~\bibnamefont {{Cardoso}}}, \bibinfo
  {author} {\bibfnamefont {Y.}~\bibnamefont {{Chen}}}, \ and\ \bibinfo {author}
  {\bibfnamefont {R.}~\bibnamefont {{Norte}}},\ }\href {\doibase
  10.1103/PhysRevD.80.124047} {\bibfield  {journal} {\bibinfo  {journal}
  {\prd}\ }\textbf {\bibinfo {volume} {80}},\ \bibinfo {eid} {124047} (\bibinfo
  {year} {2009})},\ \Eprint {http://arxiv.org/abs/0909.0287} {arXiv:0909.0287
  [gr-qc]} \BibitemShut {NoStop}%
\bibitem [{\citenamefont {{Cardoso}}\ \emph {et~al.}(2014)\citenamefont
  {{Cardoso}}, \citenamefont {{Crispino}}, \citenamefont {{Macedo}},
  \citenamefont {{Okawa}},\ and\ \citenamefont {{Pani}}}]{2014PhRvD..90d4069C}%
  \BibitemOpen
  \bibfield  {author} {\bibinfo {author} {\bibfnamefont {V.}~\bibnamefont
  {{Cardoso}}}, \bibinfo {author} {\bibfnamefont {L.~C.~B.}\ \bibnamefont
  {{Crispino}}}, \bibinfo {author} {\bibfnamefont {C.~F.~B.}\ \bibnamefont
  {{Macedo}}}, \bibinfo {author} {\bibfnamefont {H.}~\bibnamefont {{Okawa}}}, \
  and\ \bibinfo {author} {\bibfnamefont {P.}~\bibnamefont {{Pani}}},\ }\href
  {\doibase 10.1103/PhysRevD.90.044069} {\bibfield  {journal} {\bibinfo
  {journal} {\prd}\ }\textbf {\bibinfo {volume} {90}},\ \bibinfo {eid} {044069}
  (\bibinfo {year} {2014})},\ \Eprint {http://arxiv.org/abs/1406.5510}
  {arXiv:1406.5510 [gr-qc]} \BibitemShut {NoStop}%
\bibitem [{\citenamefont {{V{\"o}lkel}}\ and\ \citenamefont
  {{Kokkotas}}(2017)}]{paper1}%
  \BibitemOpen
  \bibfield  {author} {\bibinfo {author} {\bibfnamefont {S.~H.}\ \bibnamefont
  {{V{\"o}lkel}}}\ and\ \bibinfo {author} {\bibfnamefont {K.~D.}\ \bibnamefont
  {{Kokkotas}}},\ }\href {\doibase 10.1088/1361-6382/aa68cc} {\bibfield
  {journal} {\bibinfo  {journal} {Classical and Quantum Gravity}\ }\textbf
  {\bibinfo {volume} {34}},\ \bibinfo {eid} {125006} (\bibinfo {year}
  {2017})},\ \Eprint {http://arxiv.org/abs/1703.08156} {arXiv:1703.08156
  [gr-qc]} \BibitemShut {NoStop}%
\bibitem [{\citenamefont {{V\"olkel}}\ and\ \citenamefont
  {{Kokkotas}}(2017)}]{paper2}%
  \BibitemOpen
  \bibfield  {author} {\bibinfo {author} {\bibfnamefont {S.~H.}\ \bibnamefont
  {{V\"olkel}}}\ and\ \bibinfo {author} {\bibfnamefont {K.~D.}\ \bibnamefont
  {{Kokkotas}}},\ }\href {\doibase 10.1088/1361-6382/aa82de} {\bibfield
  {journal} {\bibinfo  {journal} {Classical and Quantum Gravity}\ }\textbf
  {\bibinfo {volume} {34}},\ \bibinfo {pages} {175015} (\bibinfo {year}
  {2017})},\ \Eprint {http://arxiv.org/abs/1704.07517} {arXiv:1704.07517
  [gr-qc]} \BibitemShut {NoStop}%
\bibitem [{\citenamefont {V{\"o}lkel}\ and\ \citenamefont
  {Kokkotas}(2018)}]{paper5}%
  \BibitemOpen
  \bibfield  {author} {\bibinfo {author} {\bibfnamefont {S.~H.}\ \bibnamefont
  {V{\"o}lkel}}\ and\ \bibinfo {author} {\bibfnamefont {K.~D.}\ \bibnamefont
  {Kokkotas}},\ }\href {\doibase 10.1088/1361-6382/aabce6} {\bibfield
  {journal} {\bibinfo  {journal} {Classical and Quantum Gravity}\ } (\bibinfo
  {year} {{2018}}),\ 10.1088/1361-6382/aabce6},\ \Eprint
  {http://arxiv.org/abs/1802.08525} {arXiv:1802.08525 [gr-qc]} \BibitemShut
  {NoStop}%
\bibitem [{\citenamefont {{Nollert}}(1996)}]{1996PhRvD..53.4397N}%
  \BibitemOpen
  \bibfield  {author} {\bibinfo {author} {\bibfnamefont {H.-P.}\ \bibnamefont
  {{Nollert}}},\ }\href {\doibase 10.1103/PhysRevD.53.4397} {\bibfield
  {journal} {\bibinfo  {journal} {\prd}\ }\textbf {\bibinfo {volume} {53}},\
  \bibinfo {pages} {4397} (\bibinfo {year} {1996})},\ \Eprint
  {http://arxiv.org/abs/gr-qc/9602032} {gr-qc/9602032} \BibitemShut {NoStop}%
\bibitem [{\citenamefont {{Nollert}}\ and\ \citenamefont
  {{Price}}(1999)}]{1999JMP....40..980N}%
  \BibitemOpen
  \bibfield  {author} {\bibinfo {author} {\bibfnamefont {H.-P.}\ \bibnamefont
  {{Nollert}}}\ and\ \bibinfo {author} {\bibfnamefont {R.~H.}\ \bibnamefont
  {{Price}}},\ }\href {\doibase 10.1063/1.532698} {\bibfield  {journal}
  {\bibinfo  {journal} {Journal of Mathematical Physics}\ }\textbf {\bibinfo
  {volume} {40}},\ \bibinfo {pages} {980} (\bibinfo {year} {1999})},\ \Eprint
  {http://arxiv.org/abs/gr-qc/9810074} {gr-qc/9810074} \BibitemShut {NoStop}%
\bibitem [{\citenamefont {Zhang}\ \emph {et~al.}(2011)\citenamefont {Zhang},
  \citenamefont {Wu},\ and\ \citenamefont {Leung}}]{PhysRevD.83.064012}%
  \BibitemOpen
  \bibfield  {author} {\bibinfo {author} {\bibfnamefont {Y.~J.}\ \bibnamefont
  {Zhang}}, \bibinfo {author} {\bibfnamefont {J.}~\bibnamefont {Wu}}, \ and\
  \bibinfo {author} {\bibfnamefont {P.~T.}\ \bibnamefont {Leung}},\ }\href
  {\doibase 10.1103/PhysRevD.83.064012} {\bibfield  {journal} {\bibinfo
  {journal} {Phys. Rev. D}\ }\textbf {\bibinfo {volume} {83}},\ \bibinfo
  {pages} {064012} (\bibinfo {year} {2011})}\BibitemShut {NoStop}%
\bibitem [{\citenamefont {{Andersson}}\ \emph {et~al.}(1996)\citenamefont
  {{Andersson}}, \citenamefont {{Kojima}},\ and\ \citenamefont
  {{Kokkotas}}}]{1996ApJ...462..855A}%
  \BibitemOpen
  \bibfield  {author} {\bibinfo {author} {\bibfnamefont {N.}~\bibnamefont
  {{Andersson}}}, \bibinfo {author} {\bibfnamefont {Y.}~\bibnamefont
  {{Kojima}}}, \ and\ \bibinfo {author} {\bibfnamefont {K.~D.}\ \bibnamefont
  {{Kokkotas}}},\ }\href {\doibase 10.1086/177199} {\bibfield  {journal}
  {\bibinfo  {journal} {\apj}\ }\textbf {\bibinfo {volume} {462}},\ \bibinfo
  {pages} {855} (\bibinfo {year} {1996})},\ \Eprint
  {http://arxiv.org/abs/gr-qc/9512048} {gr-qc/9512048} \BibitemShut {NoStop}%
\bibitem [{\citenamefont {Blazquez-Salcedo}\ \emph {et~al.}(2013)\citenamefont
  {Blazquez-Salcedo}, \citenamefont {Gonzalez-Romero},\ and\ \citenamefont
  {Navarro-Lerida}}]{BlazquezSalcedo:2012pd}%
  \BibitemOpen
  \bibfield  {author} {\bibinfo {author} {\bibfnamefont {J.~L.}\ \bibnamefont
  {Blazquez-Salcedo}}, \bibinfo {author} {\bibfnamefont {L.~M.}\ \bibnamefont
  {Gonzalez-Romero}}, \ and\ \bibinfo {author} {\bibfnamefont {F.}~\bibnamefont
  {Navarro-Lerida}},\ }\href {\doibase 10.1103/PhysRevD.87.104042} {\bibfield
  {journal} {\bibinfo  {journal} {Phys. Rev.}\ }\textbf {\bibinfo {volume}
  {D87}},\ \bibinfo {pages} {104042} (\bibinfo {year} {2013})},\ \Eprint
  {http://arxiv.org/abs/1207.4651} {arXiv:1207.4651 [gr-qc]} \BibitemShut
  {NoStop}%
\bibitem [{\citenamefont {Blazquez-Salcedo}\ \emph {et~al.}(2018)\citenamefont
  {Blazquez-Salcedo}, \citenamefont {Doneva}, \citenamefont {Kunz},
  \citenamefont {Staykov},\ and\ \citenamefont
  {Yazadjiev}}]{Blazquez-Salcedo:2018qyy}%
  \BibitemOpen
  \bibfield  {author} {\bibinfo {author} {\bibfnamefont {J.~L.}\ \bibnamefont
  {Blazquez-Salcedo}}, \bibinfo {author} {\bibfnamefont {D.~D.}\ \bibnamefont
  {Doneva}}, \bibinfo {author} {\bibfnamefont {J.}~\bibnamefont {Kunz}},
  \bibinfo {author} {\bibfnamefont {K.~V.}\ \bibnamefont {Staykov}}, \ and\
  \bibinfo {author} {\bibfnamefont {S.~S.}\ \bibnamefont {Yazadjiev}},\
  }\href@noop {} {\  (\bibinfo {year} {2018})},\ \Eprint
  {http://arxiv.org/abs/1804.04060} {arXiv:1804.04060 [gr-qc]} \BibitemShut
  {NoStop}%
\bibitem [{\citenamefont {Wheeler}(2015)}]{lieb2015studies}%
  \BibitemOpen
  \bibfield  {author} {\bibinfo {author} {\bibfnamefont {J.~A.}\ \bibnamefont
  {Wheeler}},\ }\href {https://press.princeton.edu/titles/861.html} {\emph
  {\bibinfo {title} {{Studies in Mathematical Physics: Essays in Honor of
  Valentine Bargmann}}}},\ Princeton Series in Physics\ (\bibinfo  {publisher}
  {Princeton University Press},\ \bibinfo {year} {2015})\ pp.\ \bibinfo {pages}
  {351--422}\BibitemShut {NoStop}%
\bibitem [{\citenamefont {Chadan}\ and\ \citenamefont
  {Sabatier}(1989)}]{MR985100}%
  \BibitemOpen
  \bibfield  {author} {\bibinfo {author} {\bibfnamefont {K.}~\bibnamefont
  {Chadan}}\ and\ \bibinfo {author} {\bibfnamefont {P.~C.}\ \bibnamefont
  {Sabatier}},\ }\href {\doibase 10.1007/978-3-642-83317-5} {\emph {\bibinfo
  {title} {{Inverse problems in quantum scattering theory}}}},\ \bibinfo
  {edition} {2nd}\ ed.,\ Texts and Monographs in Physics\ (\bibinfo
  {publisher} {Springer-Verlag},\ \bibinfo {address} {New York},\ \bibinfo
  {year} {1989})\BibitemShut {NoStop}%
\bibitem [{\citenamefont {{Bonatsos}}\ \emph {et~al.}(1992)\citenamefont
  {{Bonatsos}}, \citenamefont {{Daskaloyannis}},\ and\ \citenamefont
  {{Kokkotas}}}]{1992JMP....33.2958B}%
  \BibitemOpen
  \bibfield  {author} {\bibinfo {author} {\bibfnamefont {D.}~\bibnamefont
  {{Bonatsos}}}, \bibinfo {author} {\bibfnamefont {C.}~\bibnamefont
  {{Daskaloyannis}}}, \ and\ \bibinfo {author} {\bibfnamefont {K.}~\bibnamefont
  {{Kokkotas}}},\ }\href {\doibase 10.1063/1.529565} {\bibfield  {journal}
  {\bibinfo  {journal} {Journal of Mathematical Physics}\ }\textbf {\bibinfo
  {volume} {33}},\ \bibinfo {pages} {2958} (\bibinfo {year}
  {1992})}\BibitemShut {NoStop}%
\bibitem [{\citenamefont {{Bonatsos}}\ \emph {et~al.}(1991)\citenamefont
  {{Bonatsos}}, \citenamefont {{Daskaloyannis}},\ and\ \citenamefont
  {{Kokkotas}}}]{1991JPhA...24L.795B}%
  \BibitemOpen
  \bibfield  {author} {\bibinfo {author} {\bibfnamefont {D.}~\bibnamefont
  {{Bonatsos}}}, \bibinfo {author} {\bibfnamefont {C.}~\bibnamefont
  {{Daskaloyannis}}}, \ and\ \bibinfo {author} {\bibfnamefont {K.}~\bibnamefont
  {{Kokkotas}}},\ }\href {\doibase 10.1088/0305-4470/24/15/002} {\bibfield
  {journal} {\bibinfo  {journal} {Journal of Physics A Mathematical General}\
  }\textbf {\bibinfo {volume} {24}},\ \bibinfo {pages} {L795} (\bibinfo {year}
  {1991})}\BibitemShut {NoStop}%
\bibitem [{\citenamefont {V{\"o}lkel}(2018)}]{paper6}%
  \BibitemOpen
  \bibfield  {author} {\bibinfo {author} {\bibfnamefont {S.~H.}\ \bibnamefont
  {V{\"o}lkel}},\ }\href@noop {} {\bibfield  {journal} {\bibinfo  {journal} {to
  be submitted}\ } (\bibinfo {year} {{2018}})}\BibitemShut {NoStop}%
\bibitem [{\citenamefont {{Lazenby}}\ and\ \citenamefont
  {{Griffiths}}(1980)}]{1980AmJPh..48..432L}%
  \BibitemOpen
  \bibfield  {author} {\bibinfo {author} {\bibfnamefont {J.~C.}\ \bibnamefont
  {{Lazenby}}}\ and\ \bibinfo {author} {\bibfnamefont {D.~J.}\ \bibnamefont
  {{Griffiths}}},\ }\href {\doibase 10.1119/1.11998} {\bibfield  {journal}
  {\bibinfo  {journal} {American Journal of Physics}\ }\textbf {\bibinfo
  {volume} {48}},\ \bibinfo {pages} {432} (\bibinfo {year} {1980})}\BibitemShut
  {NoStop}%
\bibitem [{\citenamefont {{Gandhi}}\ and\ \citenamefont
  {{Efthimiou}}(2006)}]{2006AmJPh..74..638G}%
  \BibitemOpen
  \bibfield  {author} {\bibinfo {author} {\bibfnamefont {S.~C.}\ \bibnamefont
  {{Gandhi}}}\ and\ \bibinfo {author} {\bibfnamefont {C.~J.}\ \bibnamefont
  {{Efthimiou}}},\ }\href {\doibase 10.1119/1.2190683} {\bibfield  {journal}
  {\bibinfo  {journal} {American Journal of Physics}\ }\textbf {\bibinfo
  {volume} {74}},\ \bibinfo {pages} {638} (\bibinfo {year} {2006})},\ \Eprint
  {http://arxiv.org/abs/quant-ph/0503223} {quant-ph/0503223} \BibitemShut
  {NoStop}%
\bibitem [{\citenamefont {V\"olkel}(2018)}]{paper4}%
  \BibitemOpen
  \bibfield  {author} {\bibinfo {author} {\bibfnamefont {S.~H.}\ \bibnamefont
  {V\"olkel}},\ }\href {\doibase 10.1088/2399-6528/aaaee2} {\bibfield
  {journal} {\bibinfo  {journal} {Journal of Physics Communications}\ }\textbf
  {\bibinfo {volume} {2}},\ \bibinfo {pages} {025029} (\bibinfo {year}
  {2018})},\ \Eprint {http://arxiv.org/abs/1802.08684} {arXiv:1802.08684
  [quant-ph]} \BibitemShut {NoStop}%
\bibitem [{\citenamefont {Suvorov}(2018)}]{Suvorov:2018bvs}%
  \BibitemOpen
  \bibfield  {author} {\bibinfo {author} {\bibfnamefont {A.~G.}\ \bibnamefont
  {Suvorov}},\ }\href {\doibase 10.1093/mnras/sty1080} {\bibfield  {journal}
  {\bibinfo  {journal} {Mon. Not. Roy. Astron. Soc.}\ }\textbf {\bibinfo
  {volume} {478}},\ \bibinfo {pages} {167} (\bibinfo {year} {2018})},\ \Eprint
  {http://arxiv.org/abs/1804.09413} {arXiv:1804.09413 [astro-ph.HE]}
  \BibitemShut {NoStop}%
\bibitem [{\citenamefont {{Pavlidou}}\ \emph {et~al.}(2000)\citenamefont
  {{Pavlidou}}, \citenamefont {{Tassis}}, \citenamefont {{Baumgarte}},\ and\
  \citenamefont {{Shapiro}}}]{2000PhRvD..62h4020P}%
  \BibitemOpen
  \bibfield  {author} {\bibinfo {author} {\bibfnamefont {V.}~\bibnamefont
  {{Pavlidou}}}, \bibinfo {author} {\bibfnamefont {K.}~\bibnamefont
  {{Tassis}}}, \bibinfo {author} {\bibfnamefont {T.~W.}\ \bibnamefont
  {{Baumgarte}}}, \ and\ \bibinfo {author} {\bibfnamefont {S.~L.}\ \bibnamefont
  {{Shapiro}}},\ }\href {\doibase 10.1103/PhysRevD.62.084020} {\bibfield
  {journal} {\bibinfo  {journal} {\prd}\ }\textbf {\bibinfo {volume} {62}},\
  \bibinfo {eid} {084020} (\bibinfo {year} {2000})},\ \Eprint
  {http://arxiv.org/abs/gr-qc/0007019} {gr-qc/0007019} \BibitemShut {NoStop}%
\bibitem [{\citenamefont {Chandrasekhar}\ and\ \citenamefont
  {Detweiler}(1975)}]{Chandrasekhar441}%
  \BibitemOpen
  \bibfield  {author} {\bibinfo {author} {\bibfnamefont {S.}~\bibnamefont
  {Chandrasekhar}}\ and\ \bibinfo {author} {\bibfnamefont {S.}~\bibnamefont
  {Detweiler}},\ }\href {\doibase 10.1098/rspa.1975.0112} {\bibfield  {journal}
  {\bibinfo  {journal} {Proceedings of the Royal Society of London A:
  Mathematical, Physical and Engineering Sciences}\ }\textbf {\bibinfo {volume}
  {344}},\ \bibinfo {pages} {441} (\bibinfo {year} {1975})},\ \Eprint
  {http://arxiv.org/abs/http://rspa.royalsocietypublishing.org/content/344/1639/441.full.pdf}
  {http://rspa.royalsocietypublishing.org/content/344/1639/441.full.pdf}
  \BibitemShut {NoStop}%
\bibitem [{\citenamefont {Barausse}\ \emph {et~al.}(2014)\citenamefont
  {Barausse}, \citenamefont {Cardoso},\ and\ \citenamefont
  {Pani}}]{Barausse:2014tra}%
  \BibitemOpen
  \bibfield  {author} {\bibinfo {author} {\bibfnamefont {E.}~\bibnamefont
  {Barausse}}, \bibinfo {author} {\bibfnamefont {V.}~\bibnamefont {Cardoso}}, \
  and\ \bibinfo {author} {\bibfnamefont {P.}~\bibnamefont {Pani}},\ }\href
  {\doibase 10.1103/PhysRevD.89.104059} {\bibfield  {journal} {\bibinfo
  {journal} {Phys. Rev.}\ }\textbf {\bibinfo {volume} {D89}},\ \bibinfo {pages}
  {104059} (\bibinfo {year} {2014})},\ \Eprint {http://arxiv.org/abs/1404.7149}
  {arXiv:1404.7149 [gr-qc]} \BibitemShut {NoStop}%
\bibitem [{\citenamefont {{Konoplya}}\ \emph {et~al.}(2018)\citenamefont
  {{Konoplya}}, \citenamefont {{Stuchl{\'{\i}}k}},\ and\ \citenamefont
  {{Zhidenko}}}]{2018arXiv181001295K}%
  \BibitemOpen
  \bibfield  {author} {\bibinfo {author} {\bibfnamefont {R.~A.}\ \bibnamefont
  {{Konoplya}}}, \bibinfo {author} {\bibfnamefont {Z.}~\bibnamefont
  {{Stuchl{\'{\i}}k}}}, \ and\ \bibinfo {author} {\bibfnamefont
  {A.}~\bibnamefont {{Zhidenko}}},\ }\href@noop {} {\  (\bibinfo {year}
  {2018})},\ \Eprint {http://arxiv.org/abs/1810.01295} {arXiv:1810.01295
  [gr-qc]} \BibitemShut {NoStop}%
\end{thebibliography}%
%

\section{Appendix}\label{Appendix}
%
In this appendix we provide additional information about the new Bohr-Sommerfeld rule \ref{AGBS}, the linear tortoise transformation \ref{tortoise} and the approximated axial perturbation potential \ref{approximate_pot}.
\subsection{New Bohr-Sommerfeld Rule}\label{AGBS}
In this section we outline the derivation of the new Bohr-Sommerfeld rule eq. \eqref{new_BS}. We start from the one-dimensional wave equation
\begin{align}\label{wave_equation}
\psi^{\prime \prime }(x) + \left(E_n -V(x) \right) \psi(x)  = 0,
\end{align}
where $E_n = \omega_n^2$ are the QNMs of the potential $V(x)$. We assume that the potential diverges at $x=0$ and has one discontinuity at $x_\text{s} > 0$, where the potential drops by the constant
\begin{align}\label{Delta V}
\Delta V \equiv \lim_{x_{-} \rightarrow x_\text{s}} V(x)- \lim_{x_{+} \rightarrow x_\text{s}} V(x).
\end{align}
The standard WKB solutions left and right of $x_\text{s}$ can be written as
\begin{align}
\psi_{1}(x) &= \frac{A_\text{1}}{p_{1}(x)} \exp\left(i \int_{x_{0, {1}}}^{x} p_{1}(u) \text{d}u \right) +\frac{B_{1}}{p_{1}(x)} \exp\left(-i \int_{x_{0, {1}}}^{x} p_{1}(u) \text{d}u \right),
\\
\psi_{4}(x) &= \frac{A_{4}}{p_{4}(x)} \exp\left(i \int_{x_{0, {4}}}^{x} p_{4}(u) \text{d}u \right) +\frac{B_{4}}{p_{4}(x)} \exp\left(-i \int_{x_{0, {4}}}^{x} p_{4}(u) \text{d}u \right),
\end{align}
with
\begin{align}
p_{{1}}(x) &= \sqrt{E_n-V_1(x)}, \qquad 
p_{{4}}(x) = \sqrt{E_n-V_4(x)}.
\end{align}
$(A_1, A_4, B_1, B_4)$ and $(x_{0,1}, x_{0,4})$ are constants being determined by matching the solutions and initial conditions. The WKB solution $\psi_1(x)$ is not valid at $E_n = V(x)$, but it can be shown by using Kramer's relations, that the WKB solution in the classically allowed region,  $E_n > V(x)$ can be written as
\begin{align}
\psi_{{1}}(x) &= \frac{A_{1}}{p_{1}(x)}  \sin\left( \int_{x_{0, {1}}}^{x} p_{1}(u) \text{d}u + \frac{\pi}{4} \right),
\end{align}
which is in agreement with a regular solution at $x=0$. $x_{0,1}$ is the classical turning point where $E_n = V(x)$. The WKB solutions are not valid at the discontinuity itself. In order to nevertheless match the internal and external WKB solution, we find exact solutions of the wave equation eq. \eqref{wave_equation} on a small interval around the discontinuity and match them with the WKB solutions. On a small interval around the discontinuity, we can make a Taylor expansion of the potential, defined on the left and right of $x_\text{s}$ as
\begin{align}
V_2(x) \approx \lim_{x_{-}\rightarrow x_\text{s}} V(x), \qquad V_3(x) \approx  \lim_{x_{+}\rightarrow x_\text{s}} V(x),
\end{align}
and stop at zero order, since this is the interesting property of the discontinuity and the interval can be made arbitrarily small.
In contrast to the standard matching of WKB solutions with exact solutions on some finite interval where WKB is not valid\footnote{Typically the WKB solution is not valid on a finite region around a turning point, where $E_n \approx V(x)$. In this case one matches the WKB solutions with an exact solution on this interval.}, the WKB solutions are valid arbitrary close to the discontinuity. By demanding the usual conditions that $\psi(x)$ and $\psi^{\prime}(x)$ are continuous throughout the discontinuity $x_\text{s}$, one finds from the Wronskian
\begin{align}
W(x) = \psi_1(x) \psi_2^{\prime}(x) - \psi_1^{\prime}(x)\psi_2(x)=0,
\end{align}
now evaluated in the limit $x \rightarrow x_{s}$
\begin{align}\label{meq_full}
\tan\left(\int_{x_{0, {1}}}^{x_\text{s}} \sqrt{E_n-V_1(u)} \text{d}u + \frac{\pi}{4} \right) = i \frac{\sqrt{E_n-V_1(x_\text{s})}}{\sqrt{E_n-V_4(x_\text{s})}},
\end{align}
where $V_1(x)$ and $V_4(x)$ are the internal and external potential, respectively. We can make two comments. First, if the potential is smooth there are no solutions. Second, if the potential has a discontinuity there are discrete solutions, which are determined by
\begin{align}\label{meq_BS}
\int_{x_{0, {1}}}^{x_\text{s}} \sqrt{E_n-V_1(u)} \text{d}u  
=\pi \left(n-\frac{1}{4}\right) + i \tanh^{-1}\left(\frac{\sqrt{E_n-V_1(x_\text{s})}}{\sqrt{E_n-V_4(x_\text{s})}} \right).
\end{align}
This relation looks like a modified Bohr-Sommerfeld rule for a potential well between $(x_{0,1}, x_\text{s})$ with an additional imaginary part described by $\tanh^{-1}(\phi)$ and $n \in \mathbb{Z}$. However, as we will argue in the subsequent section, there is a further restriction for this set that demands $n \in \mathbb{N}$.
\par
We can further simplify our result, by assuming that $E_n$ is either much larger than $(V_1(x), V_4(x))$ or  that $\Delta V$ is small compared to $E_n$. Both assumptions are valid for the asymptotic behavior of the kind of spectra that we are interested in. Thus one can approximate the argument of $\tanh^{-1}(\phi)$ and find
\begin{align}\label{expansion_arg}
\frac{\sqrt{E_n-V_1(x_\text{s})}}{\sqrt{E_n-V_4(x_\text{s})}}
\approx
1-\frac{1}{2} \frac{\Delta V}{E_n-V_4(x_\text{s})},
\end{align}
where we have used $\Delta V$ given by eq. \eqref{Delta V}. We can further expand $\tanh^{-1}(\phi)$ in the case of $\Delta V \ll E_n$ and obtain
\begin{align}\label{meq_BS_app}
\int_{x_{0, {1}}}^{x_\text{s}} \sqrt{E_n-V_1(u)} \text{d}u  
=\pi \left(n-\frac{1}{4}\right) + i \left[\frac{1}{2}\ln\left(\frac{4 (E_n-V_4(x_\text{s}))}{\Delta V} \right) -  \frac{\Delta V}{8(E_n-V_4(x_\text{s}))} \right]
\approx \pi \left(n-\frac{1}{4}\right) +  \frac{i}{2}\ln\left(\frac{4 (E_n-V_4(x_\text{s}))}{\Delta V} \right).
\end{align}
From here we see that for high eigenvalues $E_n$, which correspond to large $n$, the rhs. of eq. \eqref{meq_BS_app} has a large real part, but the imaginary part only grows logarithmic. Thus, we expect that the complex eigenvalues for large $n$ can be written in the form
\begin{align}
E_n = E_{0n} + i E_{1n},
\end{align}
with $E_{1n} \ll E_{0n}$. Inserting this in the lhs. of eq. \eqref{meq_BS_app} and expanding the argument of the integral, we find two equations. The first one determines the real part $E_{0n}$
\begin{align}\label{real_BS}
\int_{x_{0, {1}}}^{x_\text{s}} \sqrt{E_{0n}-V_1(u)} \text{d}u  = \pi \left(n-\frac{1}{4}\right),
\end{align}
while the imaginary part $E_{1n}$ follows from the second one
\begin{align}\label{imag_BS}
E_{1n} = \left(\int_{x_{0, {1}}}^{x_\text{s}}\frac{1}{\sqrt{E_{0n}-V(u)}} \text{d}u \right)^{-1}  \ln\left(\frac{4 (E_{0n}-V_4(x_\text{s}))}{\Delta V} \right).
\end{align}
In the last step we neglected small contributions to $E_{0n}$ that are related to $E_{1n}$. The turning point $x_{0,1}$ is with respect to $E_{0n}$ and real, which simplifies the otherwise complex integration. Note that the real part $E_{0n}$ is closely related to the classical Bohr-Sommerfeld rule. For an ordinary potential well, the phase of the classical Bohr-Sommerfeld rule is $\pi(n+1/2)$, for $n=(0,1,2,\dots)$. Since the integral eq. \eqref{real_BS} is greater or equal to zero, it now follows that $n \in \mathbb{N}_{> 0}$. One can rewrite the phase
\begin{align}
n - \frac{1}{4} \rightarrow n + \frac{3}{4},
\end{align}
and start from $n = 0$ to match standard notation. In order to find $E_{1n}$, one first has to solve eq. \eqref{real_BS} for $E_{0n}$ and insert it into eq. \eqref{imag_BS}. This is similar to the case of the Gamow formula for the transmission through a potential barrier. 
\subsection{Approximate Tortoise Transformation}\label{tortoise}
\begin{figure}[H]
\centering
	\includegraphics[width=0.5\linewidth]{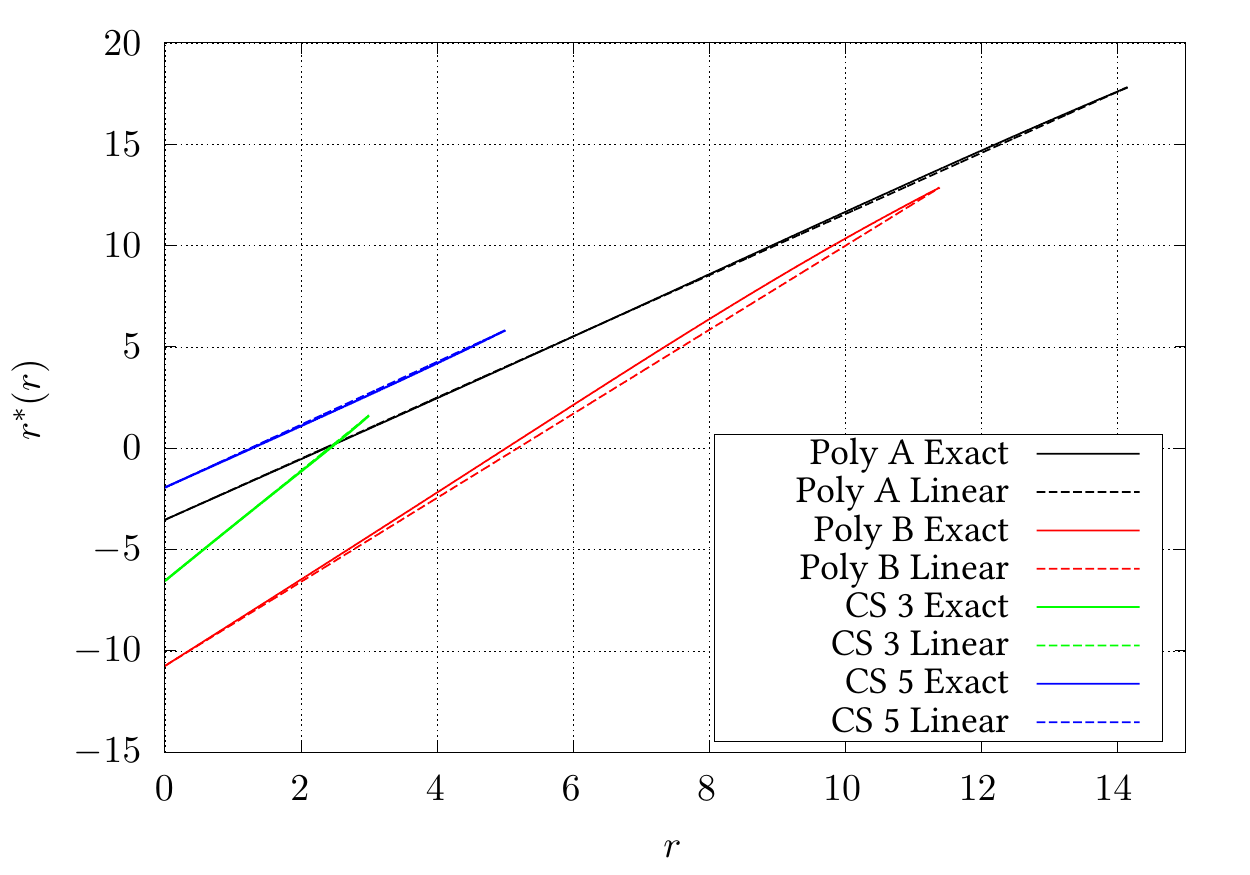}
	\caption{Here we compare the exact tortoise transformation (solid lines) inside the star with the linear approximation (dashed lines). We show results for the two polytrope models A and B along with constant density stars with compactness $R/M=$ 3 and 5 with $M=1$. \label{nur_tort_app}}
\end{figure}

The tortoise coordinate transformation $r^{*}(r)$ is essential in obtaining the simple structure of the wave equation that appears in the perturbations of spherically symmetric and non-rotating space-times. Although its details depend on the metric functions $g_{00}$ and $g_{11}$ via the integral
\begin{align}\label{tortoise_app}
r^{*}(r)  = \int_{}^{r} \sqrt{\frac{g_{11}(r^{\prime})}{g_{00}(r^{\prime})}}\text{d} r^\prime,
\end{align}
it is possible to make some general remarks for stars that are not ultra compact. In this case the ratio in eq. \eqref{tortoise_app} is approximatively constant inside the star and can be matched with the external exact tortoise transformation known from Schwarzschild. In the limit of $M/R \rightarrow 0$ one finds $r^{*} \approx r$. Here we show, as an educated guess, how the tortoise transformation inside the star can be written as a linear function and demonstrate that it is an accurate  description for polytropes and constant density stars, as long as one is not interested in compactnesses beyond normal neutron stars. The linear relation we propose is given by
\begin{align}\label{tortoise_app_2}
r^{*}(r) \approx \left[ R + 2M \ln\left(\frac{R}{2M} - 1\right) -\pazocal{R} \right] + \frac{\pazocal{R}}{R} r,
\end{align}
with $\pazocal{R} = r^{*}(R)-r^{*}(0)$. Note that $r^{*}(R) = R + 2M \ln\left(R/(2M) - 1\right)$ is the standard Schwarzschild result, which we use for matching at $r=R$. The relation in eq. \eqref{tortoise_app_2} recovers the exact values of $r^{*}(r)$ at $r=0$ and $r=R$. Its linearity makes it trivial to make explicit use of it in the perturbation equation, which then simplifies significantly. In Fig. \ref{nur_tort_app} we show the accuracy of the linear approximation for the two polytropes studied in this work, as well as for constant density stars with different compactness. The result clearly shows that the linear approximation is a useful for analytical studies.
\subsection{Approximate Potential Calculation}\label{approximate_pot}
Here we provide additional information for the derivation of the approximated axial mode potential in spherically symmetric and non-rotating neutron stars. As it was shown in \cite{1983ApJS...53...73L}, it is possible to expand all functions that appear in the TOV equations as series for small $r$. The expanded relations for $(\rho(r), P(r), \nu(r), m(r))$ are then given by
\begin{align}
\rho(r) &= \rho_0 + \frac{1}{2} \rho_2 r^2,
\\
P(r)    &= P_0 + \frac{1}{2}P_2 r^2 + \frac{1}{4}P_4 r^4,
\\
\nu(r)	&= \nu_0 + \frac{1}{2} \nu_2 r^2 + \frac{1}{4} \nu_4 r^4,
\\
m(r)	&= \frac{4 \pi}{3} \rho_0 r^3,
\end{align}
where $m(r)$ directly follows from $\rho(r)$. In the same paper it was shown how the series coefficients are related to the equation of state for small $r$. For our application we need the relation for $\nu_2$, which is given as
\begin{align}
\nu_2 = \frac{4 \pi}{3}\left(\rho_0 + 3P_0 \right),
\end{align}
for the notation $g_{00} = e^{2\nu(r)}$, while the factor is $8 \pi/3$ for $g_{00} = e^{\nu(r)}$, as it was used in \cite{1983ApJS...53...73L}. Inserting the series expansions into $V(r)$ and ordering in powers of $r$, one finds
\begin{align}
V(r) &= e^{2 \nu(r) } \left(\frac{L}{r^2}  + 4 \pi \left(\rho(r) - P(r)\right)- \frac{6m(r)}{r^3}\right) 
\\
&\approx \frac{e^{2 \nu_0} L}{r^2} + \exp\left(2 \nu_0 \right) \left( \nu_2 L -4\pi (\rho_0 + P_0 )  \right) + \pazocal{O}(r^2).
\end{align}
Making use of the linearized tortoise transformation presented in appendix \ref{tortoise}, it is possible to find explicitly $V(r^{*})$. To simplify further calculations we introduce a temporary coordinate $u^{*}$, which shifts $r^{*}(0)$ to $u^{*}=0$. The wave equation does not change under such a coordinate shift and the resulting potential is given as
\begin{align}
V(u^{*})  \approx \left(\frac{\pazocal{R}}{R} \right)^2 \frac{e^{2 \nu_0} L}{{u^{*}}^2} + e^{2 \nu_0 } \left[\nu_2 L -4\pi (\rho_0 + P_0 )  \right].
\end{align}
This potential can now be used in the new Bohr-Sommerfeld rule eq. \eqref{new_BS} to obtain a simple relation for the spectrum $n(E_n)$. The integration is straightforward and yields the result presented in eq. \eqref{BS_master}.
%
\end{document}